\documentclass[11pt]{article}
\usepackage{jheppub}
\usepackage{amsmath,amssymb,amsfonts,graphicx,slashed,amsthm,mathtools,upgreek,enumerate,tensor,subfig}
\usepackage[dvipsnames]{xcolor}
\usepackage{arydshln}
\usepackage{comment}
\usepackage{hyperref}
\usepackage[utf8]{inputenc}
\usepackage[titletoc]{appendix}

\numberwithin{equation}{section}
\allowdisplaybreaks

\allowdisplaybreaks

\newcommand{\be}{\begin{equation}}
\newcommand{\ee}{\end{equation}}
\newcommand{\f}{\frac}
\newcommand{\s}{\sqrt}
\newcommand{\p}{\partial}

\newcommand{\bea}{\begin{eqnarray}}
\newcommand{\eea}{\end{eqnarray}}
\newcommand{\ba}{\begin{align}}
\newcommand{\ea}{\end{align}}

\newcommand{\la}{\langle}
\newcommand{\ra}{\rangle}
\newcommand{\beq}{\begin{equation}}
\newcommand{\eeq}{\end{equation}}

\newcommand{\bz}{\bar{z}}
\newcommand{\bp}{\bar{\partial}}

\title{de Sitter space is sometimes not empty}

\author[a,b,c]{Vijay Balasubramanian}
\author[d,e,f,g]{\!, Yasunori Nomura}
\author[h]{\!, Tomonori Ugajin}

\affiliation[\,a]{David Rittenhouse Laboratory, University of Pennsylvania,
209 S.33rd Street, Philadelphia, PA 19104, USA}

\affiliation[\,b]{Theoretische Natuurkunde, Vrije Universiteit Brussel (VUB), and International Solvay Institutes, Pleinlaan 2, B-1050 Brussels, Belgium}

\affiliation[\,c]{Santa Fe Institute, 1399 Hyde Park Road, Santa Fe, NM 87501, USA}

\affiliation[\,d]{Berkeley Center for Theoretical Physics, Department of Physics, 
University of California, Berkeley, CA 94720, USA}

\affiliation[\,e]{Theoretical Physics Group, Lawrence Berkeley National Laboratory, Berkeley, CA 94720, USA}

\affiliation[\,f]{RIKEN Interdisciplinary Theoretical and Mathematical Sciences Program (iTHEMS), Wako, Saitama 351-0198, Japan}

\affiliation[\,g]{Kavli Institute for the Physics and Mathematics of the Universe (WPI), 
UTIAS, The University of Tokyo, Kashiwa, Chiba 277-8583, Japan}

\affiliation[\,h]{Center for Gravitational Physics,
Yukawa Institute for Theoretical Physics, Kyoto University,
Kitashirakawa Oiwakecho, Sakyo-ku,
Kyoto 606-8502, Japan}

\abstract{
Multiple lines of evidence suggest that the Hilbert space of an isolated de~Sitter universe is one dimensional but can appear larger when probed by a gravitating observer.
To test this idea, we compute the von~Neumann entropy of a field theory in a two-dimensional de~Sitter universe which is entangled in a thermal-like state with the same field theory on a disjoint, asymptotically anti-de~Sitter (AdS) black hole.
Previously, it was shown that the replica trick for computing the entropy of such entangled gravitating systems requires the inclusion of a non-perturbative effect in quantum gravity---novel wormholes connecting the two spaces.
Here we show that:\ (a) the expected wormholes connecting de~Sitter and AdS universes exist, avoiding a no-go theorem via the presence of  sources on the AdS boundary; (b) the entanglement entropy vanishes if the nominal entropy of the de~Sitter cosmological horizon ($S_{\rm dS} = A_{\rm horizon}^{\rm dS} / 4G_{\rm N}$) is less than the entropy of the AdS black hole horizon ($S_{\rm BH} = A_{\rm horizon}^{\rm AdS} / 4G_{\rm N}$), i.e., $S_{\rm dS} < S_{\rm BH}$; (c) the entanglement entropy is finite when $S_{\rm dS} > S_{\rm BH}$.
Thus, the de~Sitter Hilbert space is effectively nontrivial only when $S_{\rm dS} > S_{\rm BH}$.
The AdS black hole we introduce can be regarded as an ``observer'' for de~Sitter space.
In this sense, our result is a non-perturbative generalization of the recent perturbative argument that the algebra of observables on the de~Sitter static patch becomes nontrivial in the presence of an observer.
}


\begin{document}

\maketitle

\section{Introduction}

Gibbons and Hawking argued that the cosmological horizon of de~Sitter space acts as if it carries a thermodynamic entropy $S_{\rm dS}$ proportional to the horizon area~\cite{Gibbons:1977mu}.
If this entropy has a conventional interpretation in terms of the dimension of the microscopic Hilbert space, it should be possible to entangle the corresponding microstates with a reference universe, creating a state which appears, to observers in either universe, to have a large von~Neumann entropy.

Recent work~\cite{Chen:2020tes,Hartman:2020khs,Balasubramanian:2020xqf,Sybesma:2020fxg,Geng:2021wcq,Fallows:2021sge,Aalsma:2021bit,Langhoff:2021uct,Aguilar-Gutierrez:2021bns,Kames-King:2021etp,Aalsma:2022swk,Baek:2022ozg} has tested this idea using methods pioneered in Refs.~\cite{Penington:2019npb,Almheiri:2019psf,Almheiri:2019hni} for studying the entropy of Hawking radiation.
These results suggest that when the reference universe does not gravitate, the entropy of entanglement is zero, implying that the de~Sitter Hilbert space is one dimensional, so that there is no way to form an entangled state with an external reference system.%
\footnote{
 Reference~\cite{Balasubramanian:2020xqf} also proposed scenarios leading to a finite de~Sitter entropy.
 As we will discuss later, these settings involve the presence of a black hole and/or end-of-the-world brane truncating the locally de~Sitter geometry, and could be regarded as introducing ``gravitational observers'' of the de~Sitter static patch similarly to the present paper.
 For another way of seeing a finite de~Sitter entropy in a gravitational setting, see Ref.~\cite{Geng:2021wcq}.
 }
If this is true, there is a tension with naive statistical interpretations of the Gibbons-Hawking argument for the entropy of cosmological horizons~\cite{Gibbons:1977mu}.
Related arguments suggest that the Hilbert space of any closed universe is trivial~\cite{Marolf:2020xie,McNamara:2020uza}.
Another idea, called static patch holography, states that the de~Sitter static patch can be described in terms of a nontrivial Hilbert space of degrees of freedom living on the cosmological horizon (see Refs.~\cite{Nomura:2016ikr,Nomura:2017fyh,Nomura:2019qps,Murdia:2022giv} and Refs.~\cite{Susskind:2021omt,Susskind:2021esx,Shaghoulian:2021cef,Shaghoulian:2022fop}).
Earlier approaches to de~Sitter holography, some of which use structures at past and future infinity, include Refs.~\cite{Strominger:2001pn,Balasubramanian:2001nb, Balasubramanian:2002zh,Alishahiha:2004md,Alishahiha:2005dj,Dong:2018cuv,Gorbenko:2018oov,Aalsma:2020aib,Geng:2020kxh,Aalsma:2021kle,Svesko:2022txo,Franken:2023pni}.

To elucidate these issues concerning the de~Sitter Hilbert space, in this paper we apply the methods of Ref.~\cite{Balasubramanian:2021wgd} to study entanglement of a de~Sitter universe with a {\it gravitating} reference system.
We choose the reference to be a black hole in anti-de~Sitter (AdS) space and consider entangled states of the form
\be
  |\Psi \ra = \sum_{i=1}^{\infty} \s{p_{i}}\, | \psi_{i} \ra_{A}  |\psi_{i} \ra_{B},
\quad
  \sum_{i=1}^{\infty} p_{i} = 1,
\label{eq:TFD-stateonAB1}
\ee
where $A$ and $B$ are de~Sitter space and an AdS black hole, respectively.
We assume that both universes contain the same quantum field theory (QFT), whose states are being entangled.
We then compute the von~Neumann entropy by a replica trick involving gravitational path integrals (see~\cite{Penington:2019kki,Almheiri:2019qdq,Geng:2020fxl,Anderson:2020vwi,Geng:2021iyq,Balasubramanian:2021xcm,Iizuka:2021tut,Miyata:2021qsm,Geng:2021mic} for related approaches).

Recent work has proposed that when applying the replica method to the entropy of a gravitating spacetime entangled with a non-gravitating reference, we must include the contributions of novel wormholes that connect the replica copies of the gravitating system in order to obtain results consistent with unitarity~\cite{Penington:2019kki,Almheiri:2019qdq}.
This suggests that  when the reference system is also gravitating, we should include additional wormholes connecting the original system to the reference, if such wormholes exist.
Indeed, Ref.~\cite{Balasubramanian:2021wgd} showed that this procedure is necessary when both the original universe and the reference with which it is entangled are AdS black holes. In this case, the entanglement is quantified by a generalized entropy in a new spacetime constructed by appropriately gluing two AdS black holes.

In this paper, we study what happens when we replace one of the two AdS black holes with de~Sitter space.
In this case, the boundary conditions of the relevant gravitational path integrals are imposed in Lorentzian spacetime at the asymptotic boundary at spatial infinity on the AdS side and at future/past infinity on the de~Sitter part of the geometry.
While we mostly work in a two-dimensional theory of gravity for the sake of calculability, our basic argument does not seem to rely crucially on the low dimensional nature of the  model, so we expect that our findings will persist in higher dimensions.

We begin by showing that there is a Euclidean wormhole connecting de~Sitter and AdS spacetimes in our setting. This wormhole can be viewed as a Euclidean AdS black hole which contains a de~Sitter false vacuum bubble, separated from the true vacuum region by a domain wall.
Our construction evades an argument of Fu and Marolf~\cite{Fu:2019oyc} forbidding such solutions, by including the effects of sources on the AdS boundary.
These sources are required in our setting to prepare the excitations $|\psi_i\rangle$ in the entangled state (\ref{eq:TFD-stateonAB1}) of the bulk QFT.
The backreaction of the matter stress tensor coming from these entangled field theory degrees of freedom allows us to explicitly construct a wormhole solution with the desired properties in two-dimensional Jackiw-Teitelboim (JT) gravity.
Thus, the entanglement of the matter degrees of freedom between the two systems is essential to our construction.
When the entanglement is large, the effect of our boundary sources can be described as injecting localized domain walls (particles in two dimensions) at the boundary of Euclidean AdS.
These walls propagate a short distance into the bulk and then decay into the domain wall surrounding a de~Sitter bubble.
In this limit, our approach resembles the construction of de~Sitter bubbles within AdS by Mirbabayi~\cite{Mirbabayi:2020grb}.
(Also see the recent works~\cite{Sahu:2023fbx,Aguilar-Gutierrez:2023odp,Antonini:2023hdh} which construct AdS big bang-big crunch cosmologies as bubbles behind an AdS black hole horizon.)
When continued to Lorentzian signature, the resulting wormhole describes an AdS black hole with an inflating region in its interior.
Such spacetimes were introduced to study inflation in AdS/CFT~\cite{Freivogel:2005qh} and to try to find a way of creating a universe in a lab~\cite{Farhi:1986ty, Farhi:1989yr}; see also~\cite{Sato:1981bf,Maeda:1981gw,Sato:1981gv,Kodama:1982sf} for earlier work.
The authors of Refs.~\cite{Anninos:2017hhn,Witten:2020ert,Anninos:2020cwo,Chapman:2021eyy,Ecker:2022vkr} have also studied how to realize two-dimensional de~Sitter space inside AdS$_2$ space in dilaton gravity.

As we will see, in our setting the entanglement  between the two systems is quantified by a generalized entropy computed on the dominant saddle of the replica path integral. We find two phases when the QFT state is strongly entangled.
First, if the area of the cosmological horizon is smaller than that of the AdS black hole, so that the Bekenstein-Hawking entropies of de~Sitter ($S_{\rm dS}$) and the black hole ($S_{\rm BH}$) satisfy $S_{\rm dS} < S_{\rm BH}$, the saddlepoints consistent with the boundary conditions do not have wormholes between the de~Sitter and AdS components.
In this case, the entanglement entropy is computed by the same ``island formula'' that appears when the reference spacetime is non-gravitating~\cite{Balasubramanian:2020xqf}, and the entropy vanishes, suggesting a one-dimensional de~Sitter Hilbert space.
In the opposite case, when $S_{\rm dS} > S_{\rm BH}$, and the bulk QFT is strongly entangled, the dominant saddlepoint includes a wormhole between the de~Sitter and AdS universes.
The entanglement entropy is then given by the AdS black hole entropy, suggesting that the de~Sitter Hilbert space is nontrivial.

The AdS black hole that we introduced can be regarded as an observer for de~Sitter space:\ it probes the de~Sitter degrees of freedom using entanglement with its own degrees of freedom.
Our result concerning the entanglement entropy described above is consistent with the recent claim that in the presence of a gravitating observer the algebra of observables on a de~Sitter static patch becomes nontrivial~\cite{Chandrasekaran:2022cip}. In our setup, the gravitational interaction between de~Sitter space and the observer is embodied by the wormhole connecting them.
Our result implies that for an observer who consists of more than $S_{\rm dS}$ qubits, the entanglement wedge covers an entire de~Sitter time slice, and accordingly the entanglement entropy vanishes.
On the other hand, for an observer consisting of less than $S_{\rm dS}$ qubits, the static patch Hilbert space can be probed using the observer's Hilbert space, leading to a non-vanishing entanglement entropy.
This is consistent with Ref.~\cite{Chandrasekaran:2022cip} where the gravitating observer is a point particle, and hence has access to fewer than $S_{\rm dS}$ qubits.

Five sections follow.
In Section~\ref{sec:replicadSAdS}, we set up our model which involves de~Sitter space, a black hole in AdS, and an entangled QFT state defined on these two spacetimes.
We then discuss the replica calculation of the entanglement entropy using gravitational path integrals.
A novel feature of the calculation is the appearance of wormholes connecting the two universes.
In Section~\ref{sec:dS-AdS-wormhole}, we explain how to construct these wormhole solutions in the simplest case, namely Einstein gravity coupled with a codimension-one domain wall.
Additionally we include the backreaction of the QFT stress tensor of the entangled state~\eqref{eq:TFD-stateonAB1}.
We explicitly construct the Euclidean wormhole solution in two-dimensional JT gravity and show that in order to connect de~Sitter and AdS, the backreaction of the entangled state is necessary.
In Section~\ref{sec:genentropy}, we discuss the behavior of the entanglement entropy by assembling the results obtained in previous sections.
In Section~\ref{sec:algebras}, we interpret our results, and discuss the relation with algebraic perspective on de~Sitter entropy offered in Ref.~\cite{Chandrasekaran:2022cip}.

\section{Replica calculation of the entanglement entropy}
\label{sec:replicadSAdS}

We are interested in entanglement between states on de~Sitter space and those on a black hole in AdS space.
In particular, we would like to understand the role of possible wormholes connecting these two spacetimes in light of Refs.~\cite{VanRaamsdonk:2010pw,Maldacena:2013xja} when the entanglement between the degrees of freedom in the two spacetimes is large.

\subsection{Entanglement between two AdS black holes}
\label{subsec:2AdS-review}

Before discussing the entanglement between de~Sitter and AdS spaces, we review the case where both spacetimes have AdS asymptotics and contain black holes.
This situation was studied in Ref.~\cite{Balasubramanian:2021wgd}, where the following thermofield-double-like entangled QFT state on two AdS spaces was considered:
\be
  |\Psi \ra = \sum_{i=1}^{\infty} \s{p_{i}}\, | \psi_{i} \ra_{A}  |\psi_{i} \ra_{B},
\quad
  p_{i} = \f{e^{-\beta E_{i}}}{\sum_j e^{-\beta E_j}}.
\label{eq:TFD-state_AdS-AdS}
\ee
Here, $A$ and $B$ are both gravitating, asymptotically AdS spacetimes, and $\{| \psi_{i} \ra_{A,B} \}_{i=0}^{\infty}$ are energy eigenstates in QFT$_{A,B}$, which we take to be the same conformal field theory (CFT).
For calculational convenience, we will focus on the situation where both $A$ and $B$ are two dimensional.

The entanglement entropy $S(\rho_{A})$ between two ``universes'' $A$ and $B$ can be computed using the replica trick
\be
  S(\rho_{A}) = \lim_{n \rightarrow 1} \f{1}{1-n} \ln {\rm tr} \rho_{A}^{n},
\quad
  \rho_{A} = \sum_{i,j} \s{p_{i}p_{j}}\, \la \psi_{i}|\psi_{j} \ra_{B} \; | \psi_{j} \ra_{A} \la  \psi_{i} |.
  \label{eq:replicatrick1}
\ee
The trace quantity appearing here is given by 
\be
  {\rm tr} \rho_{A}^{n} = \f{1}{Z_{1}^{n}} \sum_{\{i_{k},j_{k}\}} \prod_{k=1}^{n} \s{p_{i_{k}}p_{j_{k}}}\, \la \psi_{i_{k}}| \psi_{j_{k+1}} \rangle_{A_k} \la\psi_{i_{k}}| \psi_{j_{k}} \rangle_{B_k} \equiv \f{Z_{n}}{Z_{1}^{n}},
\label{eq:Renyi}
\ee
where $| \psi_{i_{n+1}} \rangle_{A_n} \equiv | \psi_{i_{1}} \rangle_{A_n}$, and
\be
  Z_{1} = \sum_{i,j} \s{p_{i} p_{j}}\, \la \psi_{i} | \psi_{j} \ra_{A} \la \psi_{i} | \psi_{j} \ra_{B}.
\label{eq:normalization}
\ee
The overlaps $\la \psi_{i} | \psi_{j} \ra_{A,B}$ can be computed by a gravitational path integral on Euclidean AdS space (a disk in two dimensions) with two local operator insertions on the boundary
\be
  \langle \psi_{i} | \psi_{j} \rangle_{A} = \langle \psi_{i}(\infty) \psi_{j}(0) \rangle_{{\rm disk}}.
\ee
Note that the overlap between two QFT energy eigenstates $| \psi_{i} \rangle_{A}$ and $| \psi_{j} \rangle_{A}$ ($i \neq j$) does not necessarily vanish in the presence of gravity, reflecting the fact that these eigenstates are overcomplete.
Since the right-hand side of Eq.~\eqref{eq:Renyi} contains a product of $2n$ overlaps, the gravitational path integral involves $2n$ copies of the universes, i.e., $2n$ copies of the disk. The path integral can thus include contributions from saddles in which the copies of the universes are connected by wormholes.

The rule for computing the gravitational path integral \eqref{eq:Renyi} in the semiclassical approximation is to include all saddles consistent with the conditions imposed on the boundaries of each universe, $\{ \p A_{k}, \p B_{k} \}$.%
\footnote{
 This corresponds to taking an ensemble average of the quantity in question over microstates that cannot be discriminated at the semiclassical level.
 This makes, for example, the contribution to $Z_1$ in Eq.~\eqref{eq:normalization} from terms with $i \neq j$ non-vanishing, despite the fact that the same calculation for each factor provides a vanishing result~\cite{Penington:2019kki}.
 While semiclassical calculation using the gravitational path integral involves an ensemble averaging, applying it in the context of the replica trick correctly reproduces the entanglement entropy of a microstate; see Ref.~\cite{Murdia:2022giv} for discussion of this point.
\label{ft:ensemble}
}
Ref.~\cite{Balasubramanian:2021wgd} examined the possible saddles for the gravitational path integral.
For example, there are saddles in which all the boundaries are disconnected in the bulk, giving the thermal entropy $S_{\rm th}$.
Another saddle connects all copies of $A$ via a wormhole, while all copies of $B$ remain disconnected.
The authors of Ref.~\cite{Balasubramanian:2021wgd} argued that in the high temperature limit $\beta \rightarrow 0$, the dominant saddle connects all the copies of $A$ and $B$ through a single wormhole.
This is because in this saddle, the indices $\{i_{k},j_{k}\}$ on the right-hand side of Eq.~\eqref{eq:Renyi} labeling QFT excited states contract to form a single loop, giving a large combinatorial factor.
In other saddles, these indices do not form a single loop, giving at least one Kronecker delta on the right-hand side of Eq.~\eqref{eq:Renyi}, significantly reducing the value of the sum in the high temperature limit.
As a consequence, they cannot dominate the gravitational path integral.

To illustrate this argument, let us consider the simplest example, i.e.\ the calculation of $Z_{1}$ defined in Eq.~\eqref{eq:normalization}.
The gravitational path integral contains two saddles:\ the disconnected saddle consisting of two disjoint disks and the connected saddle in which two disks are connected by a wormhole.
Thus, we write $Z_{1} = Z_{1,{\rm disconn}} + Z_{1,{\rm conn}}$, where the first term is
\be
  Z_{1,{\rm disconn}} = e^{-S_{{\rm grav}}[A] - S_{{\rm grav}}[B]},
  \label{eq:zdisconn}
\ee
since on this saddle the overlaps are proportional to $\delta_{ij}$.
On the other hand, the contribution from the connected saddle $A\#B$ reads
\be
  Z_{1,{\rm conn}} = e^{-S_{{\rm grav}}[A\#B]} \sum_{i,j} \s{p_{i}p_{j}}\, \la \psi_{i}(\infty_{A}) \psi_{j}(0_{A}) \psi_{i}(\infty_{B}) \psi_{j}(0_{B}) \ra_{A\#B},
\label{eq:zconn}
\ee
where the QFT four point function on the right-hand side is evaluated on $A\#B$.
When both $A$ and $B$ are asymptotically AdS, the connected saddle $A\#B$ depends on two moduli parameters:\ the renormalized length $\ell$ and twist angle $\tau$ between the two disk boundaries $\p A$ and $\p B$.
Reference~\cite{Balasubramanian:2021wgd} argued that in the limit $\beta \rightarrow 0$, the moduli parameters of the wormhole dominating the gravitational path integral satisfy $\ell \rightarrow 0$ and $\tau \rightarrow 0$.
This corresponds to the OPE channel $\psi_{i}(\infty_{A}) \rightarrow \psi_{i}(\infty_{B})$ and $\psi_{j}(0_{A}) \rightarrow \psi_{j}(0_{B})$, which makes the above four point function factorize
\be
  Z_{1,{\rm conn}} =e^{- S_{{\rm grav}}[A\#B]} Z_{{\rm CFT}}[A/B]^{2}, \quad Z_{{\rm CFT}}[A/B] = \sum_{i} \s{p_{i}} \; \la \psi_{i} (\infty_{A}) \psi_{i} (\infty_{B}) \ra_{A/B},
\label{eq:zcft}
\ee
where $\la \psi_{i} (\infty_{A}) \psi_{i} (\infty_{B}) \ra = \la \psi_{i} | \psi_{i} \ra_{A/B}$ is a two-point correlator evaluated on a space $A/B$, called a swap wormhole in Ref.~\cite{Balasubramanian:2021wgd}, obtained by merging half of $A$ and half of $B$.
Since $Z_{{\rm CFT}}[A/B]^{2}$ grows indefinitely for $\beta \rightarrow 0$, $Z_{1,{\rm conn}}$ dominates over $Z_{1,{\rm disconn}}$ in this limit.

By including the saddle in which all the copies of $A$ and $B$ are connected by a single wormhole, we find that the entanglement entropy $S(\rho_{A})$ of universe $B$ is given by
\be
  S(\rho_{A}) = {\rm Min} \{S_{\rm th},S_{\rm swap}(\rho_{A})\}, \quad S_{\rm swap}(\rho_{A}) = \underset{I}{\rm MinExt}\left[ \f{ {\rm Area}(A/B,\partial I)}{4G_{\rm N}} + S_{\rm eff}(I) \right].
\label{eq:dSAdSentropyformula}
\ee 
This expression is closely related to the island formula for the entropy of evaporating black holes~\cite{Penington:2019npb,Almheiri:2019psf,Almheiri:2019hni}, but there is one significant difference.
In the above formula, the minimization and extremization in $S_{\rm swap}(\rho_{A})$ must be done in the swap wormhole $A/B$ in which the geometry of the universe $A$ and $B$ are glued.
Examples of such glued spacetimes are constructed in Refs.~\cite{Balasubramanian:2021wgd,Miyata:2021qsm}.

\subsection{Entangling de~Sitter and AdS spaces}
\label{subsec:dS-AdS}

The main goal of this paper is to use methods similar to those described above to compute entanglement entropy for a state similar to that in Eq.~\eqref{eq:TFD-state_AdS-AdS} when one of the disjoint universes, $A$, is de~Sitter space and the other universe, $B$, is an AdS black hole.
We begin by discussing the relevant entangled state.

To define a state of the form in Eq.~\eqref{eq:TFD-state_AdS-AdS}, suppose that the low energy effective field theories in the two universes are the same CFT.
The states $| \psi_{i} \ra_{A}$ in de~Sitter space are then prepared by performing the Euclidean path integral on half of a ($d+1$)-dimensional sphere $S^{d+1}$ with the CFT operator $O_i$ corresponding to $| \psi_{i} \ra_{A}$ inserted at a pole, e.g., on the pole of the southern ``hemisphere.''
Here, $d$ is the number of spatial dimensions.
Similarly, the states $| \psi_{i} \ra_{B}$ are prepared by a Euclidean path integral on half of a ball $B^{d+1}$ with ${\cal O}_i$ inserted at the pole.
In this paper, we mostly focus on two-dimensional spacetimes, in which case the relevant geometries are a half sphere $S^2/Z_2$ and a half disk $B^2/Z_2$.

In general, we can consider an entangled state of the form
\be
  |\Psi \ra = \sum_{i=1}^{\infty} \s{p_{i}}\, | \psi_{i} \ra_{A}  |\psi_{i} \ra_{B},
\quad
  \sum_{i=1}^{\infty} p_{i} = 1.
\label{eq:TFD-general}
\ee
For definiteness, however, we take
\begin{equation}
  p_i = \frac{e^{-\beta \Delta_i}}{\sum_j e^{-\beta \Delta_j}},
\label{eq:p_i-beta}
\end{equation}
where $\Delta_i$ is the conformal dimension of the operator $O_i$.
We are mostly interested in the ``high temperature'' limit $\beta \rightarrow 0$ of this state,%
\footnote{
 This temperature characterizes the strength of the entanglement and is not related to the temperature of de~Sitter space or the AdS black hole.
}
but the conclusions we draw for this limit also apply to more general entangled states of the form in Eq.~\eqref{eq:TFD-general} if $|\Psi \ra$ receives contributions from a sufficiently large number of $| \psi_{i} \ra$'s. 

Note that while we insert operators on a background spacetime, the excited states effectively include microstates of the geometry, not just perturbative excitations in the semiclassical theory, when gravity is turned on.  This is because the microstates can also be viewed as excitations of blueshifted, i.e., locally high energy, excitations of low energy quantum fields; see, e.g., Ref.~\cite{Murdia:2022giv}.
A related phenomenon is that in the low energy QFT description of the entanglement of a subregion of a gravitating spacetime, there is a scheme dependence in the attribution of the contribution from the region near the entangling surface to the geometrical area term or power-divergent contribution from low energy fields~\cite{Susskind:1994sm,Cooperman:2013iqr}.
See also~\cite{Goel:2018ubv,Balasubramanian:2022gmo} for related discussions for two-dimensional black holes.
When we include the effects of quantum gravity, the states $| \psi_{i} \ra$'s become overcomplete.
The information about the correct number of states, however, can still be extracted by using the island formula~\cite{Penington:2019npb,Almheiri:2019psf,Almheiri:2019hni} or equivalently including the contribution from replica wormholes in path integrals~\cite{Penington:2019kki,Almheiri:2019qdq,Chandra:2022fwi,Balasubramanian:2022gmo,Balasubramanian:2022lnw}, which we study here (see also footnote~\ref{ft:ensemble}).

Our goal is to compute the entanglement entropy of the state in Eq.~\eqref{eq:TFD-general}.
We can follow the same steps used in Section~\ref{subsec:2AdS-review} to compute the entanglement entropy between two AdS black holes using the replica trick.
The relevant expressions Eqs.~\eqref{eq:replicatrick1}--\eqref{eq:normalization} contain terms that are products of overlaps of excited states such as $\la \psi_{i} | \psi_{j} \ra_{A} \la \psi_{i} | \psi_{j} \ra_{B}$ and $\prod_{k=1}^{n} \s{p_{i_{k}}p_{j_{k}}}\, \la \psi_{i_{k}}| \psi_{j_{k+1}} \rangle_{A_k} \la\psi_{i_{k}}| \psi_{j_{k}} \rangle_{B_k}$, where the subscript $A$ indicates the de~Sitter factor and the subscript $B$ indicates the AdS factor.
We will use the saddlepoint approximation to the Euclidean gravity path integral to compute these products of overlaps.
As described above, each product will then be given by a sum of contributions of saddlepoints of different topologies, some of which may be wormholes connecting de~Sitter factors, AdS factors, or de~Sitter and AdS factors.
Below we describe this zoo of possibilities and explain which ones are expected to dominate the saddlepoint sum.

\subsection{Gravitational path integrals for state overlaps}
\label{subsubsec:infty}

We want to calculate products of overlaps like $\la \psi_{i} | \psi_{j} \ra_{A} \la \psi_{k} | \psi_{l} \ra_{B}$, and more generally products that include multiple factors of each type of universe like $\la \psi_{i} | \psi_{j} \ra_{A} \la \psi_{k} | \psi_{l} \ra_{A} \la \psi_{m} | \psi_{n} \ra_{B} \la \psi_{o} | \psi_{p} \ra_{B}$.
As we discussed above, the states on $A$ (de~Sitter space) are prepared by the Euclidean path integral over a hemisphere with an operator applied at the pole, while the states on $B$ (AdS space) are prepared by the Euclidean path integral over a half-disk with an operator placed again at the pole.
But there is a critical difference between these constructions:\ in the AdS case, the operator is placed at the spacetime boundary, but in the de~Sitter case, Euclidean de~Sitter is a compact manifold without boundary.
This difference raises a question for how we should calculate the path integrals for the overlaps above.
If both $A$ and $B$ were asymptotically AdS, then both would have spacetime boundaries, and there is a well-tested rule motivated by the AdS/CFT correspondence for the saddlepoint approximation to the Euclidean gravity path integral:\ include all saddlepoints consistent with the boundary conditions imposed at the boundary loci.
However, there is no asymptotic boundary in de~Sitter space, and we do not have the guide of a holographic dual; hence, the rule for selecting saddles is not a~priori clear.

For Lorentzian de~Sitter space, boundary conditions are naturally imposed at future and past infinities, as suggested, e.g., by the dS/CFT correspondence~\cite{Strominger:2001pn,Balasubramanian:2001nb}.
We thus interpret our state preparation algorithm as selecting a state in these asymptotic regions.
Since the gravitational path integrals computing the overlaps of interest, Eq.~\eqref{eq:Renyi}, involve both Euclidean de~Sitter and AdS spacetimes, our working hypothesis is that we should select saddles whose continuations to Lorentzian signature contains future and past infinities for de~Sitter space, in addition to the conformal boundaries for AdS space.
In the Euclidean regime, this rule requires that the saddles contain the cosmological horizon (the fixed point of the U(1) isometry of dS$_2$) so that after the continuation to Lorentzian signature the geometry will contain a future/past infinity.

With this hypothesis, the calculation of the entanglement entropy of the state in Eq.~\eqref{eq:TFD-general} with  Eq.~\eqref{eq:p_i-beta} proceeds as in Section~\ref{subsec:2AdS-review}:\ we sum over all saddlepoint topologies for the given boundary conditions, which here includes the requirement that the de~Sitter past and future infinities are contained in the Lorentzian continuation.

\begin{figure}
\centering
\hspace{0.8 cm}
\includegraphics[width=5.6cm]{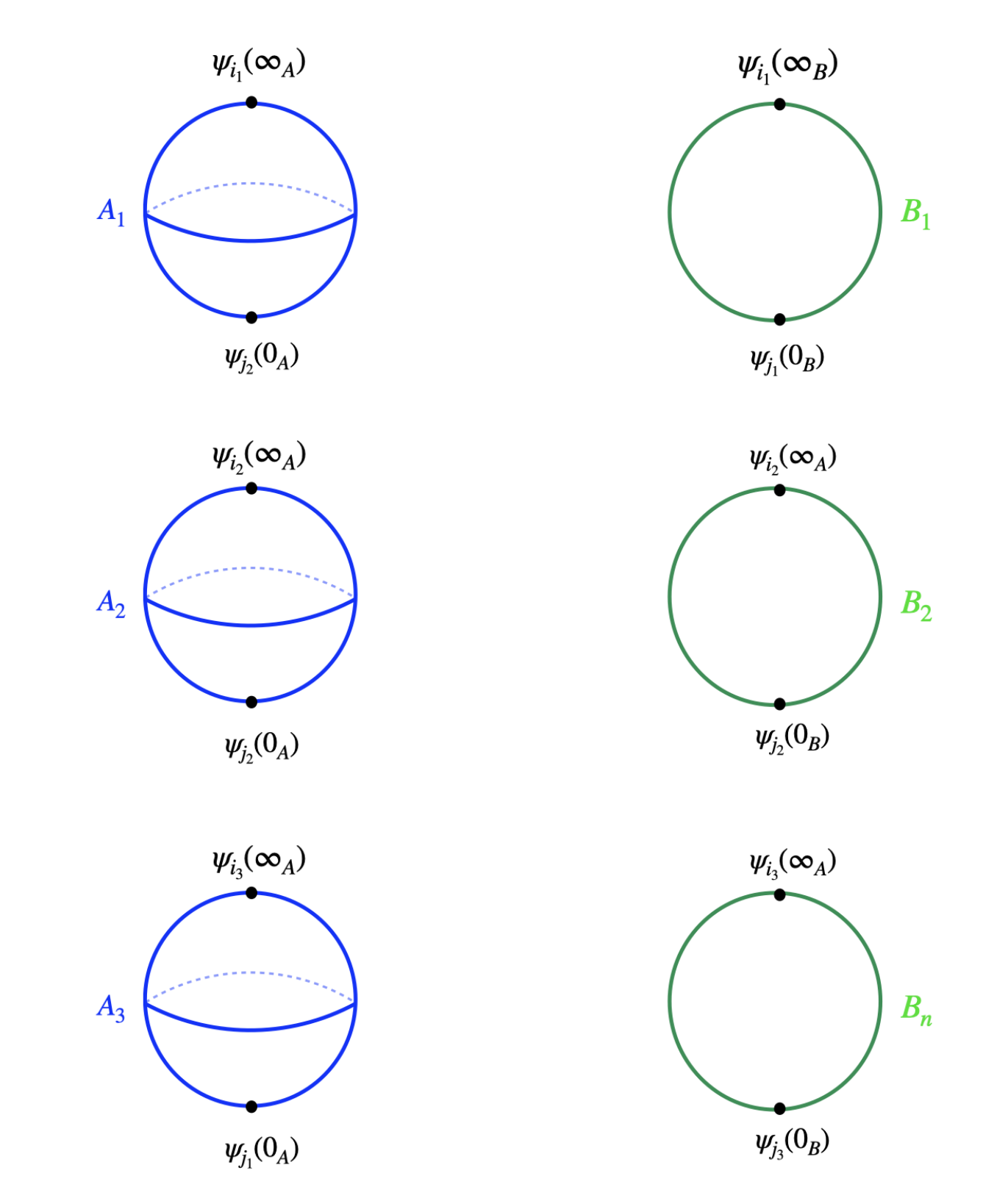}
\hspace{1 cm}
\includegraphics[width=7.2cm]{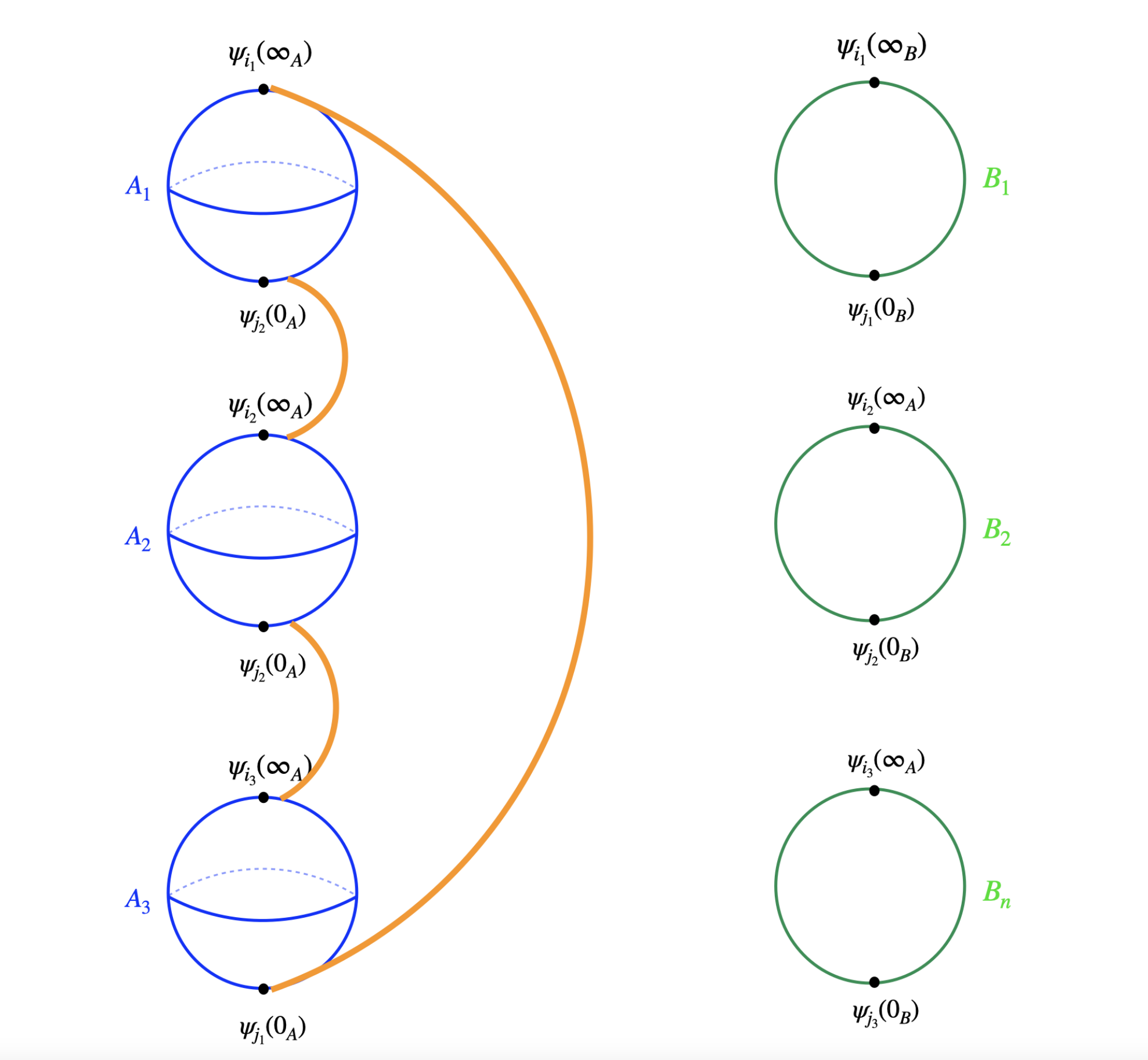}
\hspace{1 cm} \\
(a) \hspace{6.7 cm} (b) \\
\includegraphics[width=6.8cm]{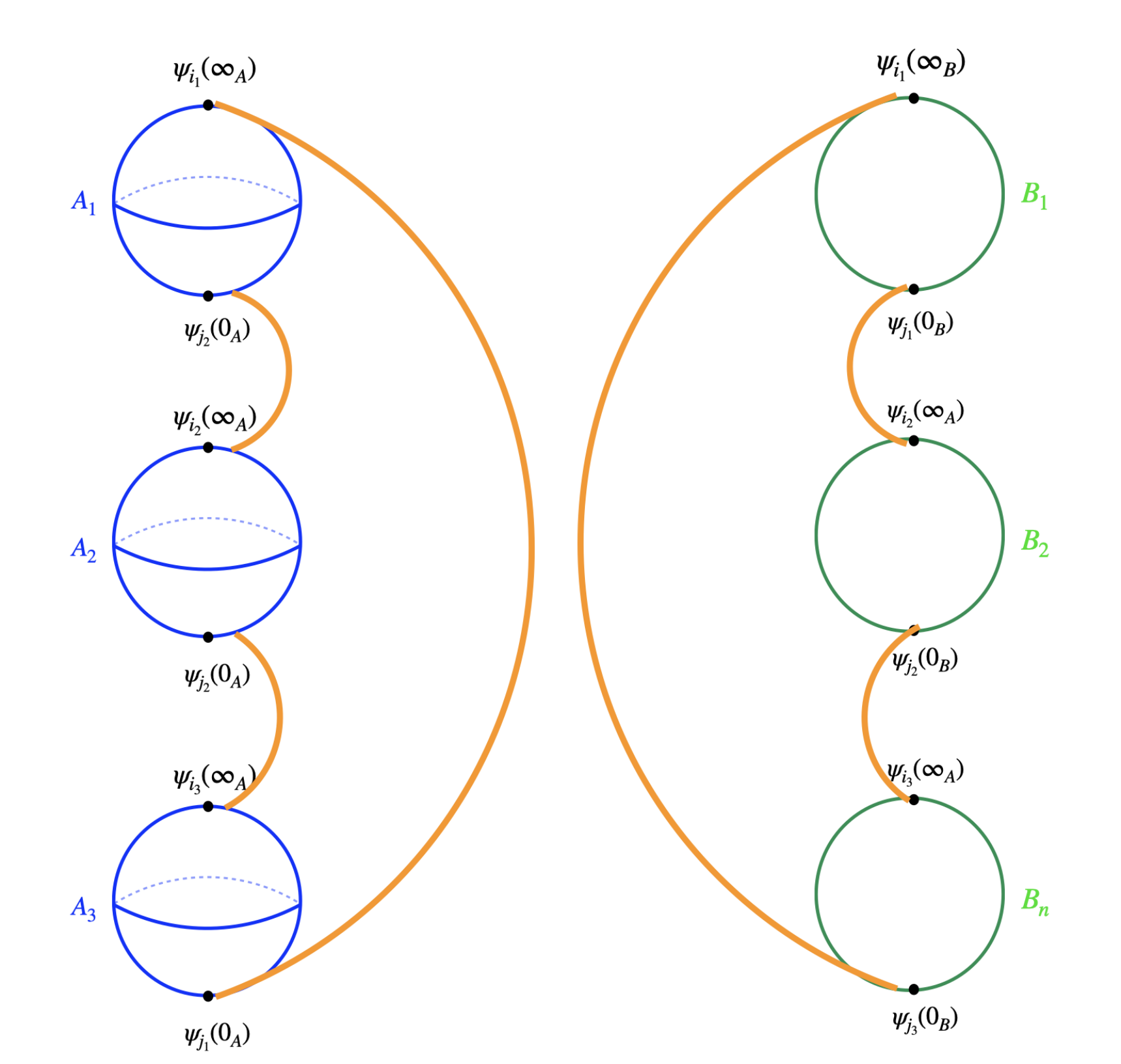}
\hspace{0.5 cm}
\includegraphics[width=6.8cm]{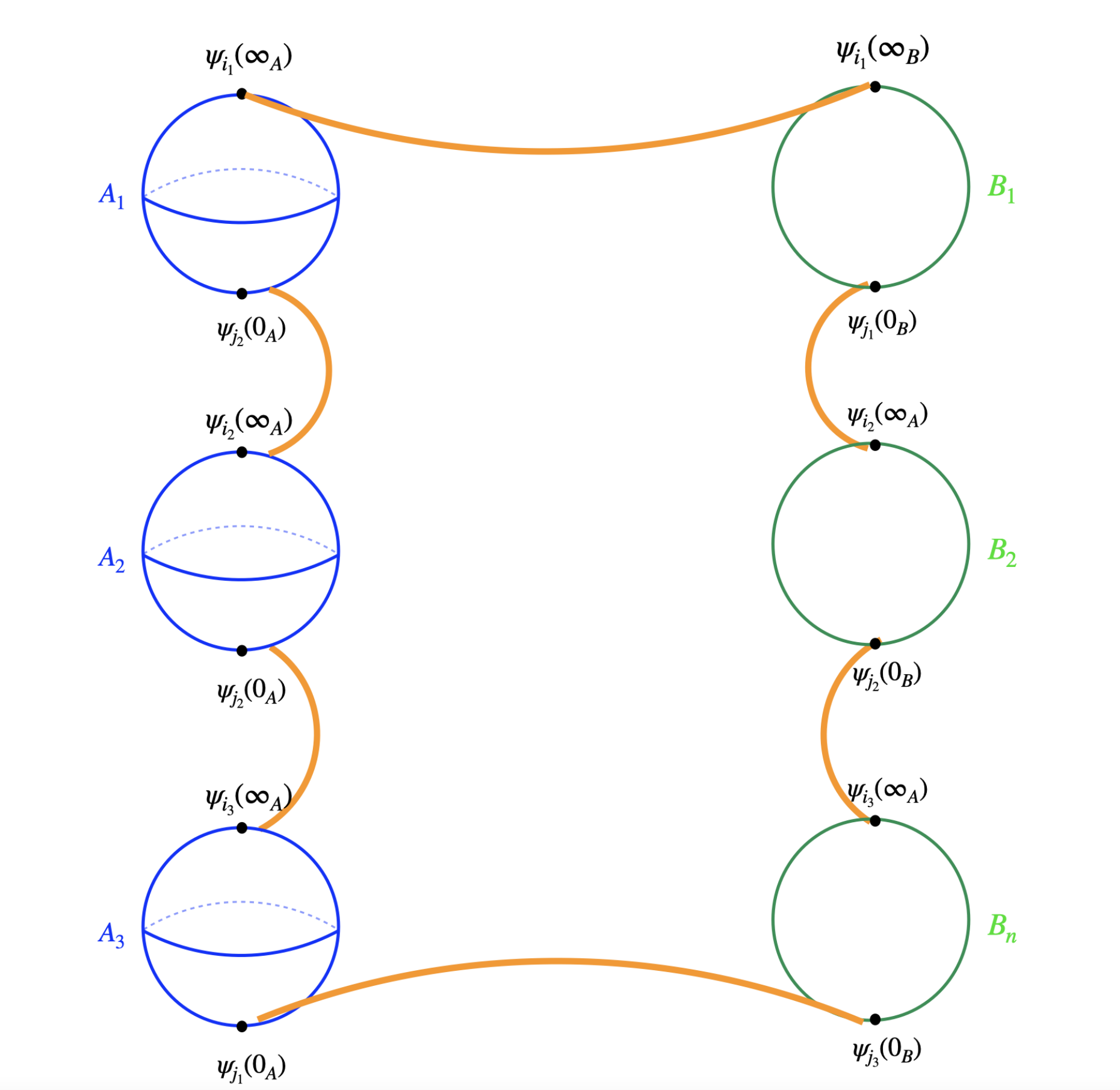} \\
(c) \hspace{6.7 cm} (d) \\
\caption{\small{
 Possible gravitational saddles of the $n=3$ R\'enyi entropy \eqref{eq:Renyi} connecting copies of the universes $A$ (de~Sitter) and $B$ (an AdS black hole).
 (a)~Totally disconnected saddle where all copies of $A$ and $B$ are disconnected.
 (b)~Type~II$_A$ configuration where all copies of the universe $A$ are connected by a replica wormhole, but the copies of the universe $B$ are disconnected.
 In the same way, we have Type~II$_B$ configurations where all copies of universe $B$ are connected by a replica wormhole.
 (c)~In the Type~III configuration, each copy of $A$ is linked by a replica wormhole, and each copy of $B$ is linked by a separate replica wormhole, but the two wormholes are not interconnected.
 (d)~In a Type~IV configuration, all copies are connected through a single wormhole.
}}
\label{fig:wormholes}
\end{figure}

One saddlepoint topology that is always present consists of disconnected geometries computing each of the overlaps in Eq.~\eqref{eq:Renyi} (Fig.~\ref{fig:wormholes}a).
This contribution treats the overlaps as independent and does not allow for topology-changing contributions, i.e., wormholes, in the product of overlaps.
This class of saddlepoints obviously exists since each factor exists.
A second class of saddlepoints includes wormholes connecting the replicas of the $A$ universes (de~Sitter space in our case).
The maximally symmetric member of this class is labeled Type~II$_A$ in Fig.~\ref{fig:wormholes}b.
Similarly, there are wormholes between the $B$ universes and we will call the maximally symmetric one Type~II$_B$.
These Type~II saddlepoints have been already shown to exist in the JT gravity setting in Refs.~\cite{Balasubramanian:2020coy,Balasubramanian:2020xqf,Balasubramanian:2021wgd}.
Next, we can have simultaneous wormholes between the $A$ universe replicas and between the $B$ universe replicas---we call these Type~III saddlepoints in Fig.~\ref{fig:wormholes}c.
If Type~II saddlepoints exist, then so do Type~III because the wormholes between $A$ universes and between $B$ universes are solved for separately.   

Finally, there may be wormholes that connect all replica copies of the $A$ and $B$ universes together (Type~IV wormholes in Fig.~\ref{fig:wormholes}d).
As described in Section~\ref{subsec:2AdS-review}, such wormholes exist when both $A$ and $B$ are asymptotically AdS~\cite{Balasubramanian:2021wgd} and dominate the path integral at high temperatures.
The main task of the present paper, which we will take up in the next section, is to demonstrate the existence of such wormholes when the two universes have cosmological constants of opposite sign.
As we will see, if such wormholes exist, they will dominate the gravitational path integrals that compute $Z_{1}$ and $Z_{n}$ in Eq.~\eqref{eq:Renyi} in the high entanglement temperature ($\beta \to 0$) limit.
This is for the same reason that they dominate the high entanglement limit when both $A$ and $B$ are asymptotically AdS \cite{Balasubramanian:2021wgd}:\ as described in Section~\ref{subsec:2AdS-review}, increasing the entanglement in Eq.~\eqref{eq:TFD-state_AdS-AdS} or \eqref{eq:TFD-general} by taking $\beta \to 0$ increases the number of matter excitations $|\psi_i\rangle$ with significant support, and on a fully connected wormhole geometry these states are able to contract to form a single loop giving a large combinatorial factor in the overlap.
This factor is controlled parametrically by $\beta$, and so when the other parameters (e.g.\ cosmological constants) are fixed, there is always a $\beta$ that is small enough so that this contribution dominates, if the saddlepoint exists.
However, we will also show in Section~\ref{sec:dS-AdS-wormhole} that if the size of the cosmological horizon is smaller than that of the AdS black hole horizon, i.e.\ $S_{\rm dS} < S_{\rm BH}$, then the Type~IV---or a de~Sitter/AdS (dS/AdS)---wormhole cannot exist.

\subsection{Entanglement entropy vanishes without a dS/AdS wormhole}
\label{sec:vanish}

In the absence of a dS/AdS wormhole, or if it is sub-dominant because the entanglement temperature is low, we can assemble results from previous work to argue that the entanglement entropy vanishes.
To do this, we will first discuss the path integral for the normalization $Z_1^n$ in the denominator of (\ref{eq:Renyi}), and then the path integral for the $Z_n$ factor in the numerator.

First, $Z_1$ in Eq.~\eqref{eq:normalization} is a sum of terms each of which is a product of one de~Sitter overlap and one AdS overlap.
In the absence of dS/AdS wormholes, each term is computed by a disconnected diagram of the form in Fig.~\ref{fig:wormholes}a, with one de~Sitter path integral and one AdS path integral, which evaluate the vacuum 2-point functions of the operators at the poles creating states on the sphere and disk respectively.
These two point functions will be proportional to the identity, $\langle \psi_i | \psi_j\rangle \propto \delta_{ij}$.
If we choose the normalization of the operators so that the proportionality constant is 1, we see that the sum in Eq.~\eqref{eq:normalization} equals 1.
More generally, even if we do not fix the normalization of operators in this way, $Z_1$ normalizes the R\'enyi entropy and will not grow with the entanglement temperature.

For $Z_{n}$, let us consider the saddle in which all copies of de~Sitter space are connected by a replica wormhole, while all copies of the AdS black hole are disconnected.  
In this case, the AdS universes have no wormholes connecting them, so for the purpose of calculating the R\'enyi entropy in Eq.~\eqref{eq:Renyi} the AdS universes in this saddlepoint behave in the same way as a non-gravitating system.%
\footnote{
 Strictly speaking, the R\'enyi entropy is the logarithm of \eqref{eq:Renyi} divided by $1-n$, but we loosely call the expression in Eq.~\eqref{eq:Renyi} the R\'enyi entropy.
}
Thus, the computation and resulting value of the entanglement entropy will agree with the results in Ref.~\cite{Balasubramanian:2020xqf} for de~Sitter space entangled with a non-gravitating reference space.
As discussed there, the result will be $S(\rho_{A}) = 0$~\cite{Balasubramanian:2020xqf} unless we include additional ingredients like end-of-the-world branes that effectively introduce boundaries into the spaces, which we do not want to do here.
Since zero is the smallest possible value for the entanglement, the dominance of this saddle is guaranteed, implying that the entanglement entropy vanishes.

This conclusion changes if a dS/AdS wormhole exists.
This is because in this case, as discussed in the previous section, $Z_{1}$ will be dominated by the connected saddle at high entanglement temperature, because of the contribution of the matter excitations.
These contributions grow as the entanglement temperature increases, thereby increasing the number of states $|\psi_i\rangle$ with significant support in (\ref{eq:TFD-general}).
Similarly, when the dS/AdS wormhole exists, the fully connected saddle in Fig.~\ref{fig:wormholes}d gives the dominant contribution to $Z_n$ at high entanglement temperature, because of the matter contribution.
Taking the $n \rightarrow 1$ limit, we will see that the leading, fully connected, contributions to $Z_n$ and $Z_1$ lead to a finite entropy in the high entanglement temperature limit, which is given by the generalized entropy formula on a geometry that looks like a de~Sitter bubble behind the horizon of an AdS black hole.

\section{dS/AdS wormholes}
\label{sec:dS-AdS-wormhole}

Our goal in this section is to construct a wormhole between Euclidean de~Sitter and AdS spaces, and also to describe the analytic continuation of this wormhole to Lorentzian signature.
Our solution will have the form of a bubble of de~Sitter space behind the horizon of an AdS black hole. There is a long history of constructions of this general kind in different settings.
To our knowledge, the first of these was an exploration of the possibility of creating a universe in the lab.
Although Penrose's singularity theorem forces a singularity in a classical experiment of this kind, there may be room to realize such spacetimes via quantum tunneling~\cite{Farhi:1986ty}.

\paragraph{A constraint on de~Sitter bubbles in AdS:}
In the context of the AdS/CFT correspondence, Freivogel et al.~\cite{Freivogel:2005qh} constructed an inflating bubble separated by a domain wall from an asymptotically AdS geometry in a setup in which a false vacuum region with a positive cosmological constant is embedded in a true vacuum with a negative cosmological constant.
The analysis in Ref.~\cite{Freivogel:2005qh} emphasized an important constraint: 
the required gluing can only be performed consistently with the Israel junction conditions when  the area of the cosmological horizon $A_{\rm dS}$ is larger than the area of the  horizon of the AdS black hole $A_{\rm BH}$, or, in terms of the associated Bekenstein-Hawking entropies
\be
  S_{\rm dS} > S_{\rm BH},
\ee
if we impose reflection symmetry with respect to a Cauchy slice.
This constraint arises because the glued spacetime is effectively a ``bag of gold'' geometry---the cosmological horizon is a ``locally maximal'' surface, whereas the event horizon of the AdS black hole is ``locally minimal.''
Therefore, if we demand continuity of the metric on the separating domain wall,  we will need $A_{\rm dS} > A_{\rm BH}$.
We will find the same constraint on the existence of our wormholes.

In fact, in our context there is a natural microscopic argument explaining why we must have $S_{\rm dS} > S_{\rm BH}$ when we entangle the de~Sitter and AdS Hilbert spaces.
Recall the idea of static patch holography in de~Sitter space.
This hypothesis asserts that the degrees of freedom describing the static patches of de~Sitter space live on their stretched horizons with the Hilbert space $H_{\rm dS}$.
From this point of view, the region of global de~Sitter space that contains future infinity emerges from entanglement between the left and right static patch degrees of freedom; see, e.g., Ref.~\cite{Murdia:2022giv}.
The area of the cosmological horizon naturally evaluates the dimension of $H_{\rm dS}$.
In our setup there are two Hilbert spaces $H_{\rm dS}$ and $H_{\rm BH}$ which are entangled.
When ${\rm dim}\, H_{\rm dS} < {\rm dim}\, H_{\rm BH}$, all the basis states of $H_{\rm dS}$ are entangled with the microstates of the AdS black hole.%
\footnote{
 This $H_{\rm dS}$ is what is necessarily to construct the region outside the de~Sitter horizon.
 In a ``single-sided''---or cosmological---de~Sitter space, this is the degrees of freedom on the stretched horizon, while in a ``two-sided''---or intrinsically global---de~Sitter space, this is the degrees of freedom in the other static patch~\cite{Murdia:2022giv}.
}
Therefore, there is no room for the matter states in the static patch of de~Sitter space to be entangled with the appropriate $H_{\rm dS}$.
This means that if we realize this bipartite entangled state  in terms of semiclassical geometry, then it should not contain the future or past infinity, because of the absence of the entanglement necessary to make the horizon smooth.

\paragraph{A no-go argument for Euclidean de~Sitter bubbles: }
We are interested in constructing a Euclidean wormhole between de~Sitter and AdS spaces.
By cutting this geometry on the time-reflection slice, we could interpret such a wormhole as a saddlepoint of a Euclidean path integral preparing a de~Sitter bubble inside an AdS universe.
Fu and Marolf~\cite{Fu:2019oyc} have advanced a no-go argument showing that this is not possible in a pure gravity theory even if we allow codimension-one domain walls~\cite{Fu:2019oyc}.
The argument is roughly as follows.
The shape of the Euclidean domain wall separating the true and false vacua is determined by a radial trajectory in the Euclidean spacetime
\be
  \dot{R}^{2} + V_{E}(R) = 0,
\label{eq:EuclideanPot}
\ee
where $\dot{R} \equiv dR /d \tau_{E}$ is the derivative of the domain wall profile with respect to Euclidean time $\tau_{E}$, and $V_{E}(R)$ is an effective potential determined from Israel junction conditions.
The domain wall position turns out to oscillate in some window $r_{{\rm min}} \leq r \leq r_{{\rm max}}$.
If we take the normal vector of the domain wall to point towards the asymptotic boundary at $r_{{\rm max}}$, then the associated component (one of the transverse directions) of the extrinsic curvature $K_{o}(r_{{\rm max}})$ must be positive because the transverse area increases in this direction; the geometry resulting from gluing across the domain wall can contain the AdS boundary only in this case.
However, it turns out that if the interior is de~Sitter then $K_{o}(r_{{\rm max}}) < 0$, making the outward normal actually point towards smaller $r$ in the exterior.
As a result, the AdS part of the glued geometry cannot contain the asymptotic boundary.%
\footnote{
 While the argument of Ref.~\cite{Fu:2019oyc} was originally made in spacetimes with dimension $d \geq 3$, the analogous statement applies in $d=2$ because the component of the extrinsic curvature $K_{o}(r)$ which causes the problem has a direct counterpart in JT gravity, namely the derivative of the dilaton along the normal direction of the brane.
}

In the present paper, in addition to the domain wall, we have the entangled state of matter \eqref{eq:TFD-general}, the stress tensor of which will backreact on the geometry.
To create this excited state we must insert operators on the Euclidean AdS boundary.
We will see that these operators, and the backreaction of the matter they create, allow us to evade the no-go argument of Fu and Marolf.
When the entanglement is large, we will see that our boundary sources can be regarded as injecting energy that decays into a domain wall surrounding a de~Sitter bubble.
This resembles a construction by Mirbabayi~\cite{Mirbabayi:2020grb} which we review in the Appendix.

\subsection{Action and equations of motion}

We start from Einstein gravity coupled to a classical scalar field with a potential on a Euclidean manifold $\mathcal{M}$
\be
  I = -\f{1}{16\pi G_{\rm N}} \int_{\mathcal{M}}  dx^{d}\s{g} \left( R + (\partial \varphi)^{2} - V(\varphi) \right) +\log Z_{{\rm QFT}}.
\label{eq:originalI}
\ee
In the action, we have also added the effective action $\log Z_{{\rm QFT}}$ of a bulk QFT in which we defined the entangled state \eqref{eq:TFD-state_AdS-AdS}.
Suppose the potential has two local minima at $\varphi = \varphi_{A}$ and $\varphi_{B}$:\ one positive $V(\varphi_{A}) > 0$ and the other negative $V(\varphi_{B}) < 0$.
In this setup, a wormhole connecting de~Sitter and AdS spacetimes is a domain wall connecting the two vacua at $\varphi = \varphi_{A}$ and $\varphi_{B}$.
We will assume that the domain wall is thin, or equivalently, that the energy difference between the de~Sitter and AdS minima of $V(\varphi)$ is small compared to the barrier height.

The domain wall solution is thus constructed by first introducing a domain wall on both spacetimes $A$ and $B$ and then gluing the two along the wall.
The tension of the wall is related to the difference between the two energy values $\kappa = V(\varphi_{A})-V(\varphi_{B})$.
In the presence of the codimension-one domain wall $\mathcal{D}$, the spacetime manifold splits into two pieces $\mathcal{M} = \mathcal{M}_{+} \cup \mathcal{M}_{-}$, while the potential energy on either side of the wall is fixed.
In the following, we will use the convention that on $\mathcal{M}_{+}$ the value of the potential is $V(\varphi_{A})>0$, while it is $V(\varphi_{B})<0$ on $\mathcal{M}_{-}$.
The total action \eqref{eq:originalI} reduces to
\be
  I = I_{{\rm AdS}} + I_{{\rm dS}} + I_{{\rm domain \,  wall }} + \log Z_{{\rm QFT}},
\ee
where the first term is the action of Einstein gravity with a negative cosmological constant
\be
  I_{{\rm AdS}} = -\f{1}{16\pi G_{\rm N}} \int_{\mathcal{M}_{-}}\!\!\! d^{d}x \s{g} \left( R +\f{2d(d-1)}{L_{{\rm AdS}}^{2}} \right) + I_{{\rm boundary}},
\quad
  \f{2d(d-1)}{L_{{\rm AdS}}^{2}} \equiv V(\varphi_{B}).
\ee
Here, $I_{{\rm boundary}}$ is defined on the conformal boundary of $\mathcal{M}_{-}$ as well as on the domain wall $\mathcal{D}$
\be
  I_{{\rm boundary}} = -\f{1}{8\pi G_{\rm N}} \int_{\p \mathcal{M}_{-}}\!\!\! d^{d-1}x \s{h}\, K,
\ee
where $K$ is the extrinsic curvature of the boundary of $\mathcal{M}_{-}$.
Similarly, we have the action $I_{{\rm dS}}$ of the de~Sitter region $\mathcal{M}_{+}$ with $V(\varphi_{A}) = 2d(d-1)/L_{{\rm dS}}^{2}$.
There is also a term coming from the domain wall, proportional to its volume
\be
  I_{{\rm domain\, wall}} =\kappa \int_{\mathcal{D}} d^{d-1}x \s{h}.
\ee

The construction of the wormhole solution to the resulting equation of motion is quite involved because we have to take into account the backreaction of both domain wall and QFT degrees of freedom.
Below we will be mostly interested in the two-dimensional case, where the dS (AdS) sides is described by dS (AdS) JT gravity, so next we will summarize the basic properties of these theories.

\subsection{Stress tensors}

We want to compute the expectation value of the stress energy tensor $\la T_{\mu \nu} \ra$ on the dS/AdS wormhole geometry.
In general, the stress tensor depends on the background geometry and vice versa, so to determine it we must solve the gravitational equation of motions.
However, in JT gravity there is a simplification because the background metric on either side of the domain wall is locally dS or AdS, and the equation of motion only couples the stress tensor to the dilaton.
So we can take the metrics on the two sides to describe the vacuum dS and AdS solutions, work out the matter stress tensors on these backgrounds, backreact these stress tensors to find the dilaton profile, and then glue the solutions across the domain wall.
Below we will take the first step in this procedure:\ finding the stress tensor.

We are primarily interested in the large entanglement temperature limit.
In this limit, we argued that the correlation between degrees of freedom of the dS side, which we can think of as located at future infinity, and degrees of freedom of the AdS side, which live on the conformal boundary, becomes large.
This suggests that when the entanglement is large, the domain wall separating two sides in the Euclidean geometry approaches the AdS boundary.
We will assume this, and then show the self-consistency of the solution.

\subsubsection{Stress tensor on the connected saddle}

\begin{figure}
\vspace{-1cm}
  \begin{minipage}[b]{0.45\linewidth}
    \centering
    \includegraphics[keepaspectratio, scale=0.35]{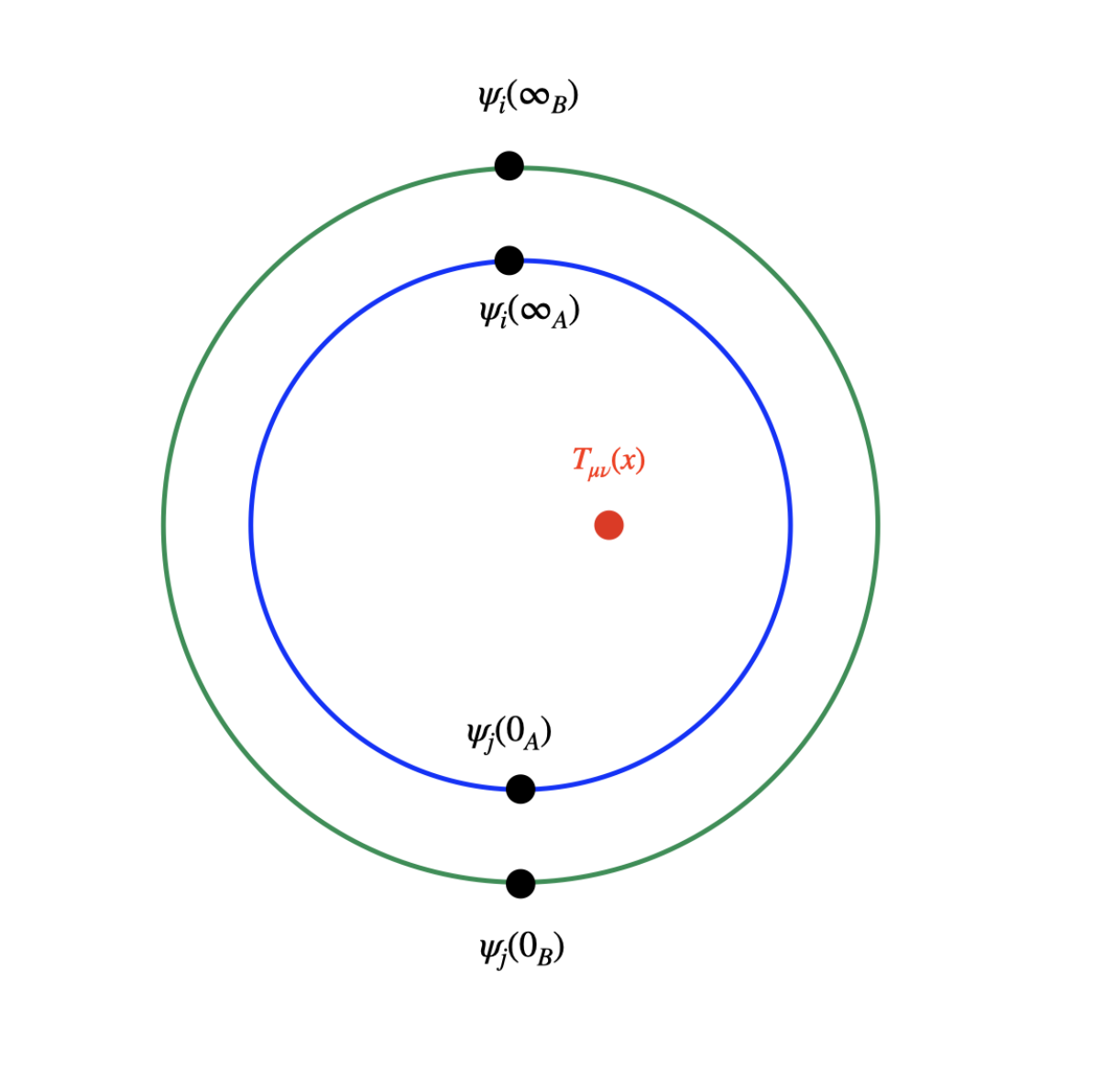}    
  \end{minipage}
  \hspace{2em}
  \begin{minipage}[b]{0.45\linewidth}
    \centering
\raisebox{1.5mm}{\includegraphics[keepaspectratio, scale=0.35]{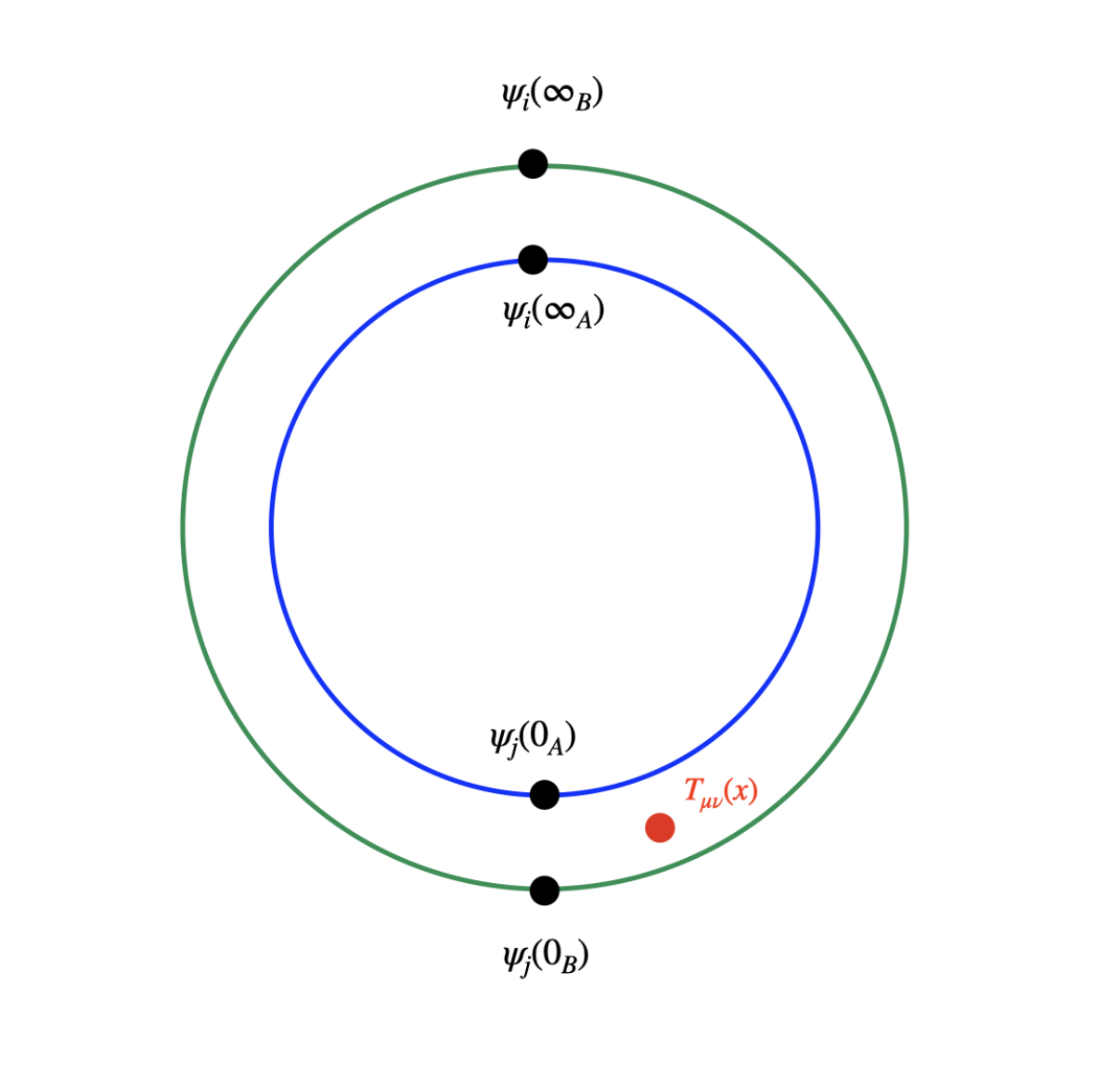} }   
  \end{minipage}
\vspace{-5mm}
\caption{
 The stress tensor expectation value $\la \Psi | T_{\mu\nu}(x) |\Psi \ra$ \eqref{eq:expectationv} on the dS/AdS wormhole.
 {\bf Left}: When it is inserted in the de~Sitter region (the red dot), its expectation value vanishes because of the OPE $\psi_i(\infty_{A}) \rightarrow \psi_i(\infty_{B})$ as well as $\psi_j(0_{A}) \rightarrow \psi_j(0_{B})$.
 {\bf Right}: When it is inserted in the AdS region, the expectation value is constant as shown in Eq.~\eqref{eq:stresstensor}.
 }
\label{fig:Wormholestress}   
\end{figure}

\begin{figure}
\vspace{3mm}
  \hspace{-2em}
  \begin{minipage}[b]{0.45\linewidth}
    \centering
    \includegraphics[keepaspectratio, scale=0.23]{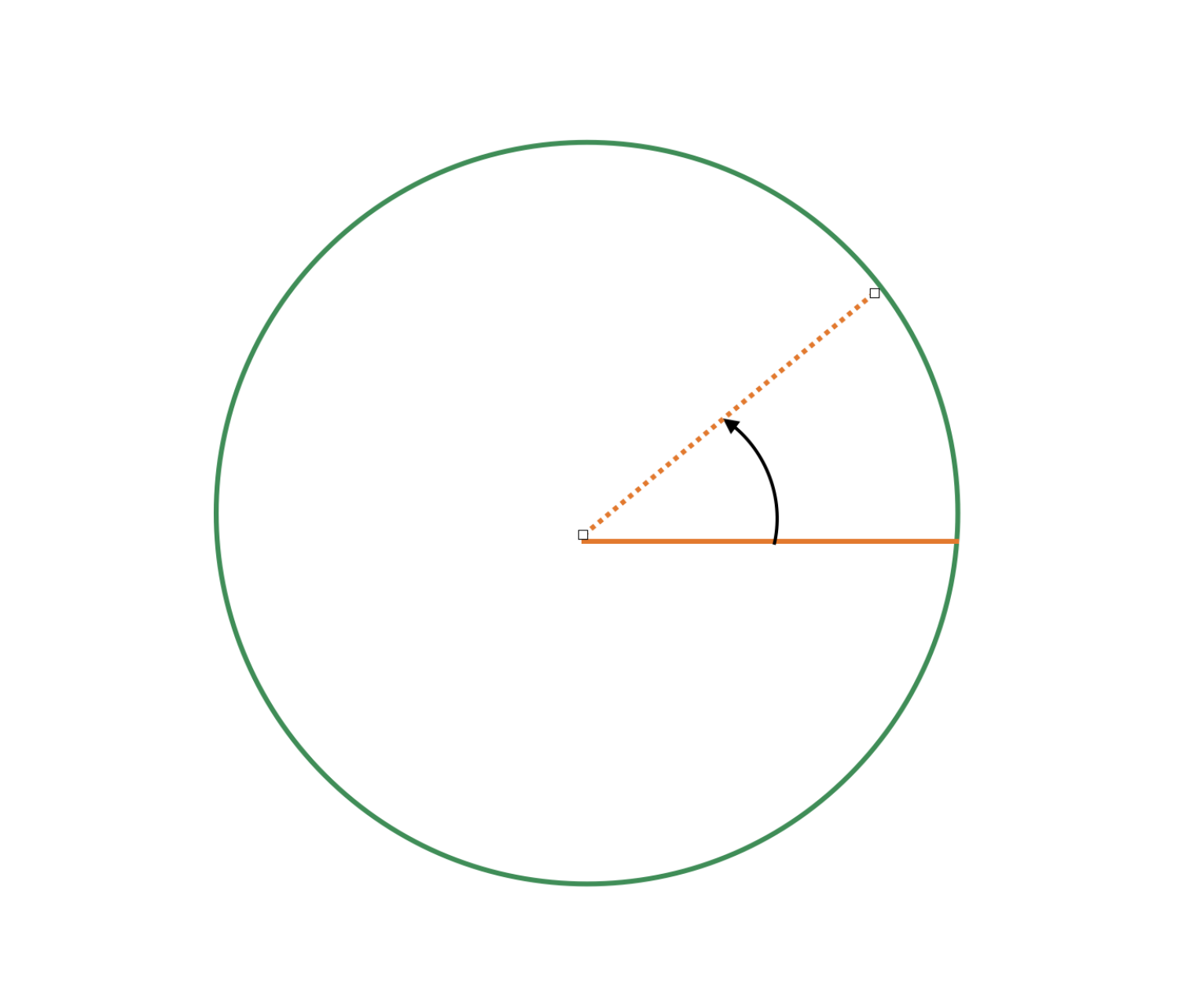}    
  \end{minipage}
  \hspace{-1em}
  \begin{minipage}[b]{0.25\linewidth}
    \centering
\raisebox{-3mm}{\includegraphics[keepaspectratio, scale=0.23]{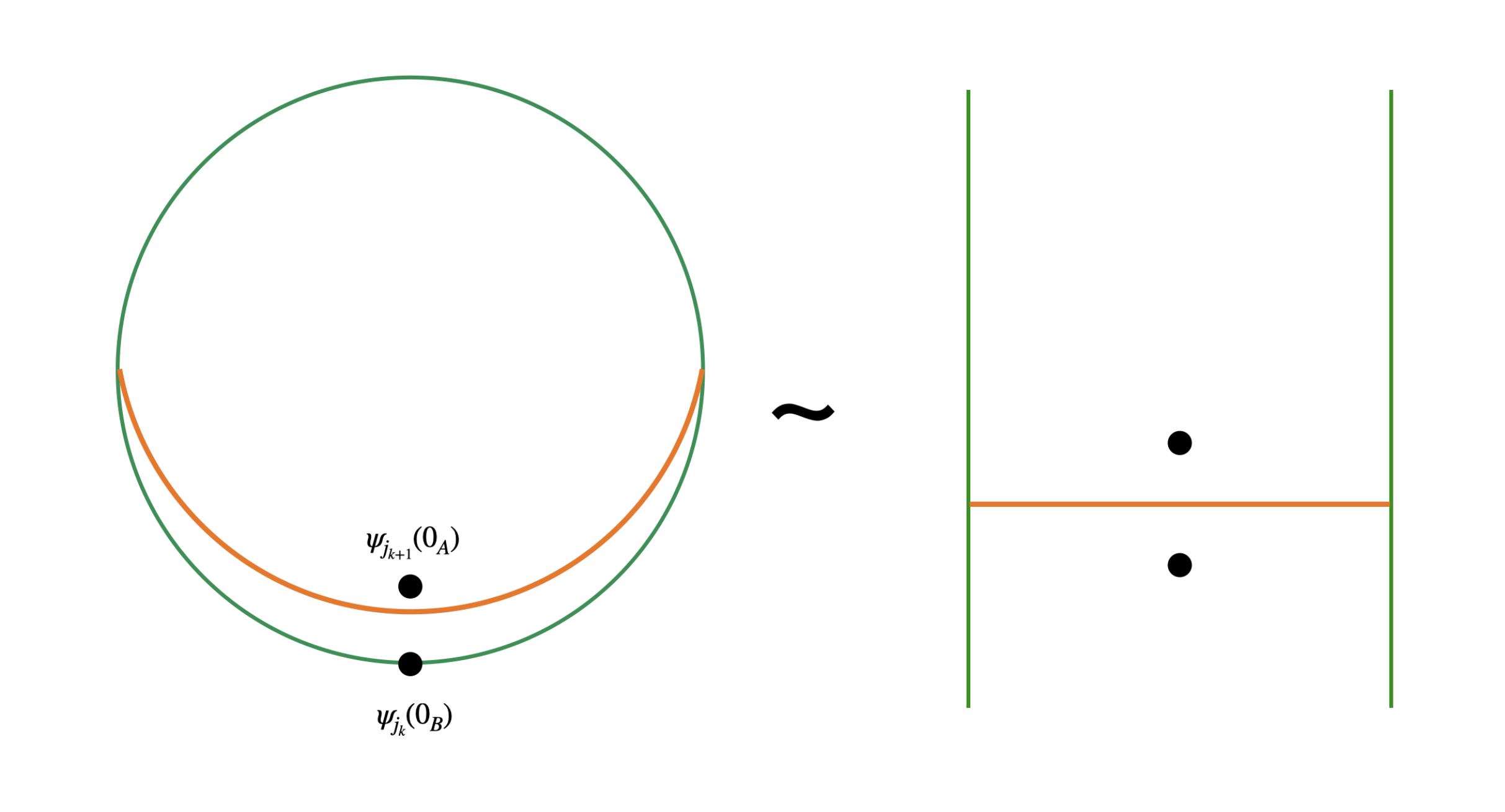} }   
  \end{minipage}
\vspace{-2mm}
\caption{
 Two kinds of Euclidean time on the disk.
 {\bf Left}: Euclidean Rindler time, defined in Eq.~\eqref{eq:adsrind}.
 We will use this when we construct the dS/AdS wormhole by gluing (two copies of) disk and Euclidean de~Sitter in Section~\ref{subsection:gluingAdS}.
 {\bf Right}: The global time on the disk induced from the time on a Euclidean strip.
 Black dots represent operators on the dS/AdS wormhole as in Eq.~\eqref{eq:stressOPE}.
}
\label{fig:AdStime}
\end{figure}

Let us first evaluate the stress energy tensor $\la \Psi | T_{\mu\nu} (x) | \Psi \ra$ on the candidate wormhole geometry.
If universes $A$ and $B$ are connected by a wormhole, the stress tensor value is evaluated by the five point function
\be
  \la \Psi | T_{\mu\nu} (x) | \Psi \ra = \f{1}{Z_{1,{\rm conn}}} \sum_{i,j} \s{p_{i}p_{j}}\; \la \psi_{i}(\infty_{A}) \psi_{j}(0_{A}) T_{\mu\nu}(x) \psi_{i}(\infty_{B}) \psi_{j}(0_{B}) \ra,
\label{eq:expectationv}
\ee
where $Z_{1,{\rm conn}}$ is defined in Eq.~\eqref{eq:zconn}, and is related to the four point functions of $\phi_{i}$'s.
As we have argued, we expect that when $\beta \rightarrow 0$, the domain wall approaches the conformal boundary of AdS, and we have an OPE limit 
$\psi_{i}(0_{B})\rightarrow \psi_{i}(0_{A})$   
and $\psi_{j}(\infty_{A})\rightarrow \psi_{j}(\infty_{B})$ as in Fig.~\ref{fig:Wormholestress}.
Consider the stress tensor in the AdS region.
Then $T_{\mu\nu}(x)$ is on a (Euclidean) time slice between $\psi_{j}(0_{A})$ and $\psi_{j}(0_{B})$.
Here the slice is taken with respect to the time derived by mapping the disk to a Euclidean strip plus points at infinities, which we refer to as global time.
Below we will also use Euclidean Rindler time.
Both these times are shown in Fig.~\ref{fig:AdStime}.

In the limit of high entanglement temperature, we thus have
\be
  \la \Psi | T_{\mu\nu} (x) | \Psi \ra = \f{\sum_{i} \s{p_{i}}\la \psi_{i}(0_{A}) T_{\mu\nu}(x) \psi_{i}(0_{B})\ra }{\sum_{i} \s{p_{i}}\la \psi_{i}(0_{A}) \psi_{i}(0_{B})\ra},
\label{eq:stressOPE}
\ee
since $\sum_{i} \s{p_{i}} \la \psi_{i}(\infty_{A}) \psi_{i}(\infty_{B}) \ra$ factors out from both numerator and denominator of the right-hand side of Eq.~\eqref{eq:expectationv}.
Moreover, since $p_{i} = e^{-\beta E_{i}}/Z$ is a Boltzmann factor, by picking up the saddlepoint in the energy spectrum, the holomorphic part of stress-energy tensor is evaluated as follows
\be
  \la \Psi | T_{zz} (x) | \Psi \ra = \f{\sum_{i} \s{p_{i}}\la \psi_{i}(0_{A}) T_{zz}(x) \psi_{i}(0_{B})\ra }{\sum_{i} \s{p_{i}}\la \psi_{i}(0_{A})  \psi_{i}(0_{B})\ra} \sim \f{E_{J}\s{p_{J}} \la\psi_{J}(\infty_{A})  \psi_{J}(0_{B})\ra}{\s{p_{J}}\la \psi_{J}(\infty_{A}) \psi_{J}(0_{B})\ra } = E_{J}(\beta),
\label{eq:stresstensor}
\ee
where $J$ denotes this saddlepoint in the sum with respect to the spectrum.
Here, $E_{J}(\beta)$ denotes the energy with respect to the global time of the Euclidean strip, which coincides with the conformal dimension of the corresponding operator via the state operator correspondence.

We have used the fact that
\be
  T_{zz}(x) \psi_{J}(0_{B}) |0 \ra = E_{J}\, \psi_{J}(0_{B})|0 \ra,
\ee
which is true because the operator is located at the south pole of the disk (which is $t= -\infty$ of the strip); see Fig.~\ref{fig:Wormholestress}.
In more detail, insertion of a local operator at the south pole of the disk is equivalent to having a globally excited state on the strip $|E_{J} \ra_{{\rm strip}}$ via the state operator correspondence:
\be
  T_{zz}(x) \psi_{J}(0_{B}) |0 \ra_{\rm disk} = T_{zz}(x)\, |E_{J} \ra_{{\rm strip}} = E_{J}\, |E_{J} \ra_{{\rm strip}}.
\ee
This shows in particular that the stress tensor is independent of the position $x$ on the time slice.
The rest of position dependence in \eqref{eq:stresstensor}  cancels between the numerator and the denominator.
Here the ($z,\bz$) coordinates are the usual holomorphic and anti-holomorphic coordinates on the AdS region of the disk, whose precise definition is given in Section~\ref{subsection:dilAdS}.

In contrast, when the stress tensor is located in the de~Sitter region  we have the factorization 
\be
  \la \psi_{i}(\infty_{A}) \psi_{j}(0_{A}) T_{\mu\nu}(x) \psi_{i}(\infty_{B}) \psi_{j}(0_{B}) \ra = \la \psi_{i}(\infty_{A}) \psi_{i}(\infty_{B}) \ra \, \la T_{\mu\nu}(x) \ra \, \la \psi_{j}(0_{A}) \psi_{j}(0_{B}) \ra.
\ee
Therefore, when the stress tensor is in the de~Sitter region, these OPEs tell us that the expectation value coincides with its vacuum value $\la \Psi | T_{\mu\nu} (x) | \Psi \ra = \la 0 |T_{\mu\nu} (x) | 0\ra$.
In a curved spacetime with the metric $ds^{2} = e^{2\omega} dz d\bar{z}$, it is given by
\be
  \la 0 | T_{zz}(x) | 0 \ra = \f{c}{12\pi} \left( \p^{2} \omega - (\p \omega)^{2} \right) + \tau_{zz},
\quad
  \la 0 | T_{\bz\bz}(x) | 0 \ra = \f{c}{12\pi} \left( \bp^{2} \omega - (\bp \omega)^{2} \right) + \tau_{\bz\bz},
\label{eq:zzcomp}
\ee
and 
\be
  \la 0 | T_{z\bz}(x) | 0 \ra = -\f{c}{12\pi} \p \bp\, \omega,
\label{eq:zbzcomp}
\ee
where $\tau_{zz} = \tau_{\bz\bz} = -c/48\pi$ are the stress tensors for the flat metric $ds^{2} = dz d\bz$.
By explicitly plugging in the Weyl factor of the de~Sitter metric~\eqref{eq:globalmetric}, we can show that the the terms in $\la 0 | T_{zz}(x) | 0 \ra$ and $\la 0 | T_{\bz\bz}(x) | 0 \ra$ coming from the Weyl factor $\omega$ cancel with the Casimir energies in $\tau_{zz}$ and $\tau_{\bz \bz}$ (also see Section~3.2 of Ref.~\cite{Balasubramanian:2020xqf}).

One might be puzzled by the fact that we start with an operator inserted at the south pole of the sphere, so it would appear that the de~Sitter region is excited by the operator and that there should thus be a non-vanishing stress tensor.
But when the de~Sitter region is connected to AdS black hole, the effect of this local insertion is absorbed by the identical operator located at the south pole of the AdS disk.
This occurs in the large entanglement limit, which is controlled by the OPEs of $\psi_{i}$.
In this OPE limit, the two operators fuse together to behave as if the identity operator has been inserted in the tip of the de~Sitter region.

\subsubsection{Stress tensor on the disconnected saddle}

In the disconnected saddle, the stress energy tensor on each side is just given by the thermal expectation value in the high entanglement temperature limit.
This is because in this limit the local energy density is dominated by high frequency modes that are insensitive to the spatial curvatures, which are larger scales.
Thus, we have
\be
  \la T_{zz} \ra_{A} = \la T_{zz} \ra_{B} = \f{c}{24\pi} \left( \f{2\pi}{\beta} \right)^{2}.
\ee
The same result hold for $\la T_{\bz\bz} \ra_{A} = \la T_{\bz\bz} \ra_{B}$.

\subsection{Solving for the dilaton}

\subsubsection{The dilaton on the dS side}

First, we summarize the properties of de~Sitter JT gravity studied in Refs.~\cite{Maldacena:2019cbz, Cotler:2019nbi} and specify the solution of our interest.
The Euclidean action is
\be
  -\ln Z = \f{\phi_{0}}{16\pi G_{\rm N}}\int\!\s{g}\, R +\f{1}{16\pi G_{\rm N}} \int\!\s{g}\, \Phi \left( R - \f{2}{L^{2}} \right) -\ln Z_{\rm CFT},
\ee
where $Z_{\rm CFT}$ denotes the partition function of the CFT which only depends on the metric $g_{\mu\nu}$.
Varying the action with respect to the dilaton, we find that the metric satisfies $R = 2/L^2$.
Below we will use global coordinates
\be
  ds^{2} = L^{2} \left(\f{d\tau^{2} + d\varphi^{2}}{\cosh^{2}\!\tau} \right),
\quad
  -\infty < \tau < \infty, \quad 0 < \varphi < 2\pi.
\label{eq:globalmetric}
\ee
If we write the metric as $ds^{2} = e^{2\omega} dz d\bar{z}$, with $z = i\tau + \varphi$ and $\bar{z} = -i \tau + \varphi$, varying the action with respect to the metric gives equations of motion
\be
  e^{2\omega} \p \left[ e^{-2\omega} \p\Phi \right] = 8\pi G_{\rm N} \la 0 | T_{zz} | 0 \ra,
\quad
  e^{2\omega} \bar{\p} \left[ e^{-2\omega} \bp \Phi \right] = 8\pi G_{\rm N} \la 0 | T_{\bz\bz} | 0 \ra,
\label{eq:dileom}
\ee
and 
\be
  e^{2\omega} \Phi + 2\p \bp \Phi = 16\pi G_{\rm N} \la 0 | T_{z \bz} | 0 \ra.
\label{eq:dileom2}
\ee

We have argued in the last section that in the de~Sitter region the stress tensor is given by \eqref{eq:zzcomp} and \eqref{eq:zbzcomp}:
\be
  \la 0 | T_{zz} | 0 \ra = \la 0 | T_{\bz\bz} | 0 \ra = 0,
\quad
  \la 0 | T_{z\bz} | 0 \ra = \f{c}{48 \pi^{2}\cosh^{2}\!\tau}.
\ee
Inserting this stress tensor in the equation for the dilaton, we find that
\be
  \Phi_{\rm dS} (\tau, \theta) =B\, \f{\cos \varphi}{\cosh \tau} + \f{c G_{\rm N}}{3},
\label{eq:dSdil}
\ee
where the constant piece is the contribution of the anomalous term $\la 0 | T_{z \bz} | 0 \ra$.

We will find later that it is useful to work in static patch coordinates when deriving the trajectory of the domain wall separating the dS and AdS parts of the geometry.
This is because, in static coordinates both the dS geometry and the AdS black hole have a U(1) symmetry which we use to define a common angular direction.
To this end, we start from the embedding space representation of dS$_2$
\be
  X_{0}^{2} + X_{1}^{2} + X_{2}^{2} = 1,
\qquad
  ds^{2} = L^{2} (dX_{0}^{2} + dX_{1}^{2} + dX_{2}^{2}).
\ee
Then global coordinates are defined by
\be
  X_{0} = \tanh\tau, \quad X_{1} = \f{\sin\varphi}{\cosh\tau}, \quad X_{2} = \f{\cos\varphi}{\cosh\tau}.
\ee
On the other hand, static patch coordinates are given by
\be
  X_{0} = \sin\theta \sin t, \quad X_{1} = \sin\theta \cos t, \quad X_{2} = \cos\theta,
\label{eq:dstcoord}
\ee
and the resulting metric is $ds^{2} = d\theta^{2} + \sin^{2}\!\theta\, dt^{2}$, with a dilaton $\Phi = B\cos\theta$.

\begin{figure}
  \begin{minipage}[b]{0.45\linewidth}
    \centering
    \includegraphics[keepaspectratio, scale=0.17]{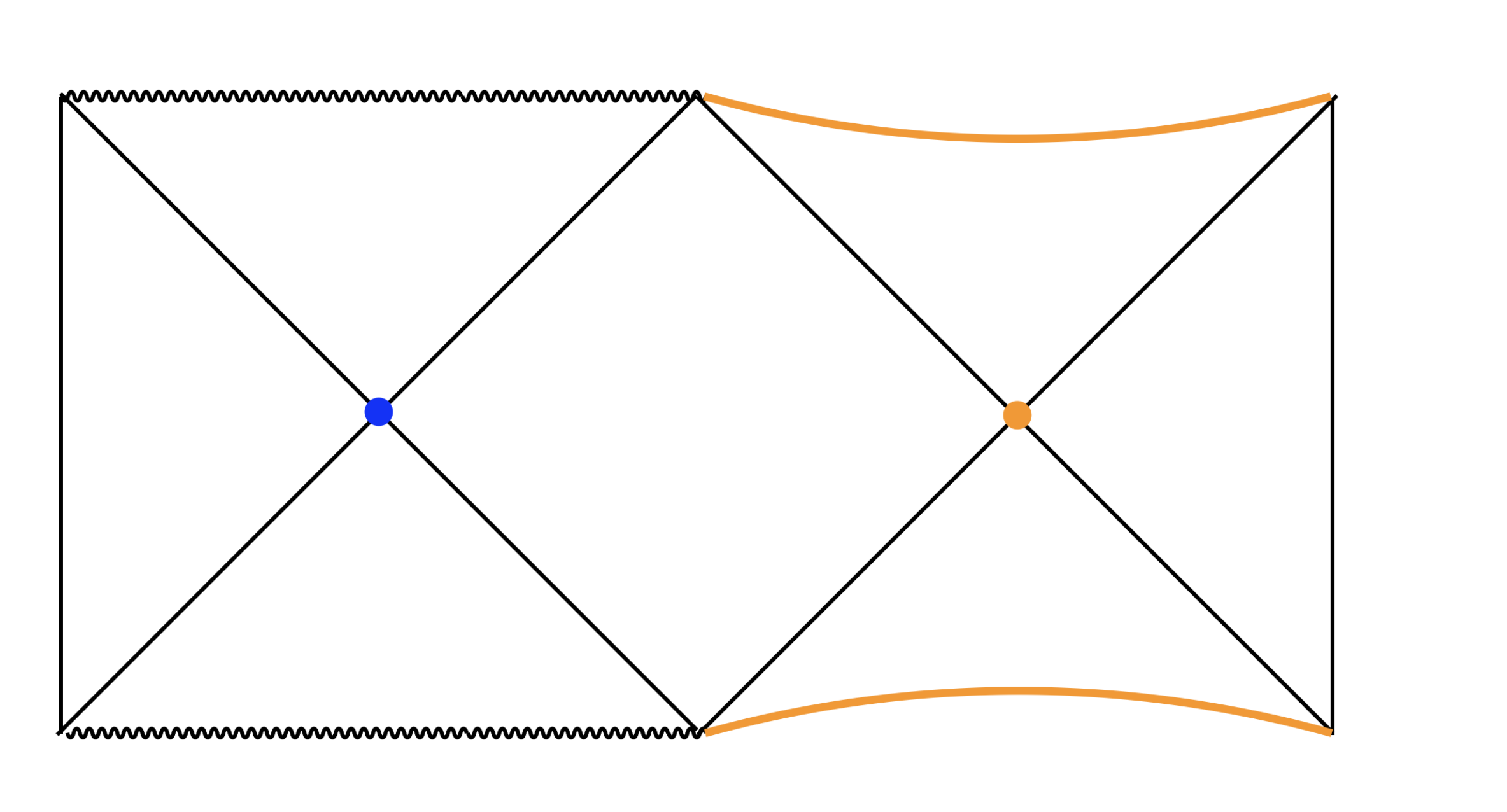}    
  \end{minipage}
  \hspace{1em}
  \begin{minipage}[b]{0.45\linewidth}
    \centering
\raisebox{0mm}{\includegraphics[keepaspectratio, scale=0.23]{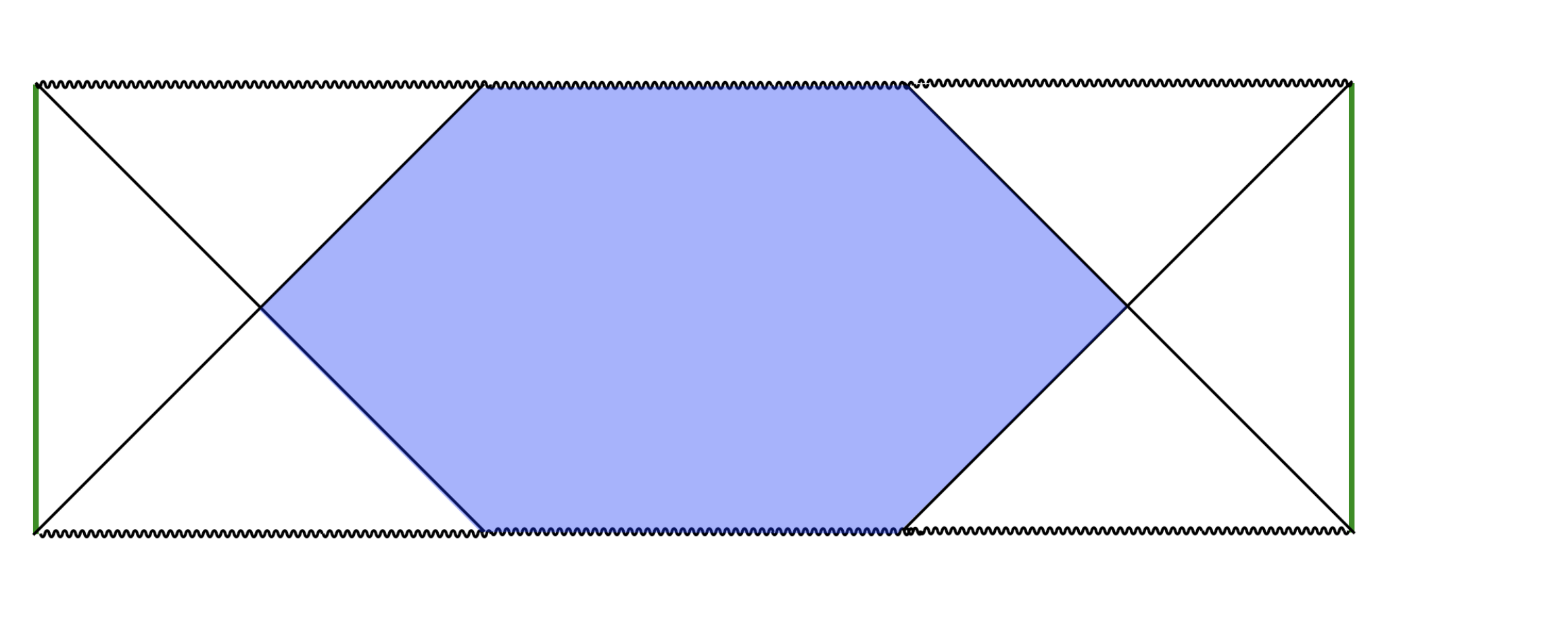} }   
  \end{minipage}
\caption{
 The Penrose diagrams for the de~Sitter black hole and the AdS black hole with backreaction, which are used to construct a dS/AdS wormhole of interest.
 {\bf Left}: de~Sitter black hole with the dilaton profile \eqref{eq:dSdil}.
 The bifurcation surface of the black hole is depicted by the blue dot, and the de~Sitter bifurcation surface by the orange dot.
 Orange curves are future and past infinity of de~Sitter.
 {\bf Right}: The two-sided black hole in AdS with the backreaction \eqref{eq:dilback}.
 The green lines are the conformal boundaries.
 The blue shaded region is in the black hole interior and is called the causal shadow because it is not causally connected with the asymptotic boundary.
}
\label{fig:PenrosedSAdS}   
\end{figure}

This dilaton profile describes the geometry of a Schwarzschild black hole in the Nariai limit~\cite{Maldacena:2019cbz}.
The value of the dilaton profile together with a constant piece $\phi_{0} +\Phi$ is the area of the manifold transverse to the 2d direction.
Therefore, an extremal surface in the geometry satisfies $\p \Phi = \bp \Phi = 0$.
For the profile \eqref{eq:dSdil}, we have two such surfaces at $(\tau,\varphi) = (0,0)$ and $(\tau,\varphi) = (0,\pi)$.
The first one corresponds to the cosmological horizon, since the dilaton is maximal at the point, and the second one is the black hole horizon.
The Penrose diagram of the geometry is depicted in the left panel of Fig.~\ref{fig:PenrosedSAdS}.

\subsubsection{The dilaton on the AdS side}
\label{subsection:dilAdS}

The action of AdS JT gravity coupled with bulk CFT degrees of freedom~\cite{Almheiri:2014cka,Maldacena:2016upp} is 
\be
  \ln Z = \f{\phi_{0}}{16\pi G_{\rm N}} \left[ \int_{B}\! \s{g}\, R + 2 \int_{\p B}\! K \right] + \f{1}{16\pi G_{\rm N}} \left[ \int\! \Phi \left( R+\f{2}{\tilde{L}^{2}} \right) + \Phi_{b}\int_{\p B}\! K \right].
\label{eq:AdSaction}
\ee
We will work in global coordinates
\be
  ds^{2} = \tilde{L}^{2} \left( \f{d\tau^{2} + d\mu^{2}}{\cos^{2}\!\mu} \right),
\quad
  -\f{\pi}{2} < \mu < \f{\pi}{2}.
\ee

Again, by writing the metric as $ds^{2} = e^{2\omega} dz d\bz$, with $z = i\tau + \mu$ and $\bz = -i\tau + \mu$, the equations of motion for the dilaton are given by \eqref{eq:dileom} and by
\be
  -e^{2\omega}\, \Phi + 2\p \bp\Phi = 16\pi G_{\rm N} \la T_{z \bz} \ra,
\label{eq:dileom3}
\ee
where the first term has the opposite sign from that in Eq.~\eqref{eq:dileom2}.
If the stress tensor vanished as it would in the vacuum, the dilaton profile would be given by
\be
  \Phi_{\rm AdS}(\tau,\mu) = A \f{\cosh\tau}{\cos\mu},
\label{eq:oridil}
\ee
as can be checked by inserting it in the dilaton equation of motion.
The coefficient $A$ will be fixed by the boundary condition.
This dilaton profile, when continued to Lorentzian regime describes an eternal black hole with a square Penrose diagram.
On the dS/AdS wormhole, however, the stress energy tensor on the AdS side is given by Eq.~\eqref{eq:stresstensor}.
We thus have to set $\la T_{zz} \ra = \la T_{\bz\bz} \ra = E_{J}(\beta)$, and we get a solution of the following form:
\be
  \Phi(\tau,\mu) = \Phi_{0}(\tau,\mu) - 16\pi G_{\rm N} E_{J}(\beta) (\mu \tan\mu + 1).
\label{eq:defdil}
\ee
Here, $\Phi_{0}(\tau,\mu)$ satisfies the equations of motion with the vanishing stress energy tensor.
This portion is fixed by imposing a boundary condition at the asymptotic boundary $\mu \rightarrow \pm \pi/2$.
In particular, we demand that \eqref{eq:defdil} approaches the vacuum dilaton profile \eqref{eq:oridil}~\cite{Bak:2018txn} at the boundary (in some SL$(2,R)$ frame; see below).

To match the boundary condition, it is convenient to write
\be
  \f{\bar{\phi}}{\pi} \left(b-\f{1}{b} \right) =16\pi G_{\rm N} E_{J}(\beta),
\label{eq:eternalBH-0}
\ee
and set
\be
  \Phi_{0}(\tau,\mu) = \f{\bar{\phi}}{2} \left( b+\f{1}{b} \right) \f{\cosh\tau}{\cos\mu}.
\label{eq:eternalBH}
\ee
As we will see, this form of the dilaton profile satisfies the boundary condition near asymptotic infinity $\mu \rightarrow \pm \pi/2$; i.e., the backreacted dilaton profile becomes \eqref{eq:oridil} near infinity in some SL$(2,R)$ frame.
To show this, we plug \eqref{eq:eternalBH-0} and \eqref{eq:eternalBH} into the backreacted solution \eqref{eq:defdil}, and obtain
\begin{align}
  \Phi_{\beta}(\tau,\mu) &= \f{\bar{\phi}}{2} \left(b+\f{1}{b}\right) \f{\cosh\tau}{\cos\mu} - \f{\bar{\phi}}{\pi} \left(b-\f{1}{b}\right) (\mu \tan\mu +1)
\label{eq:dilback}\\
  &\rightarrow \f{\bar{\phi}}{2} \left[ \left(b+\f{1}{b}\right) \f{\cosh\tau}{\cos\mu} - \left(b-\f{1}{b}\right) \tan\mu \right] \quad (\mu \rightarrow \f{\pi}{2}).
\label{eq:dilback-2}
\end{align}
The divergence as $\mu \to \pi/2$ occurs because we are approaching the AdS boundary in this limit, and the dilaton measures the asymptotic growth of the transverse sphere in the higher dimensional theory whose compactification gives rise to JT gravity.
We now show that we can bring this expression to the same form as the vacuum dilaton solution \eqref{eq:oridil} by performing an SL$(2,R)$ transformation of the geometry, which is an isometry of the space.

To specify the necessary transformation, we realize AdS$_2$ as a hyperbola
\be
  -X_{0}^{2} - X_{1}^{2} + X_{2}^{2} = 1,
\qquad
  ds^{2} = \tilde{L}^{2} \left(dX_{0}^{2} + dX_{1}^{2} - dX_{2}^{2} \right).
\ee
Global coordinates $(\tau,\mu)$ are defined by the embedding
\be
  X_{0} = \tan\mu, \quad X_{1} =\f{\sinh\tau}{\cos\mu}, \quad X_{2} =\f{\cosh\tau}{\cos\mu}.
\label{eq:coordtrans}
\ee
The SL$(2,R)$ isometry of our interest is 
\be
  \begin{pmatrix}
    X_{0} \\
    X_{2}
  \end{pmatrix}
\rightarrow
  \begin{pmatrix}
    X'_{0} \\
    X'_{2}
  \end{pmatrix} 
 =
  \begin{pmatrix}
    b_{+} & -b_{-} \\
    -b_{-} & b_{+} \\
  \end{pmatrix}
  \begin{pmatrix}
    X_{0} \\
    X_{2}
  \end{pmatrix},
\quad
  b_{\pm} = \f{1}{2} \left( b \pm \f{1}{b} \right).
\label{eq;sl2r}
\ee
Then, the following coordinate transformation
\be
  \tan\mu' = b_{+} \tan\mu - b_{-} \f{\cosh\tau}{\cos\mu},
\quad
  \f{\cosh\tau'}{\cos\mu'} = b_{+} \f{\cosh\tau}{\cos\mu} - b_{-} \tan\mu
\ee
brings \eqref{eq:dilback-2} to the vacuum form
\be
  \Phi_{\rm AdS} (\tau',\mu')  = A \; \f{\cosh \tau'}{\cos \mu'},
\label{eq:undeformed}
\ee
with $A = \bar{\phi}$.

We will later see that it is also useful to define coordinates $(\rho, t)$
\be
  X_{0} = \sinh\rho \cos t, \quad X_{1} = \sinh\rho \sin t, \quad X_{2} = \cosh\rho.
\label{eq:adsrind}
\ee
The metric is then given by $ds^{2} = d\rho^{2} + \sinh^{2}\!\rho\, dt^{2}$, and the dilaton profile is $\Phi_{\rm AdS} = A \cosh\rho$.

The dilaton profile in Eq.~\eqref{eq:dilback} describes an AdS black hole with long wormhole in its interior region.
For instance, there are two extremal surfaces at $(\tau,\mu) = (0, \mu_{R})$ and $(\tau,\mu) = (0, \mu_{L})$, and one can check $\mu_{R} \rightarrow \pi/2$ and $\mu_{L} \rightarrow -\pi/2$ in the $\beta \rightarrow 0$ limit.
The Penrose diagram of this black hole is depicted in the right panel of Fig.~\ref{fig:PenrosedSAdS}.

\subsection{Gluing dS to AdS}
\label{subsection:gluingAdS}

Having specified the dilaton profile of interest, we would like to construct the wormhole solution connecting the dilaton in the de~Sitter side $\Phi_{\rm dS}(\theta)$ in \eqref{eq:dSdil} and the backreacted solution in the AdS side $\Phi_{\beta}(\tau,\mu)$ in \eqref{eq:dilback}, by solving the Israel junction conditions
\be
  \Phi_{\rm dS}\big|_{\rm brane} = \Phi_{\beta}(\tau,\mu)\big|_{\rm brane},
\quad
  \xi^{\mu} \p_{\mu} \Phi_{\rm dS} - \xi^{\mu} \p_{\mu} \Phi_{\beta}(\tau,\mu) = \kappa.
\label{eq:Junctionbeta}
\ee
These equations are used to specify the location of the domain wall  $\tau = \tau_{\beta}(t)$, $\mu = \mu_{\beta}(t)$ written in the coordinates in the AdS side.
For this purpose, it is useful to use $t$ coordinate for the Euclidean timelike direction, commonly defined  both on the de~Sitter side \eqref{eq:dstcoord} and on the AdS side \eqref{eq:adsrind}.
In particular, on the de~Sitter side, this is the direction of the U(1) isometry.
The first equation of Eq.~\eqref{eq:Junctionbeta} relates the brane profile in the coordinates of the AdS side to the one in the de~Sitter side $\theta(t)$.
The brane profile on the de~Sitter side does not depend on the entanglement temperature because the stress tensor is vanishing on this side in the connected geometry.

Finding the solution explicitly is difficult.
However, in the high temperature limit $\beta \rightarrow 0$, the location of the domain wall approaches the asymptotic boundary of AdS, $\mu_{\beta}(t) \rightarrow \pm \pi/2$, which simplifies the task.
For instance, in the asymptotic region the backreacted dilaton $\Phi_{\beta}(\tau,\mu)$ reduces to \eqref{eq:dilback-2}, and using the coordinate transformation between the $(\mu,\tau)$ and $(\rho,t)$ coordinates, obtained by equating Eqs.~\eqref{eq:coordtrans} and \eqref{eq:adsrind}, we get
\begin{align}
  \Phi_{\beta}(\rho,t) &= \f{\bar{\phi}}{2} \left(b+\f{1}{b}\right) \f{\cosh\tau}{\cos\mu} - \f{\bar{\phi}}{\pi} \left(b-\f{1}{b}\right) (\mu \tan\mu + 1)
\\[+10pt]
  &\rightarrow
  \begin{cases}
    \Phi_{+}(\rho,t) \equiv \f{\bar{\phi}}{2} \left(b+\f{1}{b}\right) \cosh\rho + \f{\bar{\phi}}{2} \left(b-\f{1}{b}\right) \sinh\rho \cos t & \bigl(-\f{\pi}{2} \leq t \leq 0\bigr)
  \\[+10pt]
    \Phi_{-}(\rho,t) \equiv \f{\bar{\phi}}{2} \left(b+\f{1}{b}\right) \cosh\rho - \f{\bar{\phi}}{2}\left(b-\f{1}{b}\right) \sinh\rho \cos t & \bigl(-\pi \leq t \leq -\f{\pi}{2}\bigr)
  \end{cases}
\;
  \mbox{for } \rho \rightarrow \infty.
\label{eq;twodil}
\end{align}

By defining a new coordinate $\delta$ as
\be
  \delta =
  \begin{cases}
    t + \f{\pi}{2}  & \bigl(-\f{\pi}{2} < t \leq 0\bigr) \\
    -t - \f{\pi}{2} & \bigl(-\pi \leq t \leq -\f{\pi}{2}\bigr),
\end{cases}
\ee
we see that the expressions for two dilaton profiles $\Phi_{+}(\rho,\delta)$ and $\Phi_{-}(\rho,\delta)$ become identical.
This implies that near the conformal boundary, $\Phi_{\beta}(\rho,t)$ can be thought of as obtained by first preparing two identical dilaton profiles with a single bifurcation surface, and gluing them along $t =0$.
This is exactly how to treat the backreaction of the particle which starts from the boundary and then propagates into the bulk of the Euclidean black hole.
The backreacted black hole constructed in this way is called a partially entangled state (PETS)~\cite{Goel:2018ubv}.
We review its construction in the Appendix.
A notable feature of the PETS geometry is that since the dilaton profile is constructed by gluing two identical profiles $\Phi_{+}(\rho,\delta)$ and $\Phi_{-}(\rho,\delta)$ each of which has a horizon, the resulting dilaton profile, which we denote by $\Phi_{{\rm PETS}}(\rho,t)$, has two black hole horizons.

The argument here implies that near the asymptotic boundary $\rho \rightarrow \infty$, the profile $\Phi_{\beta}(\rho,t)$ obtained by solving the backreaction of the globally excited state \eqref{eq:defdil} coincides with the dilaton profile of the PETS
\be
  \Phi_{\beta} (\rho,t) = \Phi_{{\rm PETS}}(\rho,t) \quad \mbox{at } \rho \rightarrow \infty.
\ee
This is natural because we start from the excited state in the global AdS$_2$ and in the disk frame the excited state is specified by the insertion of a local operator whose backreaction is treated by the junction condition \eqref{eq:backreactionex}.

Gluing between the PETS geometry specified by $\Phi_{{\rm PETS}}$ and the de~Sitter geometry given by $\Phi_{\rm dS}$ was studied in a paper by Mirbabayi~\cite{Mirbabayi:2020grb}, whose construction is reviewed in the Appendix.
Since the de~Sitter bubble is realized in the PETS geometry, the Euclidean bulk spacetime can be regarded as describing a particle starting from the asymptotic boundary, propagating for a while in the Euclidean black hole, and then decaying into the domain wall separating the interior de~Sitter region and the exterior AdS region (the right panel of Fig.~\ref{fig:Wormhole3}).
This construction avoids the no go argument by Fu and Marolf~\cite{Fu:2019oyc}, since the domain wall profile has a kink due to the intersection with the particle trajectory in the AdS side used to construct the PETS.
The backreaction of the particle here creates a large interior region within the AdS black hole to accommodate the de~Sitter horizon behind the black hole horizon.

Each of the dilaton profiles $\Phi_{\pm}(\rho,\delta)$ in Eq.~\eqref{eq;twodil} used in constructing $\Phi_{\beta}(\rho,t)$ (or $\Phi_{{\rm PETS}}(\rho,t)$) is related to the original profile $\Phi_{\rm AdS} = A \cosh\rho$ for the Euclidean black hole by the SL$(2,R)$ transformation \eqref{eq;sl2r}.
The rest of the procedure is then parallel to that of Ref.~\cite{Mirbabayi:2020grb}.
In particular, the location of the domain wall in the original global coordinates $(\tau_{\beta}(t), \mu_{\beta}(t)) $, i.e., the solution of Eq.~\eqref{eq:Junctionbeta}, is obtained by applying the SL$(2,R)$ transformation to the Mirbabayi's solution $\rho =\rho(t)$ presented in \eqref{eq:2dJTjunction}.
More explicitly, from the relation between these two coordinates, we have
\be
  \tan\mu_{\beta}(t) = b_{+} \cosh\rho(t) + b_{-} \sinh\rho (t) \cosh t,
\quad
  \f{\cos\tau_{\beta}(t)}{\cos\mu_{\beta}(t)} = b_{-} \sinh\rho(t) \cosh t + b_{+} \cosh \rho(t). 
\ee
From the first equation, one can check that the location of the brane $\mu = \mu_{\beta}(t)$ solving \eqref{eq:Junctionbeta} indeed satisfies our ansatz, namely $\mu_{\beta}(t) \rightarrow \pi/2$ in the high-temperature limit, because in this limit the $E_{J}(\beta)$ and $b_{\pm}$ both become large.
This self-consistently justifies our assumption that in the high entanglement temperature limit the domain wall approaches the boundary.
Furthermore, in this way of treating the problem, the condition $S_{\rm dS} > S_{\rm AdS}$ for the existence of the wormhole solution is obvious.
Continuity of the dilaton profile (the first equation of \eqref{eq:Junctionbeta}) reduces to $\Phi_{\beta} = A \cosh\rho = B \cos\theta = \Phi_{\rm dS}$, implying $B > A$.
Since the black hole bifurcation surface is located at $\rho = 0$, the black hole entropy is given by $S_{\rm BH} = \phi_{0} + A$, where $\phi_{0}$ is the constant part of the dilaton profile in Eq.~\eqref{eq:AdSaction}.
Similarly, the bifurcation surface of de~Sitter is at $\theta = 0$, giving $S_{\rm dS} = \phi_{0} + B$.
Combining these, we conclude that that $S_{\rm dS} > S_{\rm BH}$ must hold.

Another notable feature of the solution described here is that the de~Sitter side only contains the cosmological horizon, and not a black hole horizon.
This can be seen by recasting the second equation of the junction conditions into the form of a one-dimensional potential problem for the domain wall trajectory $\dot{\theta}^{2} + V(\theta) = 0$.
This equation tells us the range $\theta_{{\rm min}} < \theta < \theta_{{\rm max}}$ in which the domain wall can move.
We can check from this that the black hole horizon is indeed excluded from the de~Sitter bubble region.

Note that in constructing the dS/AdS wormhole solution, we only needed an excited state on the AdS side, and not necessarily the entangled state between two sides \eqref{eq:TFD-general}.
For example, one can obtain a similar wormhole starting from a factorized state of the form $|\psi \ra_{A} |\psi \ra_{B}$ as long as both of factors are highly excited.
However, a wormhole connecting dS and AdS will only dominate the gravitational path integral for \eqref{eq:Renyi} when there is a large entanglement between the two systems.
This indeed occurs if the bulk QFT state is of the thermofield double type \eqref{eq:TFD-general} with small $\beta$, as we will discuss in the next section.

\subsection{Continuation of the dS/AdS wormhole to Lorentzian signature}

\begin{figure}
\vspace{-0.5cm}
\begin{center}
\includegraphics[scale=0.4]{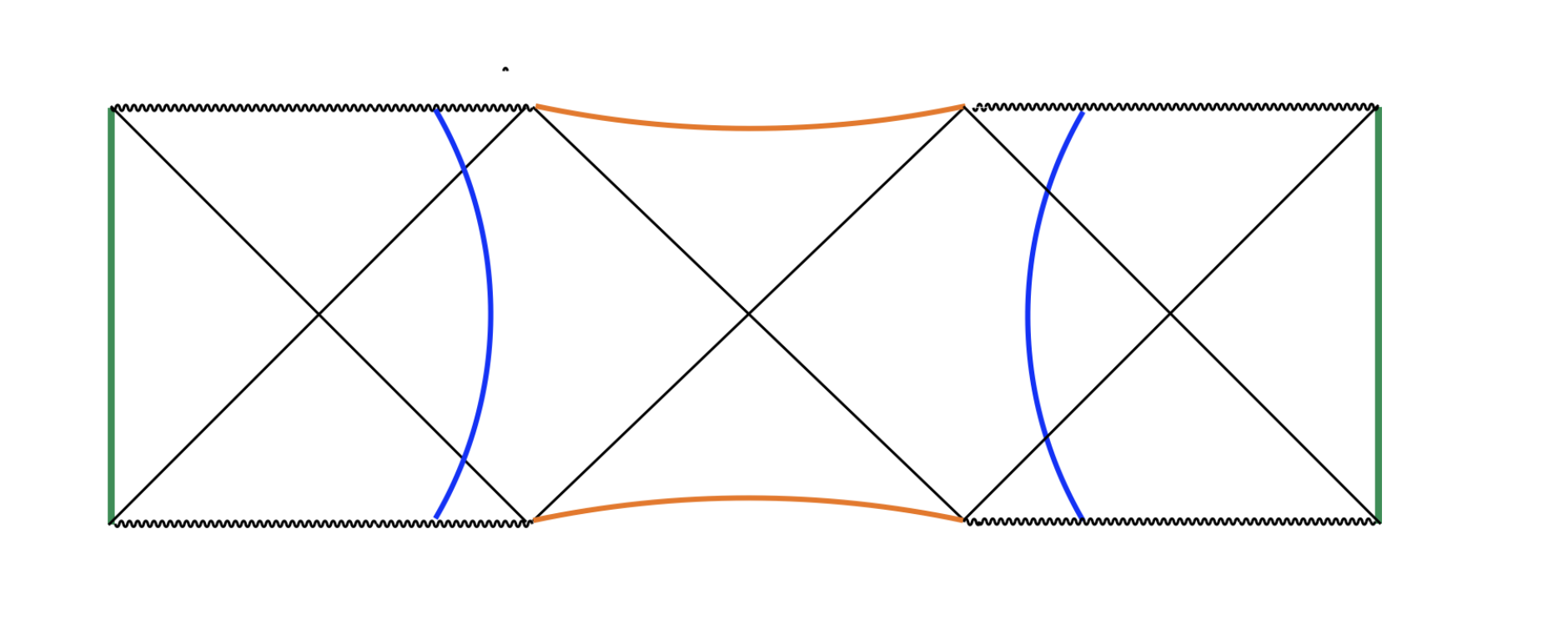}
\end{center}
\vspace{-0.5cm}
\caption{
 The Lorentzian spacetime on which we compute the generalized entropy (in Eq.~\eqref{eq:genends}).
 This spacetime is obtained by continuing the Euclidean wormhole geometry to Lorentzian signature along the reflection symmetric slice.
 This is an asymptotically AdS black hole (whose conformal boundaries are depicted in green) with an inflating de~Sitter region in its interior (with future and past infinities depicted in orange curves).
 The blue curves represent domain walls separating two geometries.
}
\label{fig:Lorentzian}
\end{figure}

The Euclidean dS/AdS wormhole constructed in this way has a time reflection symmetric slice.
Therefore, it can be analytically continued to Lorentzian regime.
The Penrose diagram of the resulting spacetime is depicted in Fig.~\ref{fig:Lorentzian}.
This geometry describes an AdS black hole with a de~Sitter bubble in its interior.
It can be explicitly checked that the de~Sitter region contains the cosmological horizon and fully contains the past and future infinity.
In Section~\ref{subsubsec:infty}, we required that all the saddles of the gravitational path integral should contain the future/past infinity of de~Sitter space when continued to Lorentzian signature.
Therefore, the dS/AdS wormhole constructed here indeed satisfies the boundary condition we demanded for the gravitational path integral.
In this construction, the de~Sitter region does not contain the bifurcation surface of a black hole.

Note that, as pointed out in Ref.~\cite{Freivogel:2005qh}, the null energy condition prohibits future infinity of de~Sitter space from being causally connected with the asymptotic boundaries of the AdS black hole, as we see in the Penrose diagram of Fig.~\ref{fig:Lorentzian}.
This is because a future-directed null congruence near future infinity is expanding, while such a null congruence has to shrink in the interior region of the AdS black hole; the Raychaudhuri equation combined with the null energy condition prohibits a smooth interpolation of these congruences.

In the next section,  we will argue that the entanglement entropy in the high temperature limit can be computed as a type of generalized entropy on the Lorentzian geometry obtained in this way.

\section{Calculation of the generalized entropy} 
\label{sec:genentropy}

\subsection{Contribution from the fully connected saddle}

We first note that in the high entanglement temperature limit $\beta \rightarrow 0$, the gravitational path integral is dominated by the contribution of the fully connected saddle even when one of the universes is closed.
This is because the argument for dominance, made in Ref.~\cite{Balasubramanian:2021wgd}, only relies on the configuration of operators in the expression \eqref{eq:Renyi} and does not depend on the global geometry of the universes.
Specifically, in the computation of the overlaps, the indices of operators form a single loop only in the fully connected saddle, making this saddle dominate in the $\beta \rightarrow 0$ limit.
This occurs through the effects of matter contributions to the path integral, no matter what the gravitational contributions are.
This argument presented in Ref.~\cite{Balasubramanian:2021wgd} goes through here despite the difference in the cosmological constant of one universe.

We thus evaluate below the contribution from the fully connected saddle to ${\rm tr} \rho_{A}^{n}$ when $A$ and $B$ are de~Sitter and AdS black hole spacetimes, respectively.
The relevant expression is
\be
  Z_{n,{\rm conn}} = e^{-S_{{\rm grav}}[\mathcal{M}_{n}]} \sum_{\{i_{k},j_{k}\}} \left( \prod^{n}_{k=1} \s{p_{i_{k}}p_{j_{k}}} \right) \left\langle \prod^{n}_{k=1} \psi_{i_{k}}(\infty_{A_{k}}) \psi_{j_{k+1}}(0_{A_{k}}) \psi_{i_{k}}(\infty_{B_{k}}) \psi_{j_{k}}(0_{B_{k}}) \right\rangle_{\!\!\!\mathcal{M}_{n}}\!\!\!,
\label{eq:fullconaction}
\ee
where $\mathcal{M}_{n}$ represents the fully connected wormhole spacetime.

\begin{figure}
\vspace{-0.2cm}
  \begin{minipage}[b]{0.38\linewidth}
    \centering
    \raisebox{7mm}{\includegraphics[keepaspectratio,scale=0.38]{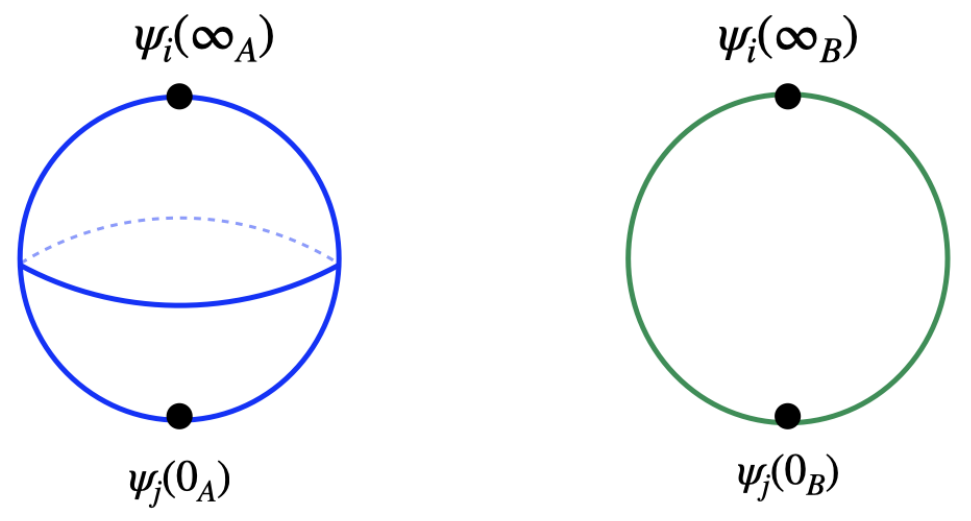}} 
  \end{minipage}
\hspace{7em}
  \begin{minipage}[b]{0.38\linewidth}
    \centering
    \includegraphics[keepaspectratio,scale=0.38]{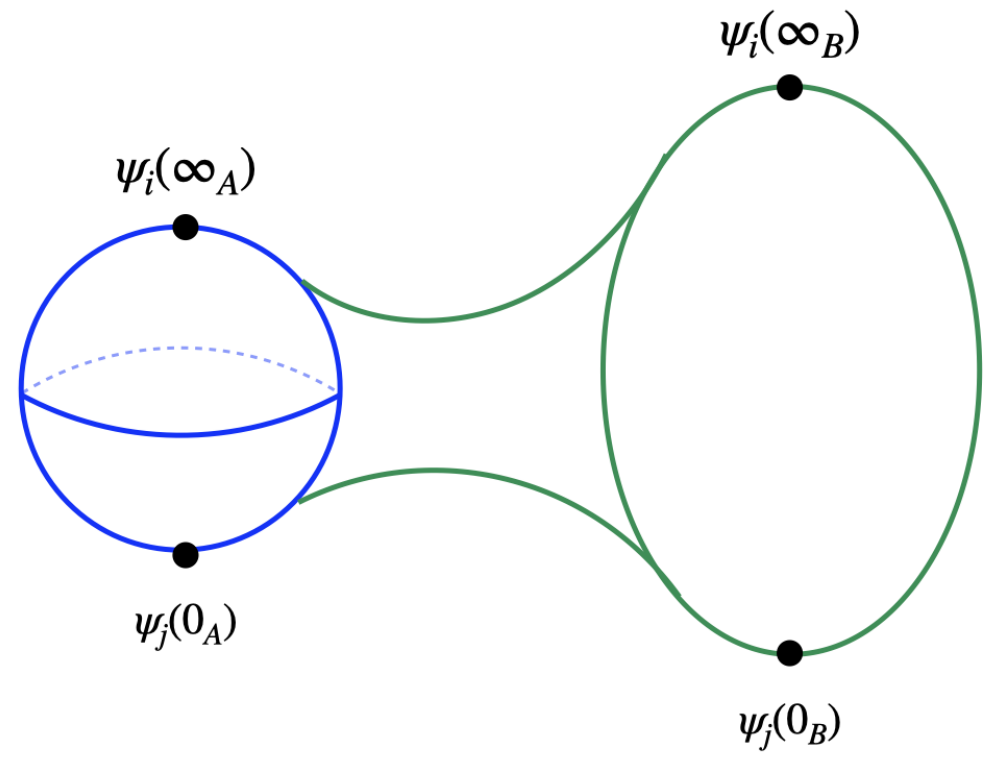}    
  \end{minipage}
\caption{
 {\bf Left}: The path integral preparations of overlaps $\la \psi_{i}| \psi_{j} \ra_{A}$ and $\la \psi_{i}| \psi_{j} \ra_{B}$ computed in Euclidean de~Sitter (blue) and AdS black hole (green) spacetimes.
 They appear in quantity~\eqref{eq:Renyi} of our interest.
 The excited states are specified by inserting corresponding operators.
 {\bf Right}: Euclidean de~Sitter and AdS black hole spacetimes are connected by a wormhole.
 Such a wormhole appears, for example, in the gravitational path integral~\eqref{eq:normalization} for the normalization of the reduced density matrix.
 We have already constructed this dS/AdS wormhole solution in JT gravity in Section~\ref{subsection:gluingAdS}.}
\label{fig:Wormhole1}
\end{figure}

\begin{figure}
\begin{center}
\includegraphics[scale=0.4]{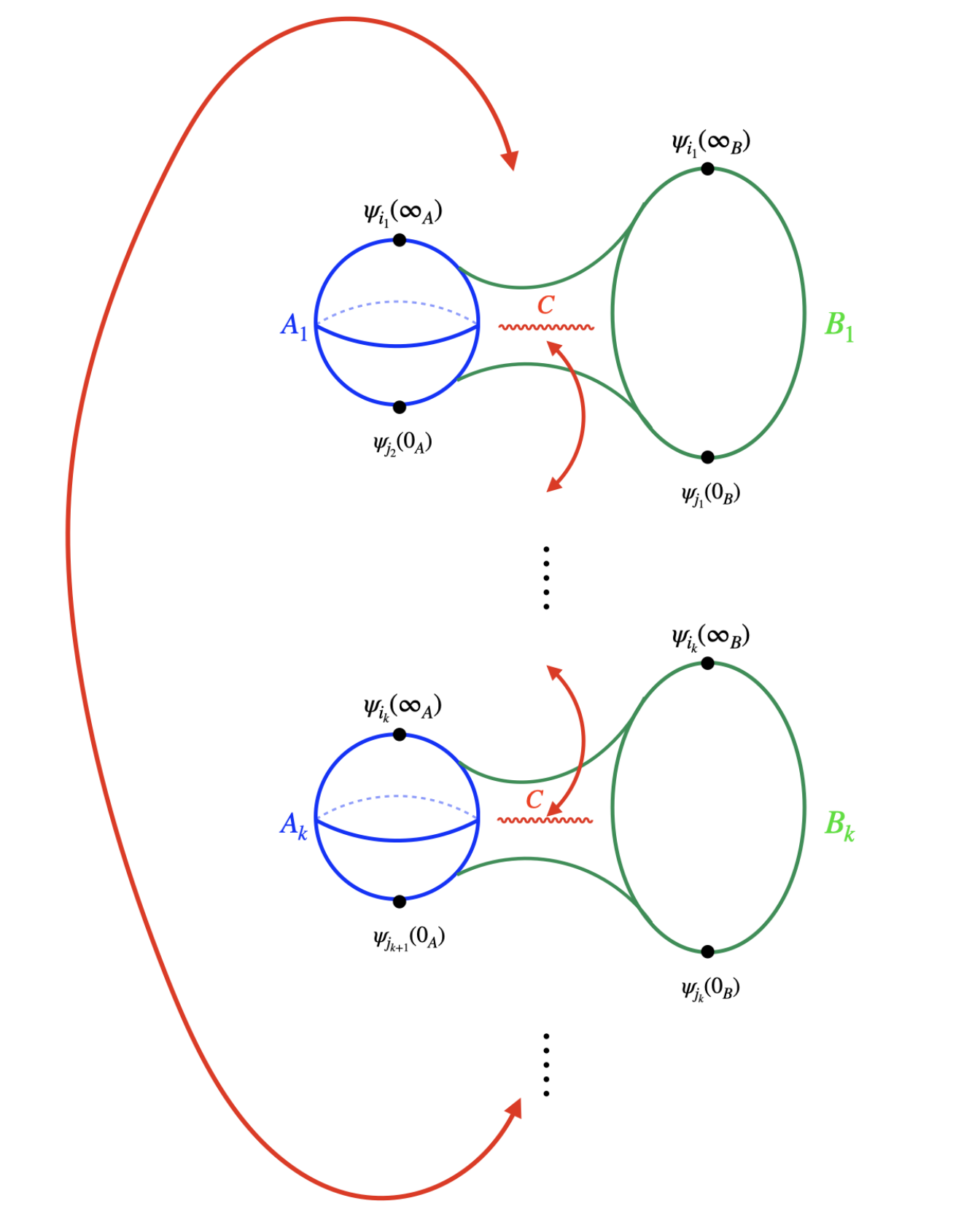}
\end{center}
\vspace{-0.3cm}
\caption{
 The fully connected saddle for the gravitational path integral, Eq.~\eqref{eq:fullconaction}, where all copies of Euclidean de~Sitter space as well as those of a Euclidean black hole are connected by a single wormhole.
 One way to think about it is that we first connect the $k$-th copy of Euclidean de~Sitter space and AdS black hole by a dS/AdS wormhole, and then we connect these wormholes by a replica wormhole (indicated by the red arrows).}
\label{fig:fullycon}
\end{figure}

One way to construct the fully connected wormhole $\mathcal{M}_{n}$ out of $2n$ universes $\{A_{k},B_{k}\}_{k=1}^{n}$ is as follows.
First, we connect $A_{k}$ and $B_{k}$ in the $k$-th replica by a wormhole as in Fig.~\ref{fig:Wormhole1}.
To do so, as we showed in Section~\ref{sec:dS-AdS-wormhole}, we poke a hole that has a circular boundary of size $b$ on $A$ (a sphere) as well as on $B$ (a disk), and then we glue $A$ and $B$ along these circular boundaries, where we place a domain wall.
We saw in Section~\ref{sec:dS-AdS-wormhole} that, after including the matter backreaction, the equations of motion can be solved to find a metric, dilaton, and domain wall trajectory consistent with the boundary conditions and the Israel junction conditions.
The resulting geometry $(A \# B)_{k}$ again has the topology of a disk, which we refer to as the dS/AdS wormhole.
Obviously, the size $b$ of the hole cannot exceed the size $R$ of the sphere.
Note that, as we showed in Section~\ref{sec:dS-AdS-wormhole}, the domain wall separating the two regions with different cosmological constants approaches the AdS boundary in the high entanglement temperature limit.
We then connect $n$ copies of these dS/AdS wormholes, $(A \# B)_{k}$ ($k = 1,\cdots,n$), by a replica wormhole.
This can be done by introducing a cut on each $(A \# B)_{k}$ and sewing these copies along the cut, as shown in Fig.~\ref{fig:fullycon}.
To emphasize that the fully connected wormhole depends on the cut $C$, we denote this spacetime by $\mathcal{M}_{n} \equiv \Sigma_{n}[C]$, and we assume $n > 1$ below.
For now we will proceed by assuming that the sewed geometry can be constructed, and we will later discuss how to select $C$ so that the equations of motion are satisfied.

\begin{figure}
\vspace{-0.8cm}
  \begin{minipage}[b]{0.45\linewidth}
    \centering
    \includegraphics[keepaspectratio, scale=0.37]{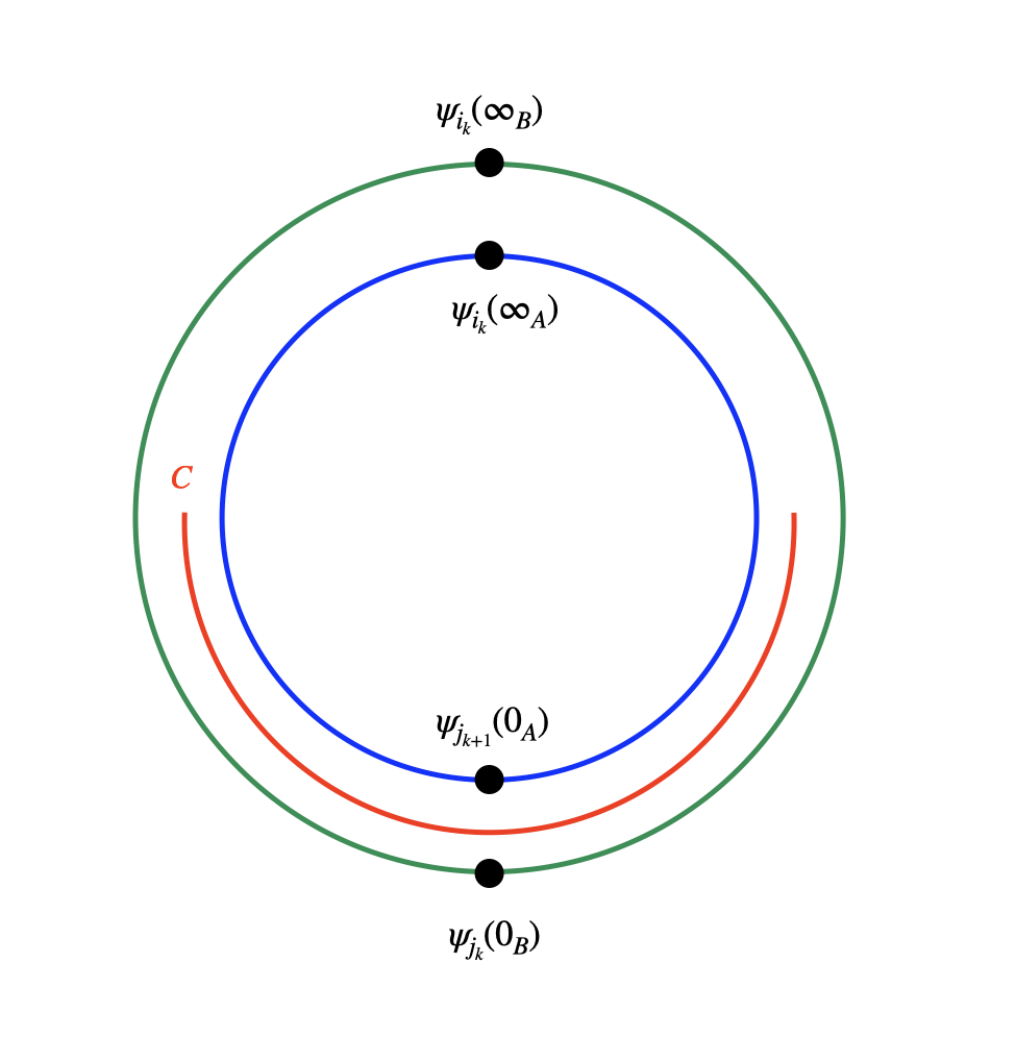}
  \end{minipage}
\hspace{2em}
  \begin{minipage}[b]{0.45\linewidth}
    \centering
    \includegraphics[keepaspectratio, scale=0.33]{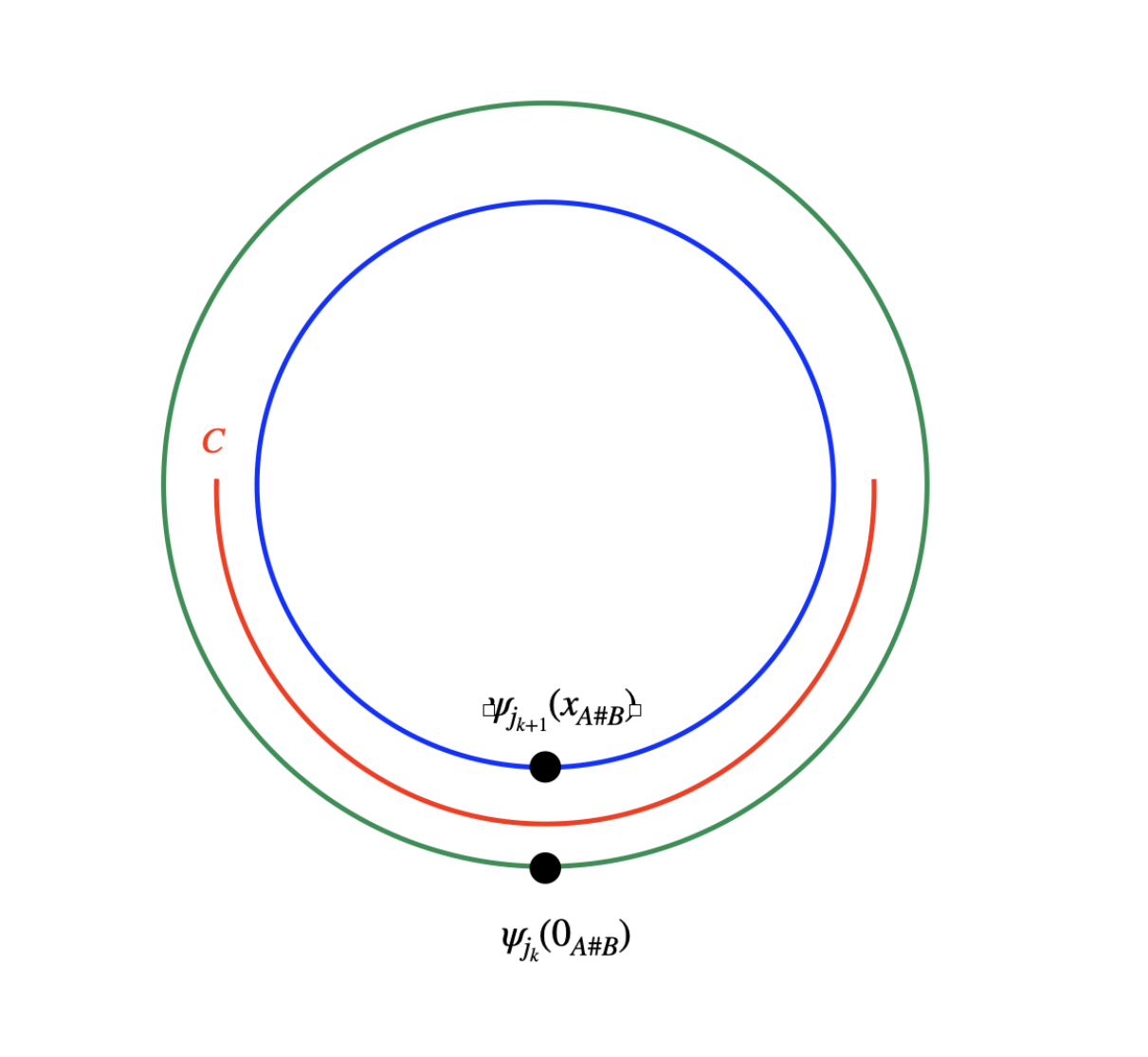}
  \end{minipage}
\vspace{-0.3cm}
\caption{
 {\bf Left}: Each of the replica copies in the fully connected wormhole depicted in Fig.~\ref{fig:fullycon} with four operator insertions, connected by cut $C$ (the red line).
 The contribution from this wormhole appears in the quantity of our interest, ${\rm tr} \rho_{A}^{n}$ in Eq.~\eqref{eq:Renyi}, and is given by Eq.~\eqref{eq:fullconaction}.
 The blue circle is the greater circle of the original Euclidean de~Sitter space with two operator insertions $\psi_{i_{k}}(\infty_{A})$ and $\psi_{j_{k+1}}(0_{A})$.
 {\bf Right}: The operator configuration in the same geometry after taking the OPE $\psi_{i_{k}}(\infty_{A_{k}}) \rightarrow \psi_{i_{k}}(\infty_{B_{k}})$ and factoring the corresponding component, which is suppressed.
 By summing over the leftover indices, this leads to Eq.~\eqref{eq:fullconsa}.
}
\label{fig:Wormhole2}   
\end{figure}

The contribution of the fully connected wormhole itself, Eq.~\eqref{eq:fullconaction}, does not in general have an interpretation in terms of a generalized entropy.
However, such an interpretation becomes available in the high entanglement temperature limit $\beta \rightarrow 0$, where the dS/AdS wormhole becomes shorter as described above.
In this limit, one can take the OPE $\psi_{i_{k}}(\infty_{A_{k}}) \rightarrow \psi_{i_{k}}(\infty_{B_{k}})$ in the correlator to get
\be
  Z_{n,{\rm conn}} = e^{-S_{{\rm grav}}[\mathcal{M}_{n}]} Z_{{\rm CFT}}[A\#B]^{n} \sum_{\{j_{k}\}} \left( \prod_{k=1}^{n} \s{p_{j_{k}}} \right) \left\langle \prod_{k=1}^{n} \psi_{j_{k+1}}(0_{A_{k}}) \psi_{j_{k}}(0_{B_{k}}) \right\rangle_{\!\!\Sigma_{n}[C]}\!\!.
\label{eq:shortlimit}
\ee
Here,
\begin{equation}
  Z_{{\rm CFT}}[A\#B] = \sum_{i} \s{p_{i}}\, \la \psi_{i} (\infty_{A}) \psi_{i} (\infty_{B}) \ra_{A\#B}.
\end{equation}
To emphasize that the operators in Eq.~\eqref{eq:shortlimit} are located on the new disk $A\#B$, made by gluing the sphere $A$ with the disk $B$, from now on we write $\psi_{j_{k+1}}(0_{A_{k}}) = \psi_{j_{k+1}}(x_{A\#B_{k}})$ and $\psi_{j_{k}}(0_{B_{k}}) = \psi_{j_{k}}(0_{A\#B_{k}})$, where $x_{A\#B_{k}}$ denotes the location of the operator $\psi_{j_{k+1}}$ in the $k$-th copy of the new disk $A\#B$, and $0_{A\#B_{k}}$ is its south pole.
Using this new notation, the correlator in Eq.~\eqref{eq:shortlimit} reads
\be
  \left\langle \prod_{k=1}^{n} \psi_{j_{k+1}}(0_{A_{k}}) \psi_{j_{k}}(0_{B_{k}}) \right\rangle_{\!\!\Sigma_{n}[C]}\!\! = \left\langle \prod_{k=1}^{n} \psi_{j_{k+1}}(x_{A\#B_{k}}) \psi_{j_{k}}(0_{A\#B_{k}}) \right\rangle_{\!\!\Sigma_{n}[C]}\!\!.
\ee
The series of operations described here is depicted in Fig.~\ref{fig:Wormhole2}.
Here, we have assumed that the cut goes between two operators.
The reason why will be explained in the next subsection.

We can now use the identity (see for example Appendix~A of~\cite{Balasubramanian:2020coy})
\be
  \left\langle \prod_{k=1}^{n} \psi_{j_{k+1}}(x_{A\#B_{k}}) \psi_{j_{k}}(0_{A\#B_{k}}) \right\rangle_{\!\!\Sigma_{n}[C]}\!\! = \left\langle \prod_{k=1}^{n} \psi_{j_{k}}(x_{A\#B_{k}}) \psi_{j_{k}}(0_{A\#B_{k}}) \right\rangle_{\!\!\Sigma_{n}[\bar{C}]}\!\!,
\label{eq:identity}
\ee
where $\bar{C}$ denotes the complement of the cut $C$ in a Cauchy slice of the disk $A\#B$ in radial quantization.
The above identity holds because, in the covering space where all the sheets are glued together along the indicated cut, the left- and right-hand sides of (\ref{eq:identity}) are actually the same correlator.
This  does not depend on the details of the geometry of each sheet, and hence holds for the dS/AdS wormhole as well.
By using this identity, we see that the sum over $\{j_{k}\}$ in Eq.~\eqref{eq:shortlimit} can be written in terms of the thermal CFT R\'enyi entropy
\be 
  \sum_{\{j_{k}\}} \left( \prod_{k=1}^{n} \s{p_{j_{k}}} \right) \left\langle \prod_{k=1}^{n} \psi_{j_{k}}(x_{A\#B_{k}}) \psi_{j_{k}}(0_{A\#B_{k}}) \right\rangle_{\!\!\Sigma_{n}[\bar{C}]}\!\! = Z_{{\rm CFT}}[A\#B]^{n}\; \f{{\rm tr}\bigl( \rho_{\f{\beta}{2},\bar{C}} \bigr)^{n}}{{\rm tr}\bigl( \rho_{{\rm vac},\bar{C}} \bigr)^{n}}.
\label{eq:corrden}
\ee
Here, we have defined
\be
  \rho_{\f{\beta}{2},\bar{C}} = \f{1}{Z_{{\rm CFT}}[A\#B]}\; \underset{C}{{\rm tr}}\!\left[ \sum_{i} \s{p_{i}}\; \psi_{i} (0_{A\#B}) |0 \ra \la 0| \psi_{i} (x_{A\#B}) \right],
\quad
  \rho_{{\rm vac},\bar{C}} = \underset{C}{{\rm tr}}\; |0 \ra \la 0|,
\label{eq:psedens}
\ee
where $|0 \ra$ is the vacuum on a time slice of the disk.
The quantity $\rho_{\f{\beta}{2},\bar{C}}$ looks like a density matrix; however, it is not Hermitian because the locations of the two operators in Eq.~\eqref{eq:psedens} are not reflection symmetric in time.
The von~Neumann entropy of such an object is sometimes called pseudo~entropy, and was studied recently in Ref.~\cite{Nakata:2020luh}.%
\footnote{
 In more details, let us consider the following density matrix like object involving two states $|\psi \ra$ and $| \phi \ra$
 \be
   \rho_{A, \psi| \phi} = \f{{\rm tr} |\psi \ra \la \phi |}{\la \psi | \phi \ra} \, .
 \ee
 Of course, this is not a density matrix; among other things, it is not Hermitian. The pseudo~entropy of $\rho_{A, \psi| \phi}$ is defined by  $S_{\rm PE} = -{\rm tr} \rho_{A, \psi| \phi} \log \rho_{A, \psi| \phi}$.
 $\rho_{\f{\beta}{2},\bar{C}}$ in Eq.~\eqref{eq:psedens} has precisely this form.}
A discussion of the associated island formula version appears in Ref.~\cite{Miyaji:2021lcq}.

Note that pseudo~entropy appears because one of the two universes here is de~Sitter space.
For instance, if $A$ and $B$ were both asymptotically AdS, the matter part of the entropy would be the usual entanglement entropy of bulk QFT.
This is because, in this case, the connected geometry is an annulus, and in the large entanglement limit the annulus pinches into two disks~\cite{Balasubramanian:2021wgd}.
In the new disks, the local operators are located at the boundaries.
This implies that the correlation functions still have interpretations in terms of CFT R\'enyi entropies.
In the current case, where one of the universes is Euclidean de~Sitter space, the complete connected geometry is a disk instead of an annulus.
As a result, the local operator originally located at the south pole of the de~Sitter sphere lies at a bulk point of the disk.
This prevents us from interpreting the correlation function as a standard R\'enyi entropy.

By inserting Eq.~\eqref{eq:corrden} into Eq.~\eqref{eq:shortlimit}, we obtain
\be
  Z_{n,{\rm conn}} = e^{-S_{{\rm grav}}[\mathcal{M}_{n}]} Z_{{\rm CFT}}[A\#B]^{2n}\; \f{{\rm tr}\bigl( \rho_{\f{\beta}{2},\bar{C}} \bigr)^{n}}
{{\rm tr}\bigl( \rho_{{\rm vac},\bar{C}} \bigr)^{n}}.
\label{eq:fullconsa}
\ee
Note that the corresponding expression for $Z_{1}$ does not have the last factor
\be
  Z_{1,{\rm conn}} = e^{-S_{{\rm grav}}[A\#B]} Z_{{\rm CFT}}[A\#B]^{2},
\ee
since there is no cut in the spacetime $A\#B$ in this case.
We thus obtain the following expression for (the exponential of) the R\'enyi entropy in the high temperature limit $\beta \rightarrow 0$:
\be
  {\rm tr} \rho_{A}^{n} \,\xrightarrow[\beta \rightarrow 0]{}\, \f{Z_{n,{\rm conn}}}{Z_{1,{\rm conn}}^{n}} = e^{-S_{{\rm grav}}[\mathcal{M}_{n}] + n S_{{\rm grav}}[A\#B]}\, \f{{\rm tr}\bigl( \rho_{\f{\beta}{2},\bar{C}} \bigr)^{n}}{{\rm tr}\bigl( \rho_{{\rm vac},\bar{C}} \bigr)^{n}}.
\label{eq:rho-n_final}
\ee

As argued in Ref.~\cite{Lewkowycz:2013nqa}, the gravitational part of Eq.~\eqref{eq:rho-n_final}, $e^{-S_{{\rm grav}}[\mathcal{M}_{n}] +n S_{{\rm grav}}[A\#B]}$, picks up, in the $n \rightarrow 1$ limit,  the area of the fixed point of $\mathbb{Z}_{n}$ replica symmetry in $\mathcal{M}_{n}$, which coincides with the boundary of $C$.
This contribution is given by $e^{-(n-1)A[\p \bar{C}]/4G_{\rm N}}$.
Finally, the location of the fixed point $\p \bar{C}$ is determined by extremizing the total gravitational path integral.
By taking the $n \rightarrow 1$ limit of the R\'enyi entropy,  we thus find a formula for the entanglement entropy of the form in Eq.~\eqref{eq:dSAdSentropyformula}:
\be
  S(\rho_A) = {\rm Min}\;\underset{\bar{C}}{{\rm Ext}} \left[ \f{A[\p \bar{C}]}{4G_{\rm N}} + S_{\rm PE}[\bar{C}] - S_{{\rm vac}}[\bar{C}]\right],
\quad
  S_{\rm PE}[\bar{C}] = -{\rm tr}\bigl(\rho_{\frac{\beta}{2},\bar{C}} \ln\rho_{\frac{\beta}{2},\bar{C}} \bigr),
\label{eq:genends}
\ee
where the extremization is performed on the spacetime $A\#B$.
The ``Min'' in the formula indicates that if there are multiple extremal surfaces, we choose the one giving the minimal value.

To evaluate the actual value of the entropy, we need to know detailed properties of $A\#B$.
While the bulk QFT entropy part of the above formula appears somewhat unusual, we will see that in the high temperature limit $\beta \rightarrow 0$, the entropy is dominated by the area term, and the bulk pseudo~entropy part does not play an important role.
This can be explicitly shown by going back to the expression of ${\rm tr}\bigl( \rho_{\f{\beta}{2},\bar{C}} \bigr)^{n}$ (whose $n \rightarrow 1$ limit yields the pseudo~entropy) written in terms of the correlator, Eq.~\eqref{eq:corrden}.
It is straightforward there to see that the correlator gets further factorized into two point functions when $\bar{C}$ is small, and in this limit it is canceled by the normalization factor of $\rho_{\f{\beta}{2},\bar{C}}$ in \eqref{eq:psedens}.
Indeed, $\bar{C}$ must become small in the large entanglement temperature limit because the entropy of entanglement cannot be greater than the entropy of the black hole or of de~Sitter; if $\bar{C}$ remained of finite size in the high temperature limit, then the bulk entropy part would become larger than the horizon areas, and hence the resulting entanglement entropy as well.
In the small $\bar{C}$ limit, the bulk entropy part as well as its variation with respect to the endpoints of the cut is almost vanishing, so that the endpoints have to be located at the classical extremal surfaces, i.e.\ the horizons.
The net result is that in Eq.~\eqref{eq:rho-n_final}, $\rho_{\f{\beta}{2},\bar{C}}$ is replaced with the vacuum reduced density matrix $\rho_{{\rm vac},\bar{C}}$, so that the contribution of the pseudo~entropy vanishes in this limit.

\subsection{Classification of possible cuts}
\label{subsection:possiblecuts}

As discussed above, in the the Lorentzian continuation of the dS/AdS wormhole, the endpoints of the cut on which the replicas are connected lie near one of the horizons when the entanglement temperature is large.
While the relevant horizon could a~priori be either the cosmological horizon or the AdS black hole horizon, in this subsection we argue that when $S_{\rm dS} > S_{\rm BH}$ (which must be the case when the wormhole exists), the cut will not end near the cosmological horizon.

\begin{figure}
\vspace{-0.5cm}
  \hspace{-1.7em}
  \begin{minipage}[b]{0.25\linewidth}
    \centering
\raisebox{3 mm}{\includegraphics[keepaspectratio, scale=0.35]{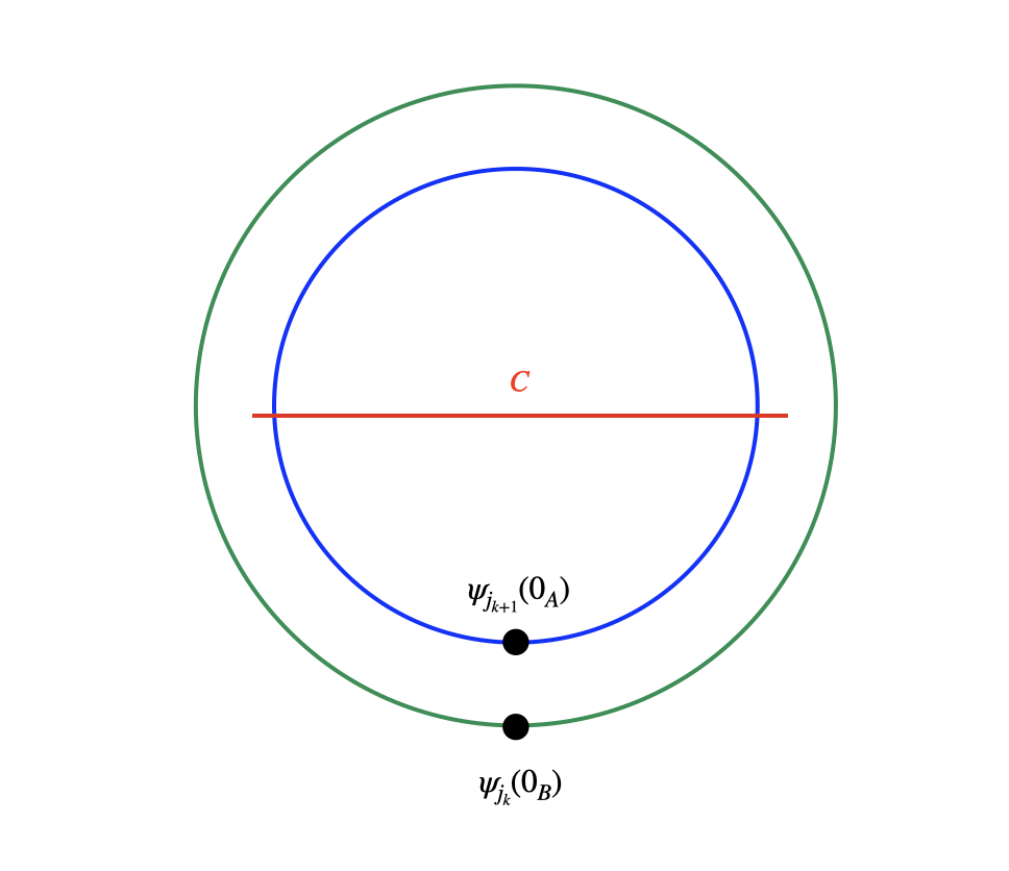} }   
  \end{minipage}
  \hspace{4.4em}
  \begin{minipage}[b]{0.25\linewidth}
    \centering
    \includegraphics[keepaspectratio, scale=0.35]{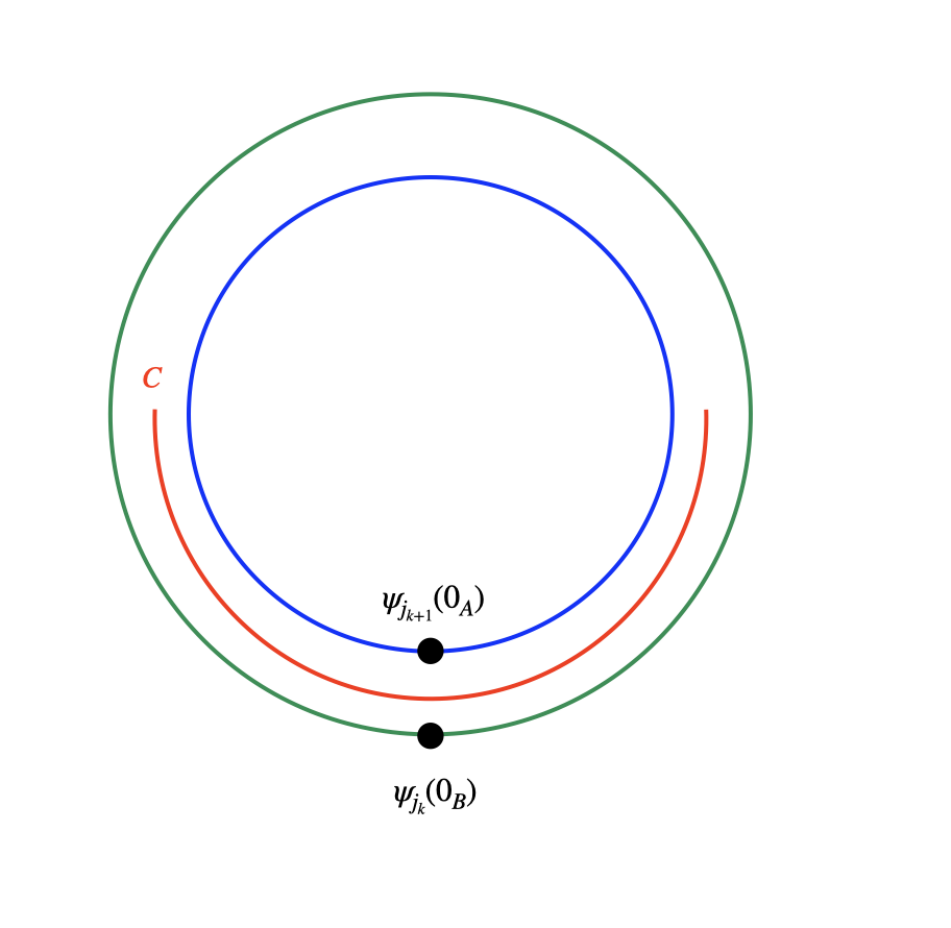}    
  \end{minipage}
  \hspace{3em}
    \begin{minipage}[b]{0.25\linewidth}
    \centering
\raisebox{4 mm}{\includegraphics[keepaspectratio, scale=0.35]{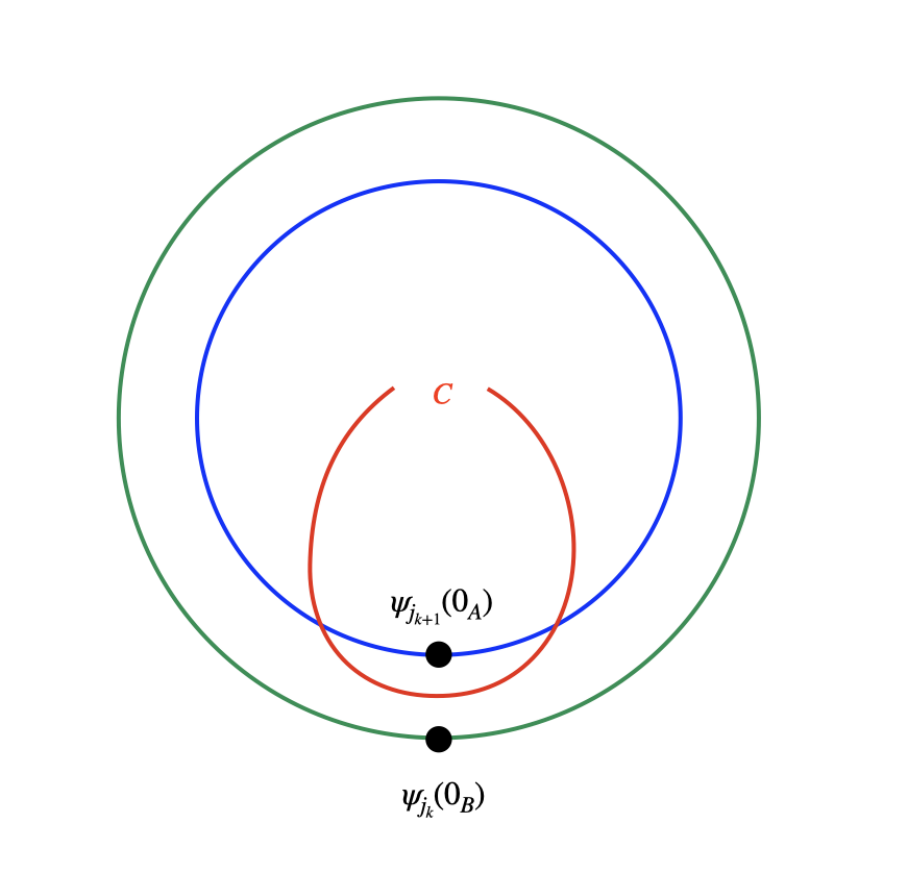}}   
  \end{minipage}
\vspace{-0.2cm}
\caption{
 Three possible cuts $C$ for the fully connected wormhole $\mathcal{M}_{n}$ which contributes to the gravitational path integral for the R\'enyi entropy as in Eq.~\eqref{eq:fullconsa}.
 {\bf Left: } type~A cut located above two operators.
 Since two operators in the disk are not always the same, this type of cuts only gives a subdominant contribution to the gravitational path integral \eqref{eq:shortlimit}.
 {\bf Center: } type~B cut whose endpoints are located near the bifurcation surfaces of the AdS black hole.
 {\bf Right:} type~B cut which ends near the cosmological horizon.
 One of the purposes of this subsection is to argue that the endpoints of the cut of this type are not precisely on the cosmological horizon.
 (If they were, it would result in a vanishing entropy $S(\rho_{A})=0$.)}
\label{fig:cuts}
\end{figure}

The fully connected saddle was constructed by gluing $n$ copies of the dS/AdS wormholes along a cut $C$.
We begin by discussing in which region of the disk the cut should be located in order to maximize the value of the gravitational action.
It is convenient to separate the discussion into three cases, depending on the location of the cut.
First, we separate two classes: (1) type~A: the cut is located above two operators, as in the left panel of Fig.~\ref{fig:cuts}, and (2) type~B: the cut is located in between two operators as in the middle and right panels of Fig.~\ref{fig:cuts}.
We would like to compute and compare the saddlepoint action of the R\'enyi entropy in Eq.~\eqref{eq:shortlimit} for these two cases.
In both cases, there are two further possibilities that we discussed above, namely that the endpoints of the cut are located near the cosmological horizon or the AdS black hole horizon.

Below we will be interested in evaluating the correlators in \eqref{eq:shortlimit} in the limit where $C$ covers most of a time slice of the Euclidean strip, so that the size of its complement in the same slice will be small $|\bar{C}| \rightarrow 0$.
For the type~A case, since the cut $C$ is located above the two  operators $\psi_{j_{k+1}} (0_{A_{k}})$ and $\psi_{j_{k}} (0_{B_{k}})$ (the left panel of Fig.~\ref{fig:cuts}) we can deform $C$ upwards to its complement $\bar{C}$ without crossing the operators.
This implies that the correlation function in \eqref{eq:shortlimit} is invariant under the deformation from $C$ to $\bar{C}$:
\be
  \left\langle \prod_{k=1}^{n} \psi_{j_{k+1}}(x_{A\#B_{k}}) \psi_{j_{k}}(0_{A\#B_{k}}) \right\rangle_{\!\!\Sigma_{n}[C]}\!\! = \left\langle \prod_{k=1}^{n} \psi_{j_{k+1}}(x_{A\#B_{k}}) \psi_{j_{k}}(0_{A\#B_{k}}) \right\rangle_{\!\!\Sigma_{n}[\bar{C}]}\!\! \, .
\ee
Notice that in the above equality the indices of the operator in the left- and right-hand sides are identical; the only difference is the replacement $C \rightarrow \bar{C}$.
The identity holds because we are fixing the endpoints of the cut.
When its size is large, $|\bar{C}| \rightarrow 0$, the right-hand side factorizes into a product of two point functions
\be
  \left\langle \prod_{k=1}^{n} \psi_{j_{k+1}}(x_{A\#B_{k}}) \psi_{j_{k}}(0_{A\#B_{k}}) \right\rangle_{\!\!\Sigma_{n}[\bar{C}]}\!\! \rightarrow \prod_{k=1}^{n} \left\langle \psi_{j_{k+1}}(x_{A\#B_{k}}) \psi_{j_{k}}(0_{A\#B_{k}}) \right\rangle_{\Sigma_{n}[\bar{C}]}
\quad
  \mbox{for } |\bar{C}| \rightarrow 0.
\ee
Since the indices of two operators in the expectation value on the right side are not generally identical, we pick up a Kronecker delta $\delta_{j_{k+1} j_{k}}$.
This Kronecker delta reduces the value of the saddlepoint action in the high temperature limit because the correlator contributes to the saddlepoint action as in Eq.~\eqref{eq:corrden}, so that the resulting sum with respect to the indices is significantly reduced in the $\beta \rightarrow 0$ limit.

On the other hand, for a type~B cut, which is located in between the two operators (the middle and right panels of Fig.~\ref{fig:cuts}), during the deformation $C \rightarrow \bar{C}$ the cut has to cross at least one of these two operators.
In this case, the relevant identity is
\be
  \left\langle \prod_{k=1}^{n} \psi_{j_{k+1}}(x_{A\#B_{k}}) \psi_{j_{k}}(0_{A\#B_{k}}) \right\rangle_{\!\!\Sigma_{n}[C]}\!\! = \left\langle \prod_{k=1}^{n} \psi_{j_{k}}(x_{A\#B_{k}}) \psi_{j_{k}}(0_{A\#B_{k}}) \right\rangle_{\!\!\Sigma_{n}[\bar{C}]}\!\!, 
\ee
where, in the correlation function on the right side, the indices of the operators on the same sheet are identical.
This implies that in the large cut limit, $|\bar{C}| \rightarrow 0$, the correlator on the right-hand side again factorizes
\be
  \left\langle \prod_{k=1}^{n} \psi_{j_{k}}(x_{A\#B_{k}}) \psi_{j_{k}}(0_{A\#B_{k}}) \right\rangle_{\!\!\Sigma_{n}[\bar{C}]}\rightarrow\prod_{k=1}^{n}  \left\langle \psi_{j_{k}}(x_{A\#B_{k}}) \psi_{j_{k}}(0_{A\#B_{k}}) \right\rangle_{\Sigma_{n}[\bar{C}]},
\ee
but in this case each two point function contains two identical operators.
Thus, the sum in Eq.~\eqref{eq:corrden} with respect to the indices is not suppressed.

By comparing these two behaviors, we conclude that type~B cuts give the dominant contribution in the gravitational path integral in the high temperature limit.
There are two possibilities for type~B cuts.
Since a cut almost entirely covers the time slice when $\beta \rightarrow 0$, as explained at the end of the previous subsection, the endpoints of the cut will be located near either the cosmological or black hole horizon.
Therefore, the resulting entropy $S(\rho_{A})$ is almost twice the horizon area.
Since the connected geometry only exists when the entropy of the cosmological horizon $S_{\rm dS}$ is larger than that of the black hole horizon $S_{\rm BH}$, we expect that the cut will stretch between two black hole horizons as it gives a smaller area term in Eq.~\eqref{eq:genends}.
For this to be the case, however, we must make sure that the cut does not occupy the entire time slice of the disk; this could happen if the endpoints of the cut were located precisely on the cosmological horizon, in which case we would find $S(\rho_{A})=0$.
We will argue below that this does not happen.

\begin{figure}
\vspace{-0.5cm}
  \begin{minipage}[b]{0.25\linewidth}
    \centering
\raisebox{2.5 mm}{\includegraphics[keepaspectratio, scale=0.35]{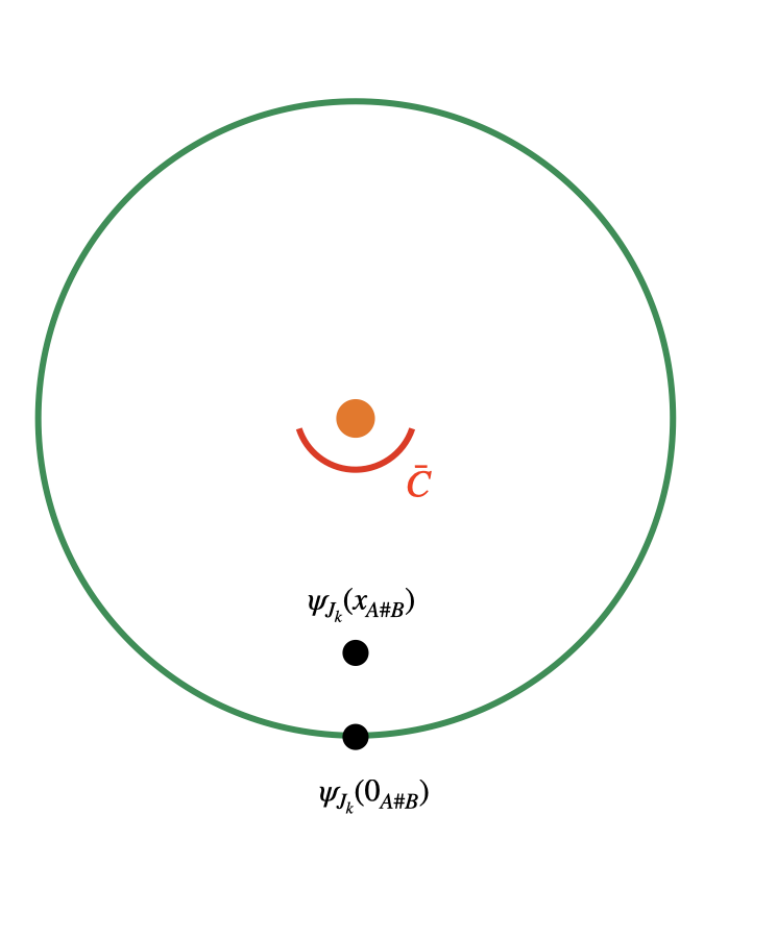} }   
  \end{minipage}
  \hspace{3em}
  \begin{minipage}[b]{0.25\linewidth}
    \centering
    \includegraphics[keepaspectratio, scale=0.35]{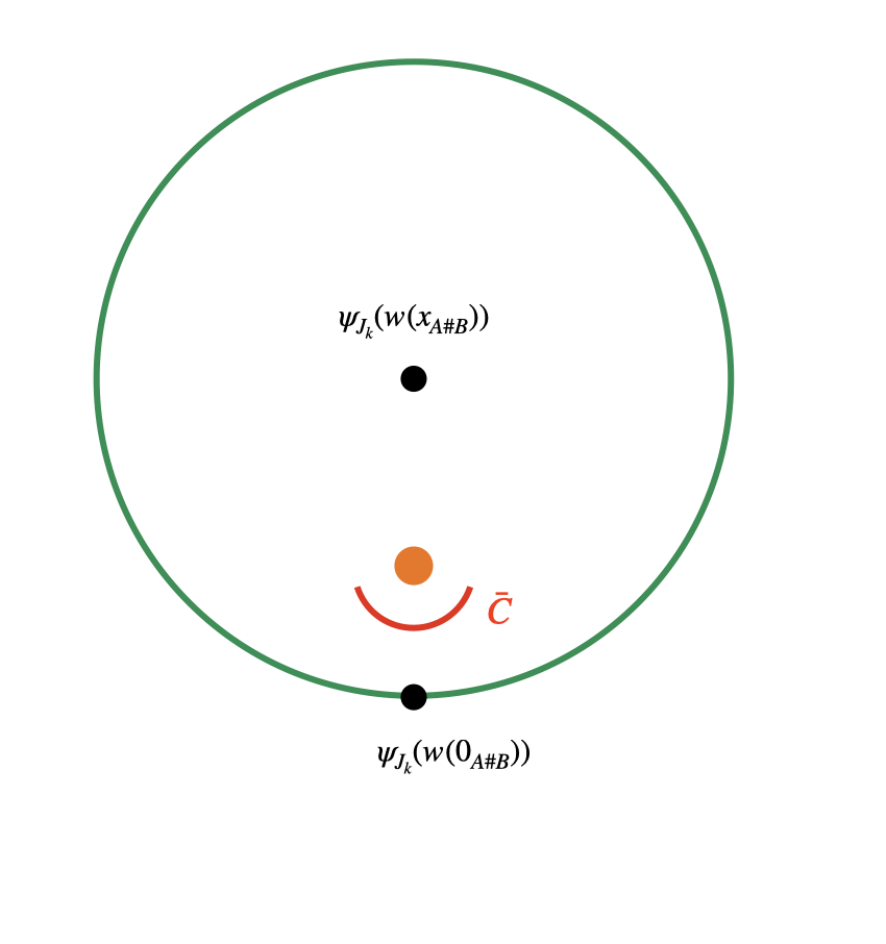}    
  \end{minipage}
  \hspace{3em}
    \begin{minipage}[b]{0.25\linewidth}
    \centering
\raisebox{4 mm}{\includegraphics[keepaspectratio, scale=0.35]{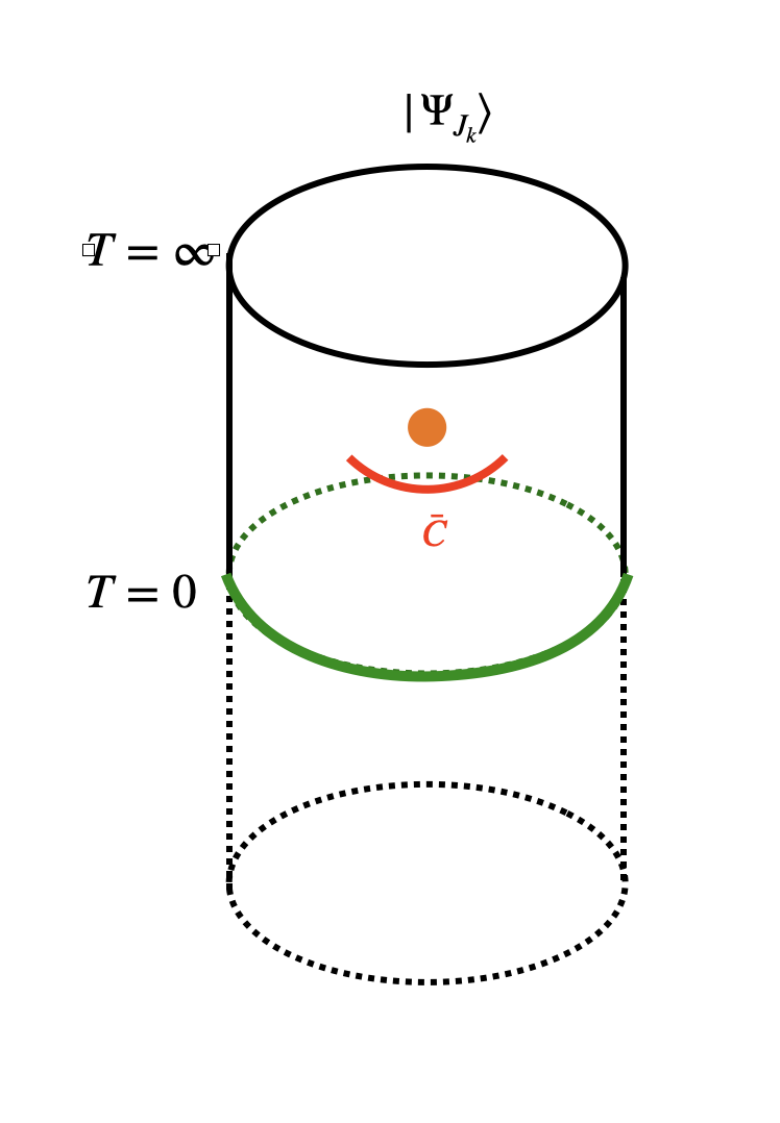}}   
  \end{minipage}
\vspace{-0.5cm}
\caption{
 Sequence of two dimensional surfaces used to compute the CFT sum \eqref{eq:CFTsum}.
 {\bf Left:} $z$ disk on which the correlation functions in \eqref{eq:CFTsum} are originally defined.
 The cosmological horizon is depicted by the orange dot at the center, and the endpoints of the cut $\bar{C}$ are located near the horizon.
 {\bf Center:} $w$ disk on which one of the operators are located at the center of this disk.
 The $w$ and $z$ disks are related by the map \eqref{eq:auto}.
 {\bf Right:} The $w$ disk is mapped to the $T>0$ part of the cylinder by an exponential map.
 The operator at the center of the $w$ disk is now at $T = \infty$ and represented by an excited state $|\psi_{J_{k}} \ra$ by the state operator correspondence.
 The $T < 0$ part of the cylinder is provided by the copy of the $w$ disk, which naturally emerges when we compute the correlation function on the disk by the doubling trick as in Eq.~\eqref{eq:spherecorr}.
 This results in the same excited state $|\psi_{J_{k}} \ra$ appearing at $T = -\infty$ as well.
 In the cylinder frame, Eq.~\eqref{eq:spherecorr} has an interpretation as a R\'enyi entropy, as in Eq.~\eqref{eq:square}.
}
\label{fig:cut2}
\end{figure}

To show this, we explicitly compute the location of the quantum extremal surface by assuming it is located near the cosmological horizon (the left panel of Fig.~\ref{fig:cut2}).
For this purpose, the central task is to evaluate the sum of the CFT correlators
\be 
  \f{1}{Z_{1,{\rm CFT}}^{n}}\! \left( \sum_{\{j_{k}\}} \prod^{n}_{k=1}\! \s{p_{j_{k}}} \right)\!\! \left\langle \prod_{k=1}^{n}\! \psi_{j_{k}}(x_{A\#B_{k}}) \psi_{j_{k}}(0_{A\#B_{k}})\!\! \right\rangle_{\!\!\!\Sigma_{n}[\bar{C}]}\!\!\!\!\!\!\!\!\!\!,
\quad
  Z_{1,{\rm CFT}} =\sum_{j}\! \s{p_{j}} \la \psi_{j} (x_{A\# B}) \psi_{j}(0_{A\# B}) \ra
\label{eq:CFTsum}
\ee
on the $n$ sheeted cover $\Sigma_{n}[C]$ of the disk (dS/AdS wormhole).

In the disk describing the original dS/AdS wormhole, in which $Z_{n,{\rm conn}}$ in \eqref{eq:shortlimit} was defined, this quantity does not itself have an interpretation as a R\'enyi entropy as one of the operators $\psi_{j_{k}}(x_{A\#B_{k}})$ is not located at a pole of the disk.
Our goal here is to map this disk to another one on which the sum of the CFT correlators \eqref{eq:CFTsum} has an interpretation as a R\'enyi entropy of an excited state $|\psi_{J} \ra$.
We can do this in several steps.
Let $z,\bz$ be the coordinates on the original disk, and map it to a second disk (which we call $w$ disk) by
\be
  w(z) = \left(\f{z-\alpha}{1+\bar{\alpha}z} \right),
\quad
  \alpha = x_{A\#B_{k}}.
\label{eq:auto}
\ee
The conformal boundary of the new disk is at $|w| = 1$.
The purpose of applying this map is to relocate $\psi_{j_{k}}(x_{A\#B_{k}})$ to the center of the new disk $w = \bar{w} = 0$.
Also, since $\psi_{j}(0_{A\# B})$ is located at the conformal boundary of the original $z$ disk, it is mapped to a point on the boundary $|w| = 1$.
See the central panel of Fig.~\ref{fig:cut2} for an illustration.

If we pick up two terms  in the sum of the CFT correlators defined in \eqref{eq:CFTsum}, then these two correlators will transform differently under the conformal map because the operators $\psi_j$'s involved have different conformal dimensions (see the definition of the entangled state in \eqref{eq:TFD-state_AdS-AdS}).
However, in the large temperature limit, the sum is dominated by correlators consisting of operators with the particular conformal dimension $\Delta =E_{J}(\beta)$ fixed by the temperature.
This results in the sum $\sum_{j}$ over all states being replaced with a sum $\sum_{J}$ over the states at the fixed energy.
In this saddlepoint approximation, the sum in \eqref{eq:CFTsum} transforms uniformly, i.e.\ just by the multiplication of a Jacobian factor of the form $|\p w /\p z|^{\Delta}$, since all the operators have the same conformal dimension.
Now, in the left expression in Eq.~\eqref{eq:CFTsum}, the total Jacobian factor from the correlators in the numerator is canceled by the analogous factor from the denominator.
Therefore, this expression is evaluated on the $w$ disk as
\be
  \f{1}{Z_{1,{\rm CFT}}^{n}} \sum_{\{J_{k}\}} \left( \prod^{n}_{k=1} \s{p_{J_{k}}} \right) \left\langle \prod_{k=1}^{n} \psi_{J_{k}}(w(x_{A\#B_{k}})) \psi_{J_{k}}(w(0_{A\#B_{k}})) \right\rangle_{\!\!\Sigma_{n}[\bar{C}]},
\label{eq:cftcorrw1}
\ee
where
\be
  Z_{1,{\rm CFT}} =\sum_{J} \s{p_{J}} \left\langle \psi_{J} (w(x_{A\# B})) \psi_{J}( w(0_{A\# B})) \right \rangle.
\label{eq:cftcorrw2}
\ee
The only differences between the above expression and \eqref{eq:CFTsum} are the replacements $x_{A\#B_{k}} \rightarrow w(x_{A\#B_{k}})$ and $0_{A\#B_{k}} \rightarrow w(0_{A\#B_{k}})$ and the restriction on the range of the sum.

We now evaluate each correlator in \eqref{eq:cftcorrw1} and \eqref{eq:cftcorrw2} defined on the disk $A \# B$ via the standard doubling trick.
To do so, we prepare a mirror copy $\widetilde{A \# B}$ of the disk with the same operator insertions and then glue $A \# B$ and $\widetilde{A \# B}$ along the conformal boundary.
The resulting manifold is a sphere $S^{2}$, and the correlator on the disk is equal to the correlator on the sphere obtained by doubling the operator insertions
\begin{align}
  I &= \left\langle \prod_{k=1}^{n} \psi_{J_{k}}(w(x_{A\#B_{k}})) \psi_{J_{k}}(w(0_{A\#B_{k}})) \right\rangle_{\!\!\Sigma_{n}[\bar{C}]}
\\
  &= \left\langle \prod_{k=1}^{n} \psi_{J_{k}}(w(x_{ A\# B}{k})) \psi_{J_{k}}(w(0_{A\#B_{k}})) \psi_{J_{k}}(\tilde{w}(x_{A\#B_{k}})) \psi_{J_{k}}(\tilde{w}(0_{A\#B_{k}})) \right\rangle_{\!\!\Sigma_{n,S^{2}}[\bar{C}\cup \tilde{\bar{C}}]}\!\!,
\label{eq:spherecorr}
\end{align}
where $\tilde{w}(x_{A\#B_{k}})$ and $\tilde{w}(0_{A\#B_{k}})$ are mirror images of $w(x_{A\#B_{k}})$ and $w(0_{A\#B_{k}})$ in the copy $\widetilde{A\#B}$.
Since we have chosen the map \eqref{eq:auto} so that $w(x_{A\#B_{k}})$ is at the center of the disk, on the sphere $\psi_{J_k}(w(x_{A\#B_{k}}))$ is at its north pole $u_{k}=\infty_{k}$ and its mirror $\psi_{J_k}(\tilde{w}(x_{A\#B_{k}})$ is at the south pole $u_{k}=0_{k}$, where $(u, \bar{u})$ are the coordinates of the sphere.
Furthermore, the remaining two operators in \eqref{eq:spherecorr} are at the identical point on the equator; so we can use the OPE to fuse them, giving the identity as the leading term which dominates as the operators actually coincide.
Also, since there is a copy of the cut $\bar{C}$ in the mirror, on the sphere we have a disjoint union of cuts.
Therefore, we denote the resulting branched sphere by $\Sigma_{n,S^{2}}[\bar{C}\cup\tilde{\bar{C}}]$, and we indicated this explicitly in \eqref{eq:spherecorr}.
In summary, by a sequence of the above operations, the CFT correlator \eqref{eq:CFTsum} becomes
\be
  I_{n} = \f{1}{Z_{1,{\rm CFT}}^{n}} \sum_{\{J_{k}\}}\! \left( \prod^{n}_{k=1}\! \s{p_{i_{k}}} \right)\! \left\langle \prod_{k=1}^{n} \psi_{J_{k}}(\infty_{k}) \psi_{J_{k}}(0_{k}) \right\rangle_{\!\!\Sigma_{n,S^{2}}[\bar{C} \cup \tilde{\bar{C}}]}\!\!\!,
\;
  Z_{1,{\rm CFT}} = \sum_{i}\! \s{p_{i}} \la \psi_{j} (\infty) \psi_{j}( 0) \ra.
\label{eq:expr}
\ee

The expression $I_{n}$ in Eq.~\eqref{eq:expr} is nothing but the thermal R\'enyi entropy of the region $\bar{C} \cup \tilde{\bar{C}}$ on the cylinder, divided by the vacuum R\'enyi entropy~\cite{Balasubramanian:2020coy}
\be
  I_{n} =\f{{\rm tr} (\rho_{\beta/2,\bar{C} \cup \tilde{\bar{C}}})^{n}}{{\rm tr} (\rho_{{\rm vac},\bar{C} \cup \tilde{\bar{C}}})^{n}}.
\label{eq:CFTans}
\ee
Concretely, the sphere and cylinder are related in the following way.
Let $(\Theta, T)$ be the coordinates on the cylinder, where $T$ is the coordinate of the Euclidean timelike direction, $-\infty < T < \infty$, and $\Theta$ is the coordinate for the spatial direction, $0 < \Theta < 2\pi$.
Then, the $w$ disk and the $T>0$ part of the cylinder are related by the map $w= e^{T+i \Theta}$, and there is a similar map between the copy of the $w$ disk (which was used to form the sphere $S^{2}$) to the $T<0$ part of the disk.
The relation is illustrated in the right panel of Fig.~\ref{fig:cut2}.

In the high temperature limit, the size of the cut $|\bar{C}|$ gets small, so the R\'enyi entropy factorizes as 
\be
  {\rm tr}(\rho_{\beta/2,\bar{C} \cup \tilde{\bar{C}}})^{n} = {\rm tr} (\rho_{\beta/2,\bar{C}})^{n}{\rm tr} (\rho_{\beta/2,\tilde{\bar{C}}})^{n} = {\rm tr} (\rho_{\beta/2,\bar{C}})^{2n},
\ee
where we have used ${\rm tr}(\rho_{\beta/2,\bar{C}})^{n} = {\rm tr}(\rho_{\beta/2,\tilde{\bar{C}}})^{n}$ because $\tilde{\bar{C}}$ is a copy of $\bar{C}$.
There is a similar factorization for ${\rm tr}(\rho_{{\rm vac},\bar{C} \cup \tilde{\bar{C}}})^{n}$.
So we have
\be
  I_{n} = \left(\f{{\rm tr}(\rho_{\beta/2,\bar{C}})^{n}}{{\rm tr}(\rho_{{\rm vac},\bar{C} })^{n}} \right)^{2}.
\label{eq:square}
\ee
In general, the R\'enyi entropy of a thermal state is a theory dependent quantity, but for a 2d CFT with a holographic dual, it has the simple form
\be
  {\rm tr} (\rho_{\beta/2,\bar{C}})^{n} = \f{1}{\left(\sinh \f{2\pi |\bar{C}|}{\beta}\right)^{\Delta_{n}}},
\qquad
  {\rm tr} (\rho_{{\rm vac},\bar{C}})^{n} = \f{1}{\left(\sin \f{ \pi|\bar{C}|}{L} \right)^{\Delta_{n}} },
\label{eq:CFT-Renyi}
\ee
where $L$ is the size of the spatial circle of the cylinder, and
\be
  \Delta_{n} = \f{c}{12} \left(n-\f{1}{n} \right).
\ee

So far, we have been computing the CFT part of \eqref{eq:shortlimit} by making use of the doubling trick, and the remaining task is to evaluate the gravitational part.
In the $n \rightarrow 1$ limit, it picks up the value of the dilaton profile at the endpoints of the cut $e^{-(n-1)\Phi[\p C] }$, as in Ref.~\cite{Almheiri:2019qdq}.
Since we are now assuming that the endpoints are located near the cosmological horizon on the dS/AdS wormhole,  the relevant dilaton profile is that of the de~Sitter side $\Phi_{{\rm dS}} = B \cos\theta$.
Since the static patch coordinates $(t, \theta)$ defined in \eqref{eq:dstcoord} and the holomorphic and anti-holomorphic coordinates $(z, \bz)$ are related at least locally near the cosmological bifurcation surface at $\theta = 0$ by the map $z = e^{it} \sin\theta$ and $\bz = e^{-it} \sin\theta$, the dilaton profile in the latter coordinates is $\Phi_{\rm dS}(z,\bar{z}) = B \s{1-z\bar{z}}$.
We further map the $z$ disk to the $w$ disk by \eqref{eq:auto}, and then to the cylinder by the exponential map $w=e^{T+i \Theta}$.
This allows us to map the dilaton profile to the cylinder.

By combining the result for the CFT part $I_{n}$ \eqref{eq:CFTans} and the one for the gravity part $e^{-(n-1)\Phi[\p C]}$, we get the full contribution of the type~B cut whose endpoints are located near the cosmological horizon (right panel of Fig.~\ref{fig:cuts}).
By taking the $n \rightarrow 1$ limit, we get the generalized entropy
\be
  S_{{\rm gen}} [\bar{C}] = \Phi_{\rm dS}[\partial \bar{C}] + 2(S_{\beta}[\bar{C}] - S_{{\rm vac}}[\bar{C}]) \, .
\label{eq:genenta}
\ee
The factor of $2$ in front of the bulk entropy part comes from the square in Eq.~\eqref{eq:square}.
For a CFT with a holographic dual we would also have
\be
  S_{\beta}[\bar{C}] = \f{c}{3} \log\left[ \f{\beta}{\pi} \sinh\f{2\pi |\bar{C}|}{\beta} \right],
\quad
  S_{{\rm vac}}[\bar{C}] = \f{c}{3} \log\left[ \f{L}{\pi} \sin\f{\pi |\bar{C}|}{L} \right].
  \label{eq:universalrenyi}
\ee
In fact, in the high temperature limit where $\beta \rightarrow 0$, the entanglement entropy of the thermal density matrix of any CFT becomes universal.
This is because the limit $\beta /L \rightarrow 0$, where $L$ is the size of the spatial slice, is equivalent for a CFT to the limit $\L \rightarrow \infty$.
So, at high entanglement temperature, the thermal density matrix of any 2d CFT acts as if it is defined on an infinitely long line.
Thermal entanglement entropy of a 2d CFT on a line is computed by applying a conformal map to the vacuum entanglement entropy~\cite{Calabrese:2004eu}, and is given by Eq.~\eqref{eq:universalrenyi}.
The remaining task is to extremize Eq.~\eqref{eq:genenta} over the cut $C$ to make the fully connected replica wormhole $\mathcal{M}_{n}$ on shell.

By denoting the location of the cosmological horizon on the cylinder by $(T,\Theta) =(T_{0},\Theta_{0})$, we make the ansatz for $\bar{C}: -x < \Theta-\Theta_{0} <x,\, T = T_{0}$.
By expanding the dilaton near the cosmological horizon, we obtain
\be 
  S_{{\rm gen}}(x) = \bar{\phi} \left( 1-\f{|\alpha|^{2} x^{2}}{(1-|\alpha|^{2})} \right) + \frac{\pi x}{\beta}.
\ee
This means that the equation $\p_{x} S_{{\rm gen}} (x) = 0$ has a solution  $x \neq 0$, and  therefore  $\bar{C}$ is non-vanishing.
We thus conclude that if the endpoints of a type~B cut are located near the de~Sitter horizon, then its contribution to the entanglement entropy is non-vanishing and given by $S(\rho_{A}) = 2 S_{{\rm dS}}$.
However, if the endpoints are located near the bifurcation surface of the AdS black hole, we get $S(\rho_{A}) =2 S_{{\rm BH}}$.
Recalling that extremization will select the smaller of these, and that the connected solution exists only when $S_{\rm dS} > S_{\rm BH}$, we find that the entropy coincides with the twice entropy of the black hole on the AdS side.

\paragraph{Summary:}
In Ref.~\cite{Balasubramanian:2020xqf}, the entanglement entropy of a thermofield double type state \eqref{eq:TFD-state_AdS-AdS} defined on gravitating de~Sitter space (universe $A$), described by the dilaton profile \eqref{eq:dSdil}, and a non-gravitating reference system  (universe $B$) was studied using the replica trick.
In that case, the cut (or the ``island'' after analytically continuing to Lorentzian signature) covers the entire time slice of de~Sitter space.
One way to interpret this is that the entanglement entropy vanishes and that the Hilbert space on the de~Sitter space is one dimensional.%
\footnote{
 In Re.~\cite{Balasubramanian:2020xqf}, a scenario was also proposed in which the inclusion of end-of-the-world branes on the de~Sitter geometry led to a cut that did not occupy the entire Cauchy slice, and hence implied a finite entropy.
 In Section~\ref{sec:algebras}, we will comment on the relation between this scenario and the present paper.
}
It is interesting to understand the relationship between this previous finding and the current ones.

For this purpose, it is useful to recall from Ref.~\cite{Balasubramanian:2020xqf} the possible types of islands in de~Sitter spacetime described by the dilaton profile \eqref{eq:dSdil}.
The dilaton profile has classical extremal surfaces at $\varphi = 0$ and $\varphi =\pi$.
Since the dilaton is maximized at $\varphi =0$, this corresponds to the de~Sitter cosmological horizon, and similarly $\varphi =\pi$ corresponds to the de~Sitter black hole horizon because the dilaton is minimized there.
So there are two types of possible island region $C$, namely one that ends near the cosmological horizon (called type~I in Ref.~\cite{Balasubramanian:2020xqf}; see Figure~4 there), and the other is the region whose endpoints are near the black hole horizon (called type~III islands in Ref.~\cite{Balasubramanian:2020xqf}).

The type~I island does not cover the entire Cauchy slice because the area of the cosmological horizon is locally maximum.
To see this concretely, let us choose the ansatz for the complement of the island $\bar{C}$ by $-x < \varphi < x$, and since we are interested in the high temperature limit we assume $x \ll 1$.
Then the generalized entropy for island is schematically given by $S_{{\rm gen}}(x) \sim B(1-x^{2}) + x/\beta$ when $x \ll 1$, where the first term comes from the dilaton profile \eqref{eq:dSdil} and the second term is the bulk entropy part.
Therefore, the solution of the $\p_{x} S_{{\rm gen}}(x) = 0$ condition is nontrivial, i.e., $x \neq 0$.
This means that the endpoints of the island are not precisely at the cosmological horizon.
In this argument, the crucial thing was that the dilaton is locally maximal so the sign of the coefficient of $x^{2}$ is negative.
On the other hand, the endpoints of the type~III island are precisely on the black hole horizon, because it is a minimal surface.
Again this can be seen by making the ansatz for $\bar{C}$ by $-x < \varphi - \pi < x$; then the generalized entropy for the region is schematically of the form $S_{{\rm gen}}(x) \sim B(-1+x^{2}) + x/\beta$, so the equation $\p_{x} S_{{\rm gen}}(x) = 0$ only has a solution at $x=0$.
Therefore, the type~III island always covers the entire Cauchy slice and dominates the entropy.

Now let us come back to our current case where we entangle de~Sitter space with an AdS black hole.
As we have argued, when $S_{\rm dS} > S_{\rm BH}$ and in the high temperature limit $\beta \rightarrow 0$, the generalized entropy on the dS/AdS wormhole computes the entanglement entropy of the thermofield double state.
In this case, there is no type~III island because the de~Sitter bubble region of the dS/AdS wormhole does not contain the de~Sitter black hole horizon.
Therefore, the only possibilities are that the island ends near the cosmological horizon which is an analogue of the type~I island in Ref.~\cite{Balasubramanian:2020xqf}, or the island ends near the AdS black hole horizon, and hence does not cover the entire Cauchy slice.
This leads to a finite entropy.

\subsection{Summary: Entanglement entropy and extremal surfaces}

We can now combine all our results to arrive at a formula for the entanglement entropy between the two universes \eqref{eq:replicatrick1} in various limits.
The result depends on the three parameters $\{ S_{\rm dS}, S_{\rm BH}, \beta \}$.
Since we compute the entanglement entropy by taking $n \rightarrow 1$ limit of the R\'enyi entropy which involves two types of gravitational path integrals, namely $Z_{n}$ defined in \eqref{eq:Renyi} and $Z_{1}$ coming from the normalization of the reduced density matrix \eqref{eq:normalization}, it is useful to discuss the dominant saddles for these path integrals separately.
We will arrive at a phase diagram for the entropy by dialing the values of the entanglement temperature $T = 1/\beta$ and the de~Sitter horizon entropy while fixing the value of the entropy of the AdS black hole $S_{\rm BH}$.

\paragraph{Saddles for $Z_{1}$:} 
There are two possible saddles, namely the connected saddle and the disconnected saddle as in Fig.~\ref{fig:Wormhole1}.
They are both consistent with the boundary condition for the gravitational path integral, since when they are continued to Lorentzian signature they both possess the future and past infinities of de~Sitter space and the conformal boundary of AdS space.
The contribution of each saddle is computed in almost the same manner as in Eqs.~\eqref{eq:zdisconn} and \eqref{eq:zconn}, by replacing one of the AdS universes in that example by Euclidean de~Sitter.
In particular, the connected saddle, if it exists, becomes dominant  in the high entanglement temperature limit $\beta \rightarrow 0$.
The only difference from the AdS/AdS scenario is that in the current dS/AdS case, the connected saddle exists only when the de~Sitter entropy is larger than the entropy of the AdS black hole as shown in Section~\ref{subsection:gluingAdS}.
In summary, when $S_{\rm dS} < S_{\rm BH}$, the disconnected saddle is always the dominant one, but when $S_{\rm dS} > S_{\rm BH}$ the connected saddle becomes the dominant one above a critical temperature.
 
\paragraph{Saddles for $Z_{n}$:}
We have listed all the saddles for $Z_{n}$ in Section~\ref{subsubsec:infty} (also see  Fig.~\ref{fig:wormholes}).
Among them, the relevant saddles are (a) the fully connected one whose contribution is given by \eqref{eq:fullconaction}, (b) the replica wormhole that only connects the de~Sitter replicas, drawn in top right of Fig.~\ref{fig:wormholes}, and (c) the fully disconnected saddle in top left of the same figure.

\subsection*{Case 1: $S_{\rm dS} < S_{\rm BH}$}

Again when $S_{\rm dS} < S_{\rm BH}$, there is no wormhole connecting the two universes in a manner consistent with the boundary conditions, since, as explained above, any wormhole connecting de~Sitter and AdS has to satisfy the condition $S_{\rm dS} > S_{\rm BH}$ because the cosmological horizon is a locally maximal surface.
Therefore, in all the saddles of the R\'enyi entropy \eqref{eq:Renyi}, de~Sitter space and the AdS black hole must be disconnected when $S_{\rm dS} < S_{\rm BH}$.
However, it is still possible to connect copies of de~Sitter by a replica wormhole.
As explained in Section~\ref{sec:vanish}, the result is that the dominant saddle for the R\'enyi entropy includes this replica wormhole connecting copies of de~Sitter, leaving copies of the AdS black hole disconnected.
The entropy is computed by the same island formula for states on de~Sitter entangled with a non-gravitating reference system found in Ref.~\cite{Balasubramanian:2020xqf}.
Therefore, as discussed earlier, the entanglement entropy vanishes regardless of the temperature,%
\footnote{
 The contribution of the fully disconnected saddle gives the QFT result for the entanglement entropy which coincides with the thermal entropy.
 Therefore, compared with the contribution from the de~Sitter only replica wormhole saddle, it is always subdominant.
}
and the de~Sitter Hilbert space $H_{\rm dS}$ looks one dimensional.
In this case, the entanglement wedge of the AdS black hole, or more accurately the dual CFT Hilbert space, covers the entire Cauchy slice.
Thus the de~Sitter region is reconstructable from the Hilbert space of the asymptotically AdS universe.
This makes sense as the black hole’s Hilbert space has room to accommodate all the states in de~Sitter.

\subsection*{Case 2: $S_{\rm dS} > S_{\rm BH}$}

When $S_{\rm dS} > S_{\rm BH}$, there is a dS/AdS wormhole.
As a result, the fully connected wormhole where all copies of universes $A$ and $B$ are connected by a single wormhole exists as well (bottom right of Fig.~\ref{fig:wormholes}).
In the low temperature regime, $\beta \gg 1$, the disconnected saddle dominates for $Z_{1}$.
Meanwhile for $Z_{n}$ the replica wormhole just connecting the de~Sitter factors dominates.
Thus, just as in the $S_{\rm dS} < S_{\rm BH}$ case, the entanglement entropy is vanishing.

However, in the high temperature regime where the connected saddle dominates $Z_{1}$, the fully connected wormhole dominates $Z_{n}$ because of the matter contribution as explained in Section~\ref{sec:vanish}.
In Section~\ref{sec:genentropy}, we showed that when the fully connected wormhole dominates $Z_{n}$, and the dS/AdS wormhole dominates $Z_{1}$, the entanglement entropy is given by a formula which almost look like a generalized entropy \eqref{eq:genends}, except that the bulk entropy part is replaced by a pseudo~entropy on the dS/AdS wormhole.
When the connected wormhole geometry is continued to Lorentzian signature, its Penrose diagram looks like Fig.~\ref{fig:Lorentzian}.
As we discussed, there are then two possibilities for the location of the quantum extremal surface, one near the de~Sitter horizon and the other near the AdS black hole horizon.
However, as we argued in Section~\eqref{subsection:possiblecuts}, the quantum extremal surface has to be always located near the AdS black hole horizon, as this gives the dominant answer.
Therefore, the entanglement entropy is equal to twice the entropy of the AdS black hole.

Thus, when $S_{\rm dS} > S_{\rm BH}$ our analysis predicts that the von~Neumann entropy goes from zero to twice the black hole entropy as we go from low to high entanglement temperature.

\begin{figure}
\begin{center}
  \includegraphics[scale=0.4]{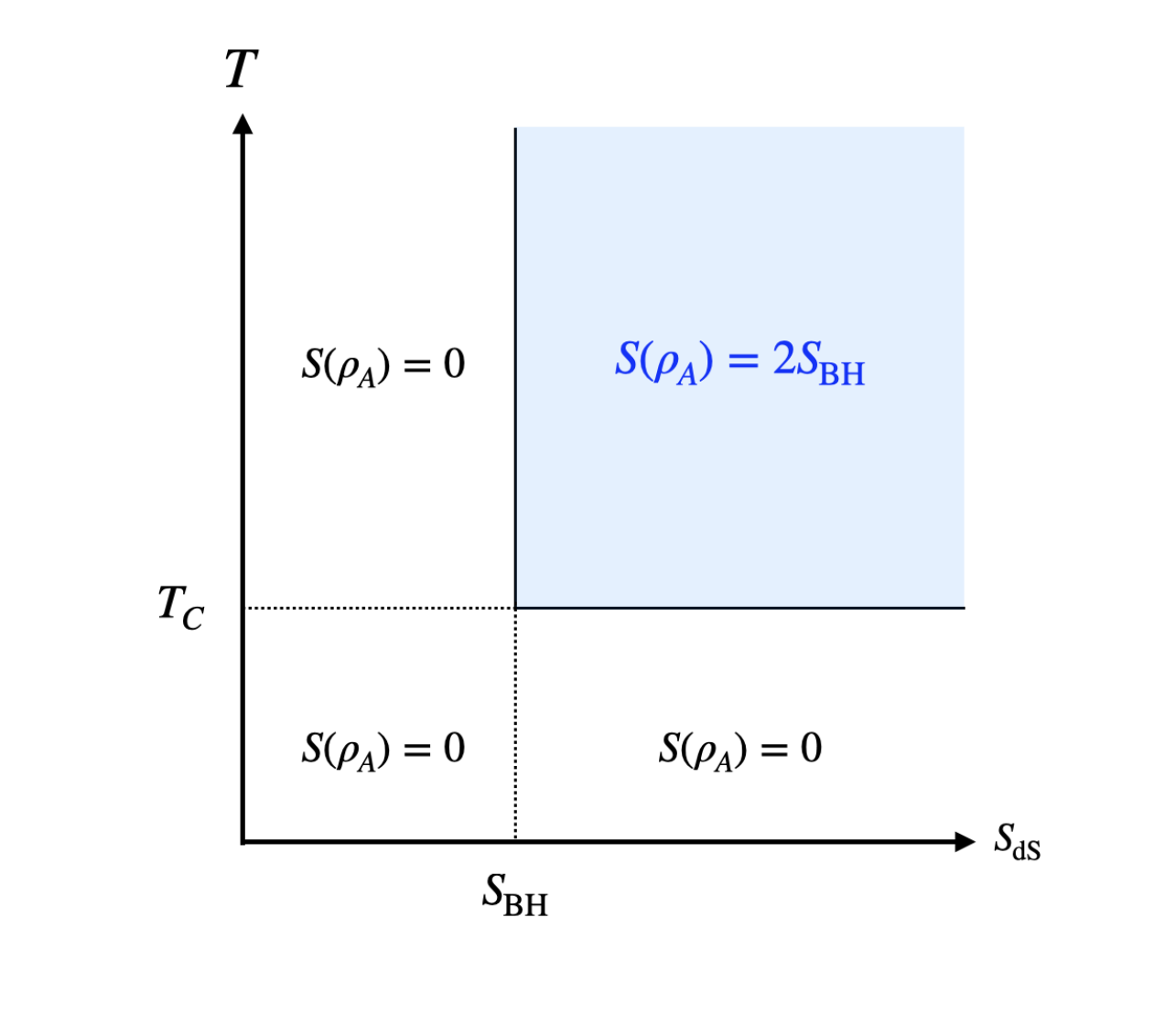}
\end{center}
\vspace{-0.8cm}
\caption{
 Phase diagram for the entanglement entropy $S(\rho_A)$ (which is equal to $S(\rho_B)$ as the entire state is pure).
 In drawing this diagram, we fix the entropy of the AdS black hole and change the entanglement temperature and the de~Sitter entropy.
 The entanglement entropy is non-vanishing only when $S_{\rm dS} > S_{\rm BH}$ and temperature is large, where the fully connected saddle dominates the gravitational path integral for the R\'enyi entropy.
}
\label{fig:phased}
\end{figure}

\paragraph{Phases of the von~Neumann entropy:}
The behavior of the von~Neumann entropy in the parameter space $\{ S_{\rm dS}, S_{\rm BH}, \beta \}$ is depicted in Fig.~\ref{fig:phased}.
We have not studied the various transitions in detail, and hence have not established whether they are sharp or whether there is a smooth crossover.
Indeed, we expect that when we go beyond the saddlepoint approximation, the transitions will be smoothed out, at least with respect to the temperature.

\section{Interpretation of our results}
\label{sec:algebras} 

The authors of Ref.~\cite{Chandrasekaran:2022cip} studied the von~Neumann algebra for the static patch of de~Sitter space in the presence of gravity in a weak coupling limit, where matter does not backreact on the geometry.
In a gravitating systems all symmetries must be gauged, in particular when the time slice is compact and without  boundary.
The von~Neumann algebra of excitations in the static patch is therefore obtained from the algebra of the matter QFT as the subalgebra that commutes with all symmetry generators of the static patch.
Naively, the commutant with the static patch symmetries is trivial because the elements of the algebra that commute with the charges are c-numbers~\cite{Chandrasekaran:2022cip}.
To get a nontrivial algebra, we thus need to introduce an observer of the static patch with its own Hilbert space and associated algebra.
We can then impose the constraint that the combined algebra of de~Sitter space and the observer commutes with the generators of the de~Sitter isometries.
The algebra obtained in this way supports the reduced density matrix on the static patch  and is of type II$_{1}$~\cite{Chandrasekaran:2022cip}.
A type II$_{1}$ algebra has a maximally entropic state which naturally corresponds to empty de~Sitter space with the Hartle-Hawking QFT state on it.
Furthermore, the fact that the density matrix which maximizes the entropy is proportional to the identity operator naturally realizes the expectation that a reduced density matrix on the static patch has a flat spectrum~\cite{Dong:2018cuv}.

These ideas align naturally with our analyses.
Although the authors of Ref.~\cite{Chandrasekaran:2022cip} study the weak gravity limit ($G_{\rm N} \ll 1$) while we account for gravitational backreaction, our conclusions are similar---de~Sitter space acts as if it has a finite entropy only when viewed by a gravitating observer.
In our case the observer is an AdS black hole, and the observation in question occurs through quantum entanglement.
In our case, if the entanglement is too weak, de~Sitter space acts as if it has vanishing entropy, and it seems that there is a threshold beyond which the effects associated to strong entanglement are sufficient to constitute a ``gravitational observer,'' at least in the saddlepoint approximation.
Note that the wormhole saddlepoint is present even at weak entanglement---it just does not dominate the path integral.
So we should really expect a steep crossover of some kind, or a sharp phase transition, with the entanglement entropy ramping from near zero to a plateau set by the entropy of the observer as the entanglement strength is increased (see Fig.~\ref{fig:phased}).

One interesting feature of our result is that de~Sitter space acts as if it has a finite entropy only when the observer, an AdS black hole, has a {\it lower} Bekenstein-Hawking entropy than the nominal entropy of the cosmological horizon.
Note that the Bekenstein-Hawking formula $S_{\rm BH} = A_{\rm BH}/4G_{\rm N}$ can be regarded as bounding the logarithm of the Hilbert space dimension at a given energy in the presence of gravity.
If there is no gravity in the AdS space, i.e.\ $G_{\rm N} \rightarrow 0$ in $S_{\rm BH}$, there is no such bound, and our results would suggest that de~Sitter space entangled with a non-gravitating observer should have vanishing entropy.
This is consistent with earlier work in which de~Sitter space seems to have vanishing entropy when ``observed'' by entanglement with a non-gravitating observer~\cite{Almheiri:2019hni,Balasubramanian:2020xqf,Hartman:2020khs,Chen:2020tes}.

In fact, the authors of Ref.~\cite{Balasubramanian:2020xqf} also proposed ways in which an apparently non-gravitating observer could nevertheless lead to a finite de~Sitter entropy.
One approach suggested there was to imagine decompactifying de~Sitter space and adding end-of-the-world branes to terminate the geometry at, say, the poles of global de~Sitter space.
From the perspective of Ref.~\cite{Chandrasekaran:2022cip} and the current paper, these branes which are coupled to the background geometry, act as gravitational observers.
Indeed, they play a role similar to the domain walls separating the de~Sitter and AdS regions of the wormholes described here.

In this paper, we studied the properties of such wormholes in the simplest setting:\ two-dimensional dS and AdS JT gravity with scalar matter producing a domain wall separating the regions with different cosmological constants.
The low dimensionality and simple action made it possible to work out the backreaction of the stress tensor of an entangled state on the background geometry.
There are pathways to generalize our analysis to higher dimensions and more realistic theories of gravity.
For example, excited states on the AdS black hole would still be prepared by a Euclidean path integral with operators inserted on the boundary.
In this paper, the domain wall separating the dS and AdS parts of the geometry was nucleated by energy density created by such injected particles that propagate behind the black hole horizon, recalling the construction of Ref.~\cite{Mirbabayi:2020grb} (reviewed in the Appendix).
In higher dimensions we could similarly create shells of matter that propagate behind the horizon~\cite{Balasubramanian:2022gmo,Balasubramanian:2022lnw} and decay to form domain walls with a different interior cosmological constant.
From this perspective, and following the reasoning of Refs.~\cite{Balasubramanian:2022gmo,Balasubramanian:2022lnw}, the de~Sitter bubbles we have described can be thought of exotic microstates of the exterior AdS black hole.

We have shown that the wormhole geometry exists only when the area of the de~Sitter horizon is larger than the area of the horizon of the observing AdS black hole.
In this regime, the reduced density matrix of the AdS black hole side obtained by tracing out the degrees of freedom in the de~Sitter side is almost maximally mixed at high entanglement temperature since the von~Neumann entropy coincides with black hole horizon entropy.
If we regard the AdS black hole as a gravitating observer of de~Sitter space, this means that the observer cannot get easily information about the de~Sitter microstates through its quantum entanglement, and hence will view it as an ensemble whose size is quantified by the entropy.%
\footnote{
 Note that this is different from saying that microstates of a black hole are highly complex and hence are effectively inaccessible to asymptotic observers using simple probes~\cite{Balasubramanian:2005kk}.
 Here, the black hole, with all of its complexity, is the device which we are using to probe de~Sitter space.}
One might speculate that when the $S_{{\rm dS}} < S_{\rm BH}$, the observer has enough ``resolving power'' to actually sense the de~Sitter microstate, thereby leading to a vanishing entropy.

Finally, in this paper we modeled a gravitating observer by an AdS black hole.
However, if our general paradigm is valid, we should be able to use any gravitating observer.
For example we could model the observer as a black hole in flat space (see~\cite{Miyata:2021ncm} for related work), or perhaps even another de~Sitter space, although in the latter case we will have to confront the absence of any asymptotic regions on a Cauchy slice.
It would be interesting to analyze these cases in detail.

\section*{Acknowledgments}

We thank Arjun Kar and Edgar Shaghoulian for discussions during the early stage of this research.
VB was supported by the Department of Energy through DE-SC0013528 and QuantISED DE-SC0020360, as well as by the Simons Foundation through the It From Qubit Collaboration (Grant No.\ 38559).
YN was supported in part by the Department of Energy, Office of Science, Office of High Energy Physics under QuantISED award DE-SC0019380 and contract DE-AC02-05CH11231 and in part by MEXT KAKENHI grant number JP20H05850, JP20H05860.
TU was supported in part by JSPS Grant-in-Aid for Young Scientists 19K14716 and in part by MEXT KAKENHI Grant-in-Aid for Transformative Research Areas~A ``Extreme Universe'' No.21H05184.
VB thanks the Santa Fe Institute and the Aspen Center for Physics, which is supported by National Science Foundation grant PHY-2210452, for hospitality as this work was completed.

\appendix

\section{Creating de~Sitter bubbles in AdS}
\label{app:mirbabayi}

In this appendix, we briefly review a Euclidean path integral preparation of an AdS black hole containing a de~Sitter bubble in its interior in the presence of a source on the AdS boundary, as discussed by Mirbabayi~\cite{Mirbabayi:2020grb}.
In this work, it was pointed out that an excitation within a Euclidean AdS black hole can decay into a domain wall whose interior has a de~Sitter geometry.
Here we will explain the detailed properties of this configuration, and its relation to our analysis.

The construction begins with the gravitational description of an excitation in Euclidean AdS black hole emanating from a point in the asymptotic boundary, traveling in the bulk, and ending at another point on the boundary.
The backreaction of such a particle is treated by Israel junction conditions.
In detail, the recipe is as follows:
\begin{itemize}
\item Step 1: Prepare two identical Euclidean disks, each of which  is continued to a Lorentzian eternal AdS black hole containing only one bifurcation surface (say the left AdS and the right AdS).
\item Step 2: Then introduce a brane in each of these two disks in the same way, and glue them along the brane in a $\mathbb{Z}_{2}$ symmetric manner (see the left panel of Fig.~\ref{fig:Wormhole3}).
\begin{figure}
\vspace{-3mm}
  \begin{minipage}[b]{0.45\linewidth}
    \centering
\raisebox{5mm}{\includegraphics[keepaspectratio, scale=0.40]{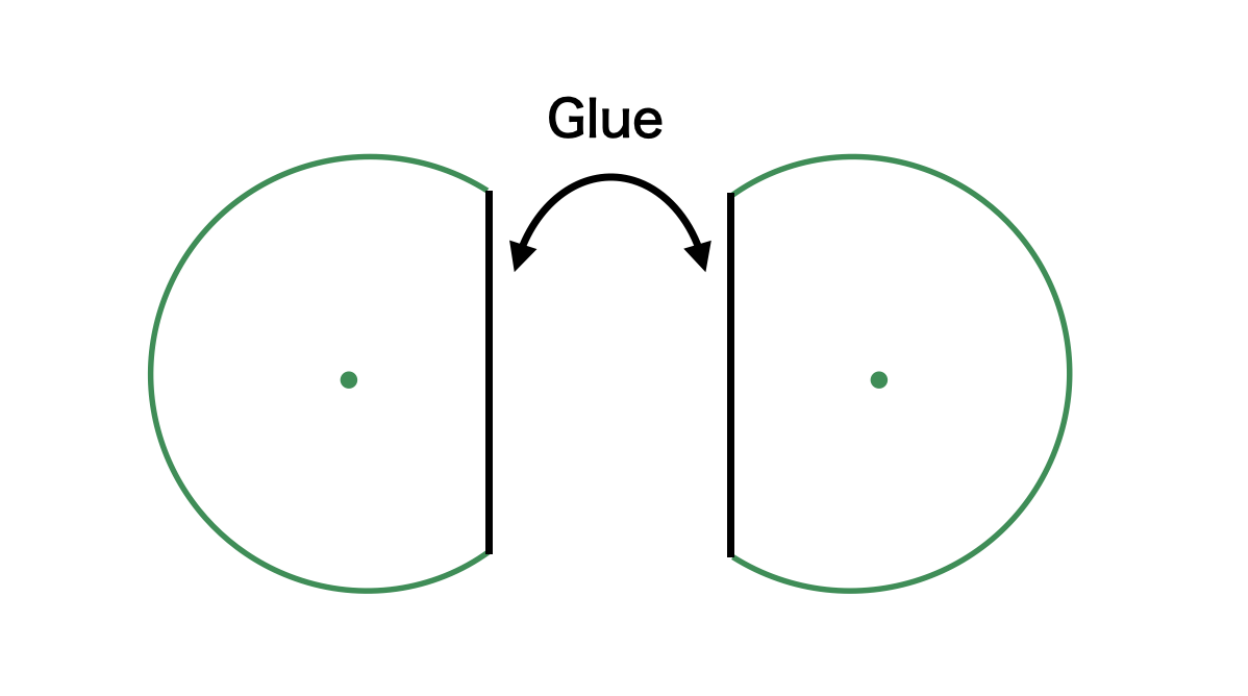}    }
  \end{minipage}
  \hspace{2em}
  \begin{minipage}[b]{0.45\linewidth}
    \centering
\includegraphics[keepaspectratio, scale=0.40]{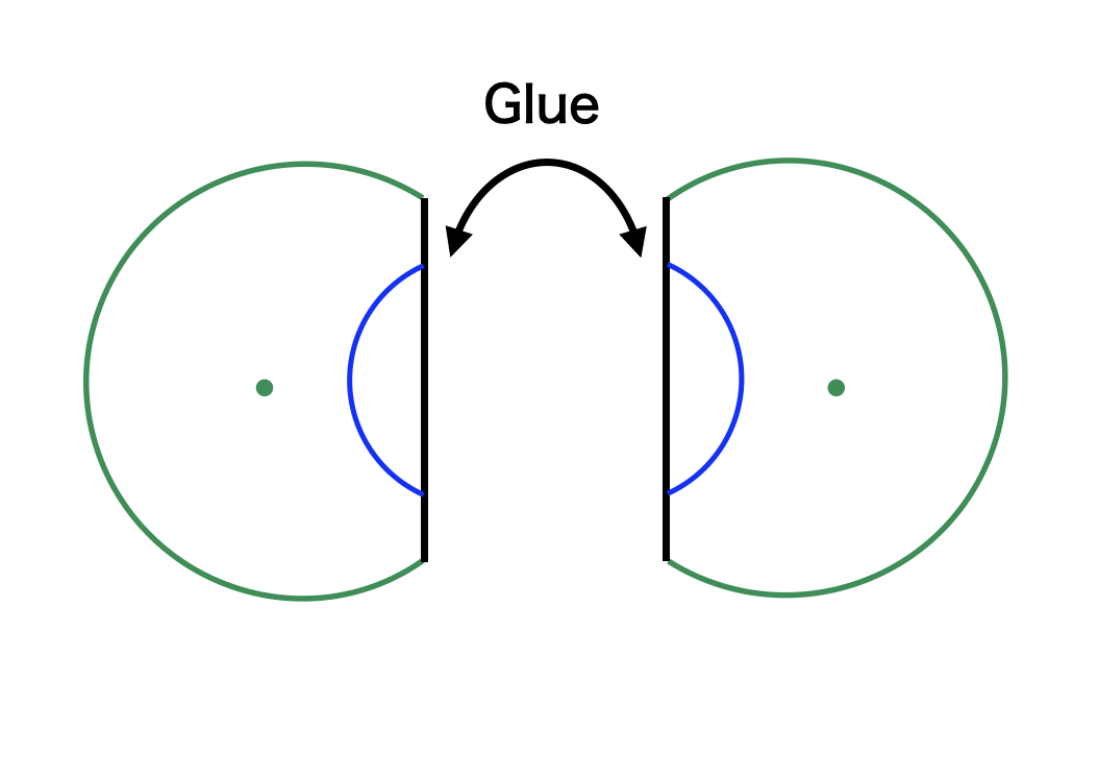} 
  \end{minipage}
\vspace{-0.7cm}
\caption{
 The construction of the de~Sitter bubble in the presence of matter backreaction, which consists of two steps.
 {\bf Left}: The first step is to construct the Euclidean AdS black hole in the presence of the matter backreaction, which is modeled by a particle propagating in the bulk.
 It is constructed by preparing two copies of a disk, introducing a particle trajectory in each of them (black lines), and gluing the two along the trajectory.
 The green dot in each disk represents the fixed point of Euclidean time translation, which after continuation to Lorentzian becomes the horizon of the black hole.
 {\bf Right}: The second step is to introduce a domain wall (the blue curve) separating the de~Sitter region and AdS in each disk. The domain wall starts and ends on the particle trajectory, satisfying the junction condition \eqref{eq:2dJTjunction}.}
  \label{fig:Wormhole3}   
\end{figure}
Choosing the form of the Euclidean AdS metric to be $ds^{2} = d\rho^{2} + \sinh^{2}\!\rho\, d\varphi^{2}$, the dilaton profile on each disk is $\Phi_{\rm AdS} = A \cosh\rho$.
The black hole horizon is located at $\rho = 0$.
The brane profile satisfies the equation
\be
  \xi^{\mu} \p_{\mu} \Phi_{\rm AdS} = \kappa_{0},
\label{eq:backreactionex}
\ee
where $\kappa_{0}$ denotes the tension of the $L$-$R$ brane, and $\xi$ is its normal vector (the normal vector of the left $\xi_{L}$ and right $\xi_{R}$ differ by a sign $\xi \equiv \xi_{R} = -\xi_{L}$).
\end{itemize}

In this construction, we need two copies of the AdS black hole because we would like to realize a de~Sitter bubble behind the black hole horizon, so the Euclidean black hole has to have two bifurcation surfaces.
(Such black holes are sometimes described as partially entangled states~\cite{Goel:2018ubv}.)
These black holes, when continued to the Lorentzian regime, have a long wormhole-like interior which can host an inflating region, as we will see below.
In the main text, we saw that in our case this first step is automatically implemented by the backreaction of the CFT stress energy tensor \eqref{eq:stresstensor}.

The Euclidean black hole with two bifurcation surfaces prepared in this way is glued with a de~Sitter dilaton profile $\Phi_{\rm dS}$ for the Euclidean de~Sitter space.
It satisfies another junction condition
\be
  \xi^{\mu} \p_{\mu} \Phi_{\rm dS} - \xi^{\mu} \p_{\mu} \Phi_{\rm AdS} = \kappa,
\label{eq:2dJTjunction}
\ee
where $\kappa$ denotes the tension of the domain wall connecting the de~Sitter region and the AdS region.
This is distinct from the tension associated with the excitation used to glue two copies of a Euclidean AdS black hole, and $\kappa$ in Eq.~\eqref{eq:2dJTjunction} is in general different from the mass of the excitation $\kappa_{0}$ in Eq.~\eqref{eq:backreactionex}.

Denoting the metric of the Euclidean de~Sitter (sphere) by $ds^{2} = d\theta^{2} + \sin^{2}\!\theta\, d\varphi^{2}$, the dilaton profile of the de~Sitter side is $\Phi_{\rm dS} = B_{0} + B \cos\theta$.%
\footnote{
 In this appendix, we include the constant part to the dilaton profiles.
}
It has two horizons, the cosmological horizon at $\theta=0$ and the black hole horizon at 
$\theta=\pi$.
Let $\theta = \theta(\varphi)$ be the brane trajectory in the de~Sitter side.
Then, the junction condition \eqref{eq:2dJTjunction} is again recast into the equation of motion of particle in one dimension
\be
  \dot{\theta}^{2} + V(\theta) = 0
\ee
with
\be
  V(\theta) = \left(\f{(B \cos\theta + B_{0})^{2} -\kappa^{2} - B^{2} \sin^{2}\!\theta}{2\kappa B \sin\theta} \right)^{2} - 1,
\quad
  A \ll B, B_{0}.
\ee
This is the two-dimensional analog of the Euclidean potential problem for higher dimensions, Eq.~\eqref{eq:EuclideanPot}.
Again, the trajectory oscillates in the bounded region $\theta_{{\rm min}} \leq \theta \leq \theta_{{\rm max}}$.

\begin{figure}
\vspace{-3mm}
\begin{center}
  \includegraphics[scale=0.32]{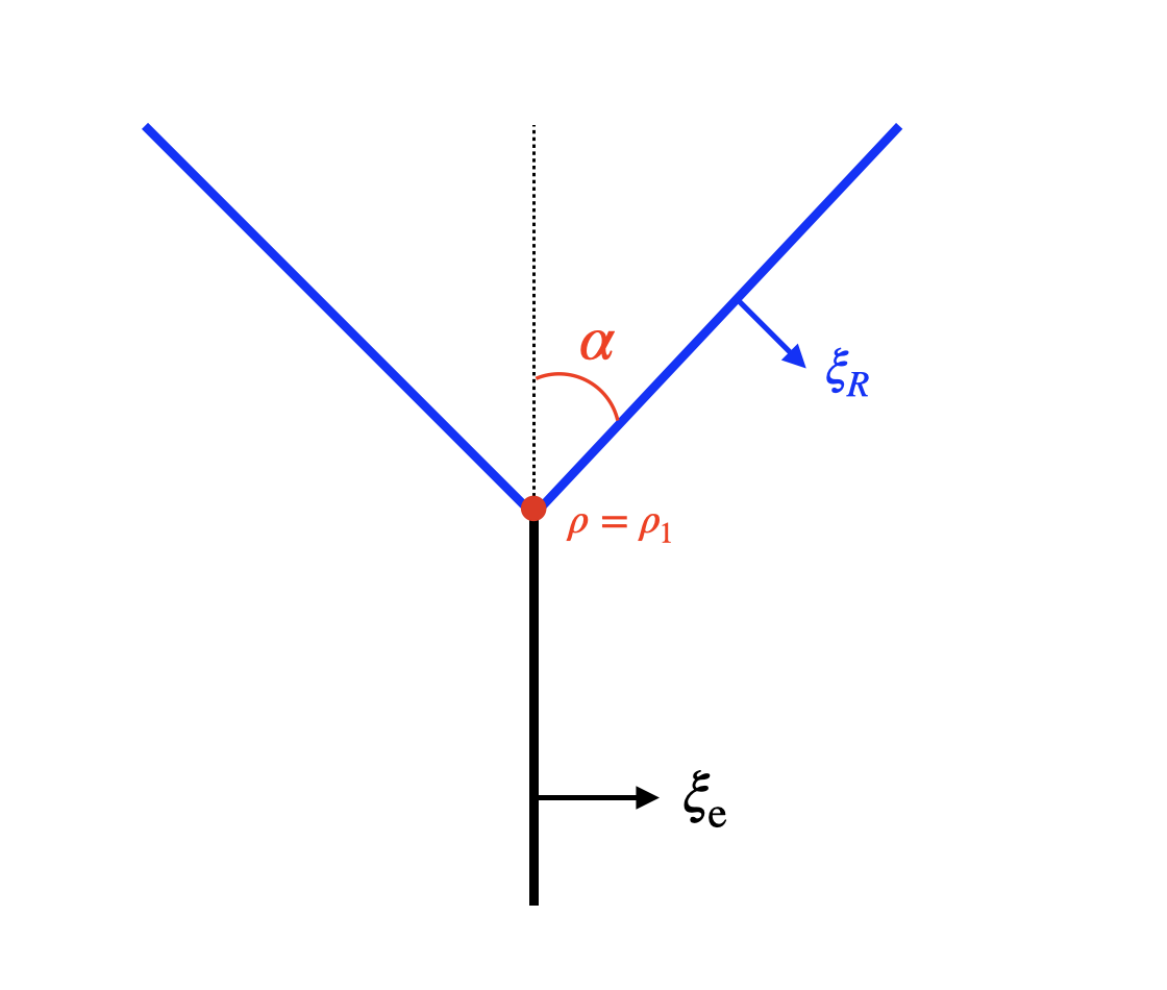}
\end{center}
\vspace{-0.6cm}
\caption{
 The branching point $\rho = \rho_{1}$ of the excitation in AdS$_2$ (black line) with the domain wall (two blue lines).
 We denote the angle between the blue and the black line by $\alpha$.} 
\label{fig:angleds}
\end{figure}

It remains to specify the location of the branching $\rho = \rho_{1}$.
Let $\xi_{R}$ be the outward point normal of the right part of the domain wall at the branching point and $\xi_{L}$ be the similar normal vector for the left part, as in Fig.~\ref{fig:angleds}.
Denoting by $\dot{\rho}_{+}$ the velocity of the domain wall in $\rho$ direction at the branching point, its normal vector is
\be
  \xi_{R} =\left( \s{1-\dot{\rho}_{+}^{2}},\; \f{ \dot{\rho}_{+}}{\sinh\rho_{1}}\right).
\ee
The normal vector for the excitation trajectory is
\be
  \xi_{{\rm e}} = \left( \f{\sinh\rho_{m}}{\sinh\rho_{1}}, \; \s{1-\left(\f{\sinh\rho_{m}}{\sinh\rho_{1}}\right)}\right),
\quad
  \sinh\rho_{m} = \f{\kappa_{0}}{2A},
\ee
where $\rho = \rho_{m}$ is the closest approach to the horizon, located at $\rho = 0$.
If we denote the angle between these two normal vectors by $\pi - \alpha$, then the angle between two normal vectors $\xi_{R}, \xi_{L}$ in the de~Sitter side is $2\alpha$.
This is because there is no conical singularity in the geometry at the branching point, a conclusion coming from the equations of motion for the dilaton profile.

This yields the following equation:
\be
  \f{\sinh\rho_{m}}{\sinh\rho_{1}} \s{1-\dot{\rho}_{+}^{2}} + \left( 1 - \f{B \sin\theta_{1}}{B\cos\theta_{1}+B_{0}}\right) \s{1 - \left(\f{\sinh\rho_{m}}{\sinh\rho_{1}}\right)^{2}} = 0,
\ee
which is solved to obtain $\theta = \theta_{1}$ (or equivalently $\rho = \rho_{1}$ in the coordinate in the AdS side).
When $\rho_{1} \gg \rho_{m}$, one can easily see that $\theta_{1} \sim \theta_{{\rm min}}$.

\bibliographystyle{JHEP}
\bibliography{deSitter}

\end{document}